\documentclass[11pt,a4paper]{article}

\usepackage[T1]{fontenc}
\usepackage[utf8]{inputenc}
\usepackage[a4paper,top=27mm,bottom=30mm,left=29mm,right=29mm,headheight=22pt,headsep=12pt,footskip=15mm]{geometry}

\usepackage[osf]{newpxtext}
\usepackage{newpxmath}
\usepackage[scaled=0.92,semibold]{sourcesanspro}
\usepackage[varqu,varl]{inconsolata}
\usepackage{microtype}
\usepackage{setspace}
\usepackage[svgnames,dvipsnames]{xcolor}
\definecolor{paperInk}{HTML}{1F2937}
\definecolor{paperMuted}{HTML}{6B7280}
\definecolor{paperAccent}{HTML}{2F5D8C}
\definecolor{paperAccentLight}{HTML}{EAF1F8}
\definecolor{paperRule}{HTML}{D7DEE8}
\definecolor{paperWatermark}{HTML}{ECEFF3}
\color{paperInk}

\usepackage{graphicx}
\graphicspath{{img/}{figures_csv/}}
\usepackage{amsmath}
\usepackage{calc}
\usepackage{booktabs,longtable,array,tabularx,dcolumn,makecell,multirow}
\usepackage{threeparttable,threeparttablex}
\usepackage{float,subcaption,caption}
\usepackage{pdflscape,rotating}
\usepackage{ragged2e}
\usepackage[gen]{eurosym}
\usepackage{comment}
\usepackage{tikz}
\usetikzlibrary{arrows.meta,positioning,fit,calc,shapes.geometric}
\usepackage{enumitem}
\setlist{itemsep=0.15em,topsep=0.35em,leftmargin=1.4em}

\usepackage{titlesec}
\titleformat{\section}
  {\Large\sffamily\bfseries\color{paperAccent}}
  {\thesection}{0.75em}{}
\titleformat{\subsection}
  {\large\sffamily\bfseries\color{paperInk}}
  {\thesubsection}{0.65em}{}
\titleformat{\subsubsection}
  {\normalsize\sffamily\bfseries\color{paperMuted}}
  {\thesubsubsection}{0.55em}{}
\titlespacing*{\section}{0pt}{2.2ex plus .6ex minus .2ex}{1.0ex}
\titlespacing*{\subsection}{0pt}{1.6ex plus .4ex minus .2ex}{0.65ex}
\titlespacing*{\subsubsection}{0pt}{1.2ex plus .3ex minus .1ex}{0.45ex}

\usepackage[font=small,labelfont={sf,bf,color=paperAccent},textfont=it,justification=raggedright,singlelinecheck=false]{caption}
\newlength{\fulllength}
\newlength{\cellWidtha}\newlength{\cellWidthb}\newlength{\cellWidthc}\newlength{\cellWidthd}\newlength{\cellWidthe}\newlength{\cellWidthf}\newlength{\cellWidthg}\newlength{\cellWidthh}\newlength{\cellWidthi}\newlength{\cellWidthj}
\newcommand{\PreserveBackslash}{\let\\\tabularnewline}

\usepackage[round,authoryear]{natbib}
\usepackage[hyphens]{url}
\usepackage{hyperref}
\hypersetup{
  colorlinks=true,
  citecolor=paperAccent,
  urlcolor=paperAccent,
  linkcolor=paperAccent,
  pdfborder={0 0 0}
}

\usepackage{fancyhdr}
\renewcommand{\headrulewidth}{0.25pt}
\renewcommand{\headrule}{\hbox to\headwidth{\color{paperRule}\leaders\hrule height \headrulewidth\hfill}}
\fancypagestyle{plain}{%
  \fancyhf{}%
  \fancyfoot[C]{\sffamily\footnotesize\color{paperMuted}\thepage}%
  \renewcommand{\headrulewidth}{0pt}%
}

\makeatletter
\renewcommand{\@makefnmark}{\hbox{\textsuperscript{\normalfont\rotatebox{180}{\ensuremath{\blacktriangle}}}}}
\renewcommand{\@makefntext}[1]{\noindent\hbox{\textsuperscript{\normalfont\rotatebox{180}{\ensuremath{\blacktriangle}}}}\, #1}
\makeatother

\newcommand{\AuthorNames}[1]{}
\newcommand{\Title}[1]{}
\newcommand{\Author}[1]{}
\newcommand{\address}[1]{}
\newcommand{\corres}[1]{}
\newcommand{\firstpage}[1]{}
\newcommand{\pubvolume}[1]{}
\newcommand{\issuenum}[1]{}
\newcommand{\articlenumber}[1]{}
\newcommand{\pubyear}[1]{}
\newcommand{\copyrightyear}[1]{}
\newcommand{\externaleditor}[1]{}
\newcommand{\datereceived}[1]{}
\newcommand{\daterevised}[1]{}
\newcommand{\dateaccepted}[1]{}
\newcommand{\datepublished}[1]{}

\title{\sffamily\bfseries\color{paperInk}Price Pass-Through of Austria's Single-Use Plastics Producer Charges:\ Evidence from Retail Offer Spells}
\author{\sffamily Felix Reichel\\[-0.15em]
\small\color{paperMuted}Lund University School of Economics and Management; Johannes Kepler University Linz; University of Essex\\[-0.15em]
\small\href{mailto:kontakt@felixreichel.com}{kontakt@felixreichel.com}}
\date{\sffamily\color{paperMuted}TBP Reg. Sci. Env. Econ. 3, 2026}

\begin{document}
\maketitle
\vspace{-1.5em}
\begin{center}\color{paperRule}\rule{0.84\textwidth}{0.5pt}\end{center}
\vspace{0.2em}
\begin{abstract}
\small
Single-use plastics (SUPs) impose substantial environmental costs.
Following Directive (EU) 2019/904, Austria introduced producer charges
and mandatory participation in collection and recycling systems. This
paper exploits a monthly aggregated and disaggregated panel of retail
offer spells drawn from a price-comparison platform to estimate the
extent to which compliance costs pass through to posted online prices
in Austria. The treated sample comprises keyword-matched SUP
products---balloons, to-go cups, wet wipes, plastic bags, food
containers, tobacco-filter items, beverage bottles, and plastic
wraps---observed alongside a control group of non-SUP listings over
2020--2024. A two-way fixed-effects (TWFE) specification places the
average post-treatment price increase at approximately 4.1 percent. A
sequential TWFE model that disaggregates the administrative reporting
phase (from March 2023) from the payment-due phase (from March 2024)
reveals that the larger adjustment occurs during the earlier reporting
stage, with a reporting-only effect of approximately 8.1 percent and
an incremental payment-phase effect of 5.6 percent. For
balloons---a category subject to pronounced regulatory fee
exposure---event-study estimates exceed 50 percent in the months
immediately following the initial payment date and remain elevated
throughout most of the post-treatment window. Taken together, these
findings indicate that Austrian online retailers began adjusting prices
in advance of fee-payment deadlines, a pattern consistent with
anticipatory pass-through of expected compliance costs rather than a
discrete response to realized payments. As the data contain price
observations but not quantity data, the analysis speaks to price
incidence and does not extend to consumption or environmental
outcomes.
\end{abstract}
\normalsize



\bigskip

\section{Introduction}
\label{sec:intro}

Single-use plastics (SUPs) litter beaches, waterways, and public spaces extensively. On EU beaches alone, plastics account for around 80--85 percent of collected litter items, about half of which are single-use products \citep{eu2019sup}. Street cleaning and waste collection tied to plastic debris cost municipalities billions of euros annually---costs that, absent regulation, are largely absorbed by the public purse. Environmental policy redirects part of that burden back to producers and consumers through charges, reporting obligations, and extended producer responsibility (EPR), pushing market prices closer to social costs \citep{fullerton2001environmental}.

Whether these upstream charges actually reach the shelf is an open empirical question. Under imperfect competition, pass-through depends on demand curvature, cost shares, and market conduct \citep{weyl2013passthrough}. Existing evidence on sales and excise taxes suggests pass-through to retail prices is often substantial \citep{marion2011fueltax,besley1999salestax}, and even modest levies on plastic bags have produced large consumption responses \citep{convery2007irish,homonoff2018bags}. Deposit--refund systems form a related instrument aimed at curbing litter and improving collection rates \citep{walls2011drs}. What remains less understood is how compliance-based EPR regimes---combining reporting obligations, system-participation requirements, and per-quantity fee settlement---affect the prices of regulated products in online retail markets.

This study addresses that question for Austria. Following Directive (EU) 2019/904, Austria introduced producer charges tied to reported quantities, with fees of \euro225 per ton for food packaging, beverage cups, and plastic wraps, and \euro450 per ton for tobacco products with plastic filters \citep{bmk2024}. Producers report volumes through the ARA Online platform; the compliance sequence requires reporting of prior-year quantities by 15 March and fee settlement through participation in licensed collection and recycling systems. March 2024 is therefore the treatment date used throughout this work, marking the first instance at which a reporting year translates into an actual payment obligation.

The empirical strategy uses a disaggregated and duration-weighted monthly panel of retail offer spells from a price-comparison platform covering both stand-alone e-tailers and third-party marketplace sellers, including those operating through Amazon and eBay. The observation window runs from January 2020 through end-2024, with a small number of spells extending into early 2025. The treated sample is defined by keyword matching to SUP-relevant product titles---balloons, to-go cups, wet wipes, plastic bags, food containers, tobacco-filter items, beverage bottles, and plastic wraps. The control group consists of non-SUP items belonging to the same categories in the product tree and, alternatively, non-related graphics card listings as a stress-test benchmark, all of which were active at the start of the sample window. Two baseline specifications are estimated: a standard two-way fixed-effects (TWFE) model and a dynamic event study, both supplemented by a sequential TWFE that separates the reporting phase from the payment phase.

The main findings are threefold. First, the pooled TWFE estimate places the post-payment price increase at approximately 4.1 percent---economically modest but statistically precise. Second, the sequential model shows that this average understates the timing pattern: the report-only phase coefficient is approximately 8.1 percent, exceeding the 5.6 percent incremental effect of the payment phase, which suggests that sellers began repricing before any fee was actually due, consistent with anticipatory pass-through of expected compliance costs. Third, in the balloon category---a product group with high regulatory exposure either through input material composition or the applicable SUP fee schedule---event-study coefficients exceed 50 percent in the months immediately following the first payment date and remain elevated through most of the post-treatment window, pointing to large and persistent pass-through in highly exposed segments. Negative pre-treatment coefficients in this category likely reflect the influence of lagged energy-price shocks rather than genuine pre-trend divergence.

These results contribute to several strands from the literature. They add to the growing body of evidence on the incidence of environmental charges and EPR systems \citep{fullerton2001environmental,weyl2013passthrough,convery2007irish,homonoff2018bags,walls2011drs}, using high-frequency retail data rather than aggregate price indices such as consumer price indices (CPIs). They also speak to the behavioral economics of regulatory compliance: the finding that price adjustment precedes the first payment date is consistent with anticipatory firm behavior in response to legal and administrative onset, rather than realized costs alone. Finally, they provide early evidence on the Austrian SUP regime, complementing the related analysis in \citet{reichel2025sup}, who studies price responses in both Austria and Germany using a constructed price-index approach and a broader product control group, and finds pooled Austrian difference-in-differences contrasts of approximately 13 index points over a twelve-month post-policy window.

Several limitations warrant acknowledgment at the outset. The graphics card control group is conceptually distant from the treated SUP products, which means the parallel-trend assumption rests on the capacity of unit and time fixed effects to absorb differential trends and on the absence of broad substitution effects between the two groups. The data contain prices but not quantities, so the analysis speaks to price incidence and not to consumption, substitution, or environmental outcomes. The 2022--2023 European energy-price shock likely affected SUP-intensive and packaging-intensive supply chains differently, introducing a source of confounding that cannot be fully addressed through lagged input-price controls owing to collinearity with the treatment timing.

The remainder of the paper is organized as follows. Section \ref{sec:literature} reviews the relevant literature. Section \ref{sec:background} describes the EU and Austrian regulatory background. Section \ref{sec:data} presents the data. Section \ref{sec:empirical} sets out the empirical strategy and reports the results using raw offer spells. Section \ref{sec:additional} presents additional analysis from an aggregated panel, and Section \ref{sec:conclusion} concludes.

\section{Literature Review}
\label{sec:literature}
\unskip

\subsection{Pass-Through of Taxes and Environmental Charges}

The incidence and pass-through of commodity taxes has a long theoretical and empirical tradition. Under competitive markets with constant marginal costs, a per-unit tax is fully passed forward to consumers. Under imperfect competition, the pass-through rate depends on the curvature of demand and the conduct of firms, and can exceed 100 percent or fall well below it \citep{weyl2013passthrough}. Empirical estimates from fuel-tax and sales-tax settings confirm that pass-through to retail prices is often substantial, though it varies across market structures, product types, and tax designs \citep{marion2011fueltax,besley1999salestax}.

For environmental policy instruments specifically, the evidence points in the same direction. \citet{convery2007irish} study the Irish plastic bag levy---at the time a rare example of a direct charge on a single-use item at point of sale---and document a sharp reduction in consumption following its introduction, implying meaningful price transmission and a corresponding demand response. \citet{homonoff2018bags} exploits variation in bag-fee adoption across Washington, D.C. neighborhoods to show that even modest charges of five cents substantially reduce disposable bag use, consistent with a combination of price sensitivity and salience effects. \citet{walls2011drs} surveys the literature on deposit--refund systems, which combine a charge at purchase with a rebate at return, and finds that such instruments improve collection rates and reduce litter when the deposit is sufficiently salient. Taken together, this body of evidence establishes that consumer-facing charges on plastic and disposable products can generate large behavioral responses, but leaves open the question of how producer-side EPR charges---which are not directly visible at point of sale---transmit to retail prices.

\subsection{Extended Producer Responsibility and EPR Incidence}

EPR systems shift end-of-life costs for packaging and other products to producers, who typically discharge their obligations through financial participation in licensed collection and recycling schemes \citep{eu2019sup}. Unlike a simple per-unit tax, EPR charges are levied annually on reported quantities, entail administrative and compliance costs beyond the statutory fee, and may create incentives to reduce packaging intensity over time. The incidence of EPR charges is therefore not mechanically determined by the fee rate alone; it depends on how firms incorporate compliance costs into pricing decisions, on the degree of competition in affected markets, and on the transparency of the charge to downstream buyers.

Empirical evidence on EPR incidence at the retail level remains relatively sparse. Most studies focus on aggregate recycling rates, collection system performance, or per-ton cost estimates rather than on retail price outcomes \citep{walls2011drs}. The closest antecedent to the present study is \citet{reichel2025sup}, who estimates price responses to Austrian and German SUP producer charges using a constructed centered price-index event-study design with product and retailer fixed effects. That study reports a pooled Austrian difference-in-differences contrast of approximately 13 index points over a twelve-month post-policy window and approximately 20 index points over the full sample, with balloons exhibiting particularly strong and persistent effects. The present study builds on that evidence using a TWFE panel design with disaggregated and aggregated monthly log prices, a sequential treatment structure that separates the reporting phase from the payment phase, and a tighter focus on the Austrian compliance timeline. It also employs a more narrowly defined control group, matching treated SUP listings to non-SUP products within the same product-tree categories.

\subsection{Behavioral Responses to Compliance Obligations}

A growing literature in behavioral economics and public finance examines how firms respond to regulatory obligations beyond the direct cost of compliance. The present study contributes to this literature by documenting that price adjustment in the Austrian data appears to begin during the administrative reporting phase---before any fee becomes financially due---consistent with anticipatory repricing in response to expected future liabilities. This pattern is related to the broader literature on tax anticipation effects, which demonstrates that firms and households frequently adjust behavior in advance of statutory changes when future obligations are foreseeable \citep{roth_etal_2023}. It also connects to the economics of regulatory salience: once producers are required to track and report quantities, compliance costs become salient prior to financial settlement, which may be sufficient to trigger upward price adjustment even in the absence of realized payments.

\section{Background}
\label{sec:background}
\unskip

\subsection{The EU SUP Directive}

Directive (EU) 2019/904 on the reduction of the impact of certain plastic products on the environment entered into force in June 2019 and required member states to transpose it into national law by July 2021 \citep{eu2019sup}. The Directive targets ten categories of single-use plastic items that account for a disproportionate share of marine litter, including food containers, beverage cups, cutlery, plates, straws, cotton bud sticks, balloons, and cigarette filters with plastic components. Its instruments include outright bans on certain items, mandatory labeling requirements, consumption reduction targets, and EPR obligations requiring producers to finance litter clean-up, awareness campaigns, and waste collection infrastructure. Member states retained latitude over the precise design of their EPR systems; Austria and Germany both adopted producer-charge mechanisms linked to reported quantities.

\subsection{Austria: Producer Charges and the ARA System}

Austria's transposition of the Directive took effect through the packaging ordinance and the single-use plastics regime administered by the Bundesministerium f\"{u}r Klimaschutz \citep{bmk2024}. The compliance architecture is as follows. From 1 January 2023, producers and importers of SUP items covered by the Directive are required to participate in a licensed collection and recycling system---in practice, primarily through Altstoff Recycling Austria AG (ARA), the dominant licensing body \citep{ara_licensing_2026}. Participation entails annual reporting of quantities placed on the Austrian market and payment of fees to the system operator, which finances collection, sorting, and reporting infrastructure.

The fee structure is product-specific. Food packaging, beverage cups, and plastic wraps carry a charge of \euro225 per ton, while tobacco products containing plastic filters face a higher charge of \euro450 per ton, reflecting greater associated clean-up costs \citep{bmk2024}. Micro-enterprises falling below defined quantity thresholds may opt for a flat payment of \euro13. Quantities placed on the market in a given calendar year must be reported to the system by 15 March of the following year, with fee settlement proceeding from that report within the compliance calendar. March 2024 therefore constitutes the first date on which quantities placed on the market in 2023 generate a concrete financial obligation, and serves as the primary treatment date throughout this work. Reporting obligations applied from March 2023 for the 2022 reference year, but the corresponding payment was smaller or transitional for many firms still completing system registration; March 2024 represents the first full compliance cycle \citep{wko_packaging_2026,uba2024}.

The digital reporting infrastructure is managed through the ARA Online platform, through which producers register, submit quantity data, and receive fee assessments \citep{ara_licensing_2026}. The existence of a structured reporting portal rendered compliance costs administratively salient from the first reporting period, prior to the first full payment obligation falling due. This institutional feature is consequential for interpreting the sequential TWFE results: the report-only phase from March 2023 through February 2024 captures a period during which producers were tracking and declaring quantities without yet bearing the full payment obligation for the 2023 reference year.

\subsection{The Macroeconomic Context}

The observation window from 2020 to 2024 encompasses several overlapping macroeconomic shocks of relevance to identification. The COVID-19 pandemic disrupted supply chains and suppressed consumer demand throughout 2020 and 2021, producing broad price movements across retail categories. From early 2022, the war in Ukraine generated a sharp energy and commodity price shock that raised production costs for plastics, packaging materials, and logistics across Europe \citep{packagingeurope2022commodity,plasticseurope2022energy}. These cost pressures likely bore more heavily on SUP-intensive and packaging-intensive retail segments than on less input-sensitive product groups, constituting a potential source of confounding: a divergence in prices between treated and control products that predates the payment date and may persist into the post-treatment period. The empirical strategy addresses this partially through a restricted sample window and through specifications that incorporate lagged Brent crude prices, the German energy import index, and Austrian gas prices as reduced-form cost shifters. These controls are nonetheless collinear with the regulatory timeline, so the absorption of energy-price confounding remains imperfect in the absence of granular product-level data on SUP input shares and fee exposure.

A further potential source of confounding is the growth in public awareness of the environmental costs of single-use plastics and, more recently, of per- and polyfluoroalkyl substances (PFAS) associated with certain plastic packaging materials. Heightened media attention to these issues over the observation window may have shifted consumer demand for affected product categories independently of the regulatory charge, a confound that cannot be separated from the compliance effect using the available offer-spell data.

\section{Data}
\label{sec:data}

The main analysis draws on high-frequency online offer spells from a
price-comparison platform covering the Austrian e-commerce market. Each offer spell
records a continuous listing of a product by a specific retailer, together with a
start and end timestamp, an observed price, and a set of retailer and product
attributes. The raw data are irregular in time because listings enter and exit the
platform at different moments and may remain active for varying durations. For the
additional analysis specifications in Section \ref{sec:additional}, these raw spells
are converted into a duration-weighted monthly panel in which the unit of observation
is a retailer--product pair observed in a given calendar month. The estimation window
covers January 2020 through December 2024.

\subsection{Treated Sample Construction}

\begin{figure}[H]
\hspace{-4pt}
  \includegraphics[width=\linewidth]{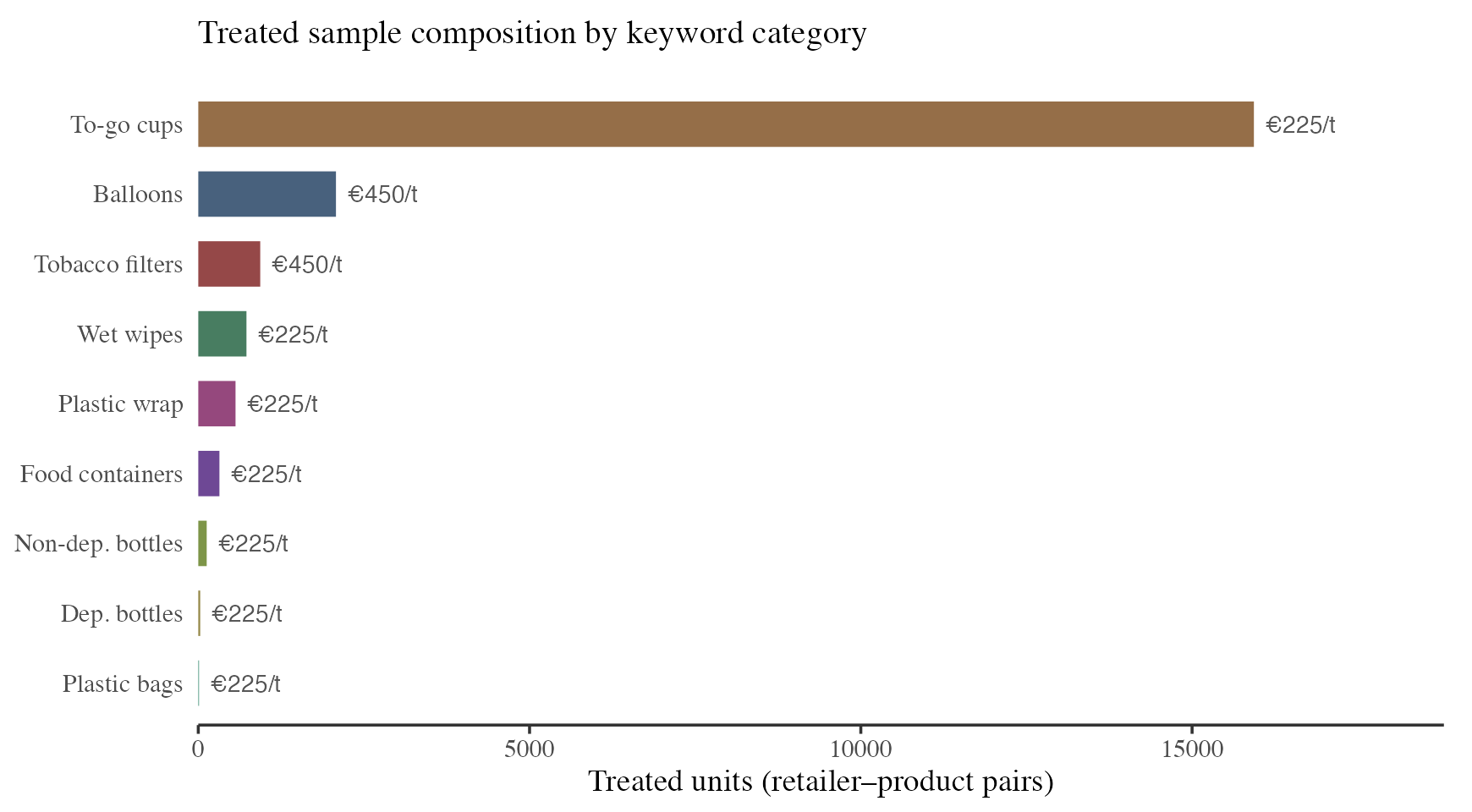}
  \caption{Composition 
of the treated sample by keyword category, with Austrian SUP
  fee tiers indicated. To-go cups dominate by unit count; balloons and tobacco
  filters constitute the next-largest categories. Categories differ in expected
  regulatory intensity, so any pooled treatment coefficient is averaged over groups
  that vary in regulatory exposure, baseline sample size, typical price level, and likely pass-through capacity.}
  \label{fig:data_category_composition}
\end{figure}

The treated sample is defined by keyword matching on product titles. Matching uses
case-insensitive string patterns designed to capture product groups plausibly exposed
to the Austrian SUP compliance regime. Table \ref{tab:treated_keywords_appendix}
lists the nine keyword categories, their German labels, and the underlying matching
patterns. The treated categories are balloons (\textit{luftballons}), 
to-go cups
(\textit{becher}), food containers (\textit{lebensmittelbehaelter}), plastic wraps
and films (\textit{tueten\_folien}), non-deposit beverage bottles
(\textit{flaschen\_ohne\_pfand}), deposit and reusable bottles
(\textit{flaschen\_mit\_pfand}), plastic carrier bags (\textit{plastiktueten}), wet
wipes (\textit{feuchttuecher}), and tobacco-filter products (\textit{tabak}). A product enters the treated sample if its title matches at least one pattern in any
of these categories. The composition of the treated sample is depicted in Figure \ref{fig:data_category_composition}.

\vspace{-6pt}
\unskip

\subsection{Control Sample Construction and the Baseline-Survivor Restriction}
\label{subsec:control}

\noindent{Intended design.
}
\vspace{3pt}

The ideal control group for this design consists of products that (i) are sold on
the same platform by the same retailers, (ii) face similar upstream cost dynamics
and demand seasonality, and (iii) fall outside the scope of the SUP regime. Let
$M_i^{C,0}\in\{0,1\}$ 
denote membership in this originally intended counterfactual.
The guiding principle was to draw the comparison group from products at the same
level of the platform's category tree as the treated items---direct substitutes or
functional complements subject to substitution effects once the SUP charge raises the
relative price of plastic alternatives---made from reusable, natural, or otherwise
non-regulated materials. This class includes reusable versions of the treated
products (stainless-steel bottles, glass food containers, textile carrier bags),
products from natural or non-plastic materials (paper cups, cardboard food trays,
wooden cutlery), and non-SUP items whose demand is plausibly linked to treated
products through substitution effects. Table \ref{tab:counterfactual_keywords_appendix}
lists the nine intended counterfactual keyword categories, their translations, and the underlying matching patterns. Had this design been implemented in full, the control group would have held fixed at least part of the retail environment in which
treated products are sold.

\vspace{3pt}\noindent{Implemented design.}\vspace{3pt}

In the platform data, the intended alternatives are sparsely listed and unevenly
matched to treated products, leaving a control group too thin for panel
fixed-effects estimation with adequate pre-period coverage. The implemented control
sample is therefore drawn from all non-SUP products in the platform data that
satisfy the life-cycle restrictions below and do not belong to the treated keyword
categories. Let $M_i^{C}\in\{0,1\}$ denote the implemented counterfactual-match
indicator. This broader control pool retains the category-tree logic of the intended
design where the data permit: it includes the natural-material and reusable
alternatives of Table \ref{tab:counterfactual_keywords_appendix} wherever spell
coverage is sufficient, and fills the comparison group with the remaining non-SUP
listings on the platform.

A separate stress-test specification restricts the control group to a single
keyword, \textit{grafikkarte} (graphics card), with $M_i^{C}=1$ if and only if the
product title matches this rule and satisfies the life-cycle restrictions below.
Graphics cards share the same platform infrastructure and price-reporting conventions
as the treated products but are driven by entirely different cost factors---semiconductor
cycles and, during part of the sample, cryptocurrency-related demand. If an effect
is detectable against this economically distant counterfactual, it is unlikely to
reflect correlated cost shocks between treated and control products. The main
estimates employ the broader non-SUP pool as well as category-matched listings in
the duration-weighted monthly aggregated panel analyzed in
Section \ref{sec:additional}.

\vspace{3pt}\noindent{Baseline-survivor restriction.}\vspace{3pt}

The control sample is further restricted by a baseline-survivor filter. Let $b_i$
denote the product birth timestamp, $d_i$ the death timestamp, and $t_0^{\text{base}} = 1{,}577{,}836{,}800$ the Unix time for
1 January 2020 00:00:00 UTC. The implemented control set is
\begin{equation}
  \mathcal{C}
  =
  \bigl\{
    i : M_i^{C}=1,\;
    b_i \leq t_0^{\text{base}},\;
    (d_i=\varnothing \;\text{or}\; d_i > t_0^{\text{base}})
  \bigr\},
\end{equation}
so that only products already active at the opening of the sample window are
retained. All post-2020 entrants and units that exited before January 2020 are
excluded by construction.

\vspace{3pt}\noindent{Implications for panel composition and identification.}\vspace{3pt}

\begin{figure}[H]
\hspace{-6pt}
  \includegraphics[width=\linewidth]{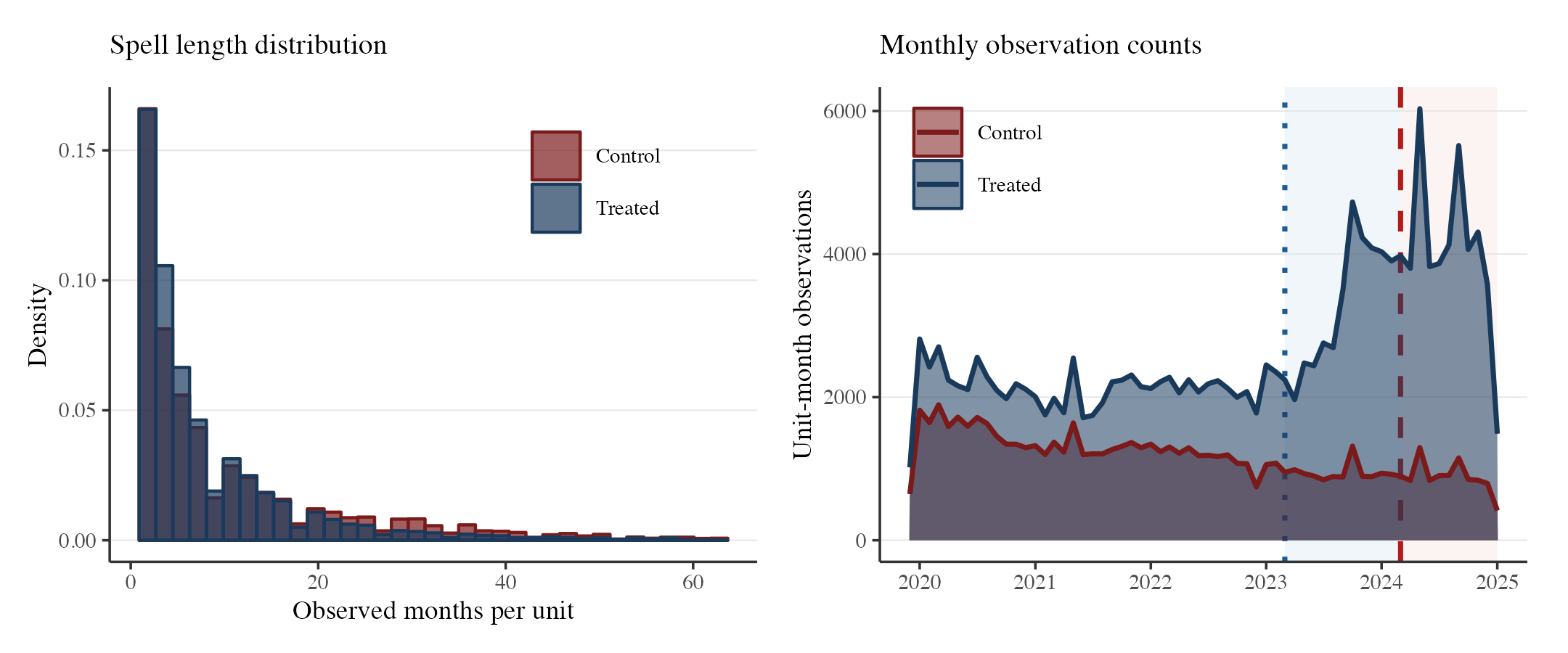}
  \caption{\textbf{Left panel}: 
Distribution of spell durations by treatment status. Both
  groups are strongly right-skewed; most units are observed for a small number of
  months while a long tail remains active throughout the window. \textbf{Right panel}:
  Monthly observation counts by treatment status over January 2020--December 2024.
  Treated counts rise sharply from 2023 onward while the control series remains
  smaller and more stable. Vertical dashed lines mark the reporting onset (blue)
  and payment onset (red).}
  \label{fig:data_spell_counts_text}
\end{figure}

Because $\mathcal{C}$ is conditioned on survival at $t_0^{\text{base}}$, the retained control sample mechanically over-represents older and more persistent
products while under-representing short-lived listings, later entrants, and products
with intermittent market presence. Formally, the observed control is not drawn from
the full untreated population $\mathcal{U}$ but from the selected subset
$\mathcal{C}\subset\mathcal{U}$ satisfying the survival condition at baseline. Three
consequences follow: the panel is tilted toward incumbent products; if turnover
patterns differ between treated and untreated goods, the restriction induces
differential sample selection unrelated to the regulation; and the final panel
represents a survivor cohort defined at baseline rather than the full market over
2020--2024. The final sample offer spell length distributions and observational unit counts are depicted in Figure \ref{fig:data_spell_counts_text}.

This is not random attrition but systematic conditioning on survival at the start of
the window. If survival correlates with price levels, retailer persistence,
assortment quality, or product type, treatment--control comparability may be
affected. The design remains valid, but the estimand is narrower: the treatment
effect should be interpreted relative to a specific untreated survivor cohort rather
than the full population of available untreated goods. The parallel-trend
diagnostics in Section \ref{subsec:pt} should be read with both the composition of
the control pool and this survival conditioning in mind.

\subsection{Summary Statistics and Price Distributions}

Given the economic distance between treated SUP products and the graphics card control documented in Section \ref{subsec:control}, cross-group-level comparisons carry no claim of
cross-group comparability; they serve only to characterize each group's price
distribution before and after policy onset.

\begin{figure}[H]
\hspace{-8pt}
  \includegraphics[width=\linewidth]{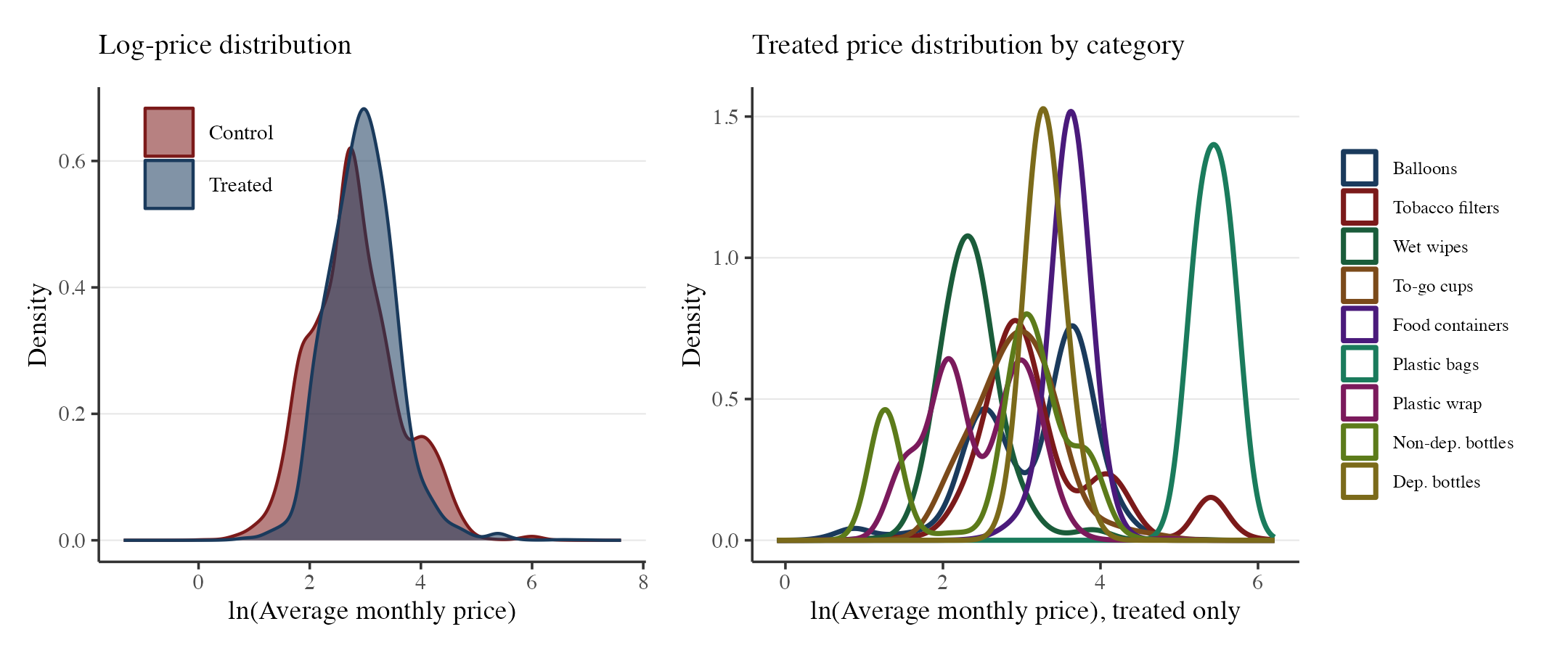}
  \caption{\textbf{Left panel}: 
Pooled log-price densities for treated and control groups.
  The two distributions overlap substantially; the treated density is shifted
  modestly rightward. \textbf{Right panel}: Log-price densities by treated category,
  illustrating pronounced within-treated heterogeneity in both central tendency
  and dispersion.}
  \label{fig:data_price_distributions_text}
\end{figure}

Table \ref{tab:spell_sumstats_appendix} reports the main figures. The treated products
have a mean monthly price of EUR 24.99 and a median of EUR 18.95; the control group
averages EUR 24.38 with a median of EUR 15.64. Average treated prices rise from
EUR 23.95 in the pre-policy period to EUR 26.86 in the post-payment period, while
control prices shift only from EUR 24.65 to EUR 24.71. Dispersion is large in both
groups. Figure 
\ref{fig:data_price_distributions_text} shows the log-price
densities: the two distributions overlap substantially but the treated density is
shifted modestly rightward. Within the treated sample, categories differ visibly in
both central tendency and dispersion, so pooled regression estimates average over
products with different baseline price levels and potentially different regulatory
incidence.

\subsection{Category-Level Descriptive Dynamics}

\begin{figure}[H]
\hspace{-3pt}
  \includegraphics[width=\linewidth]{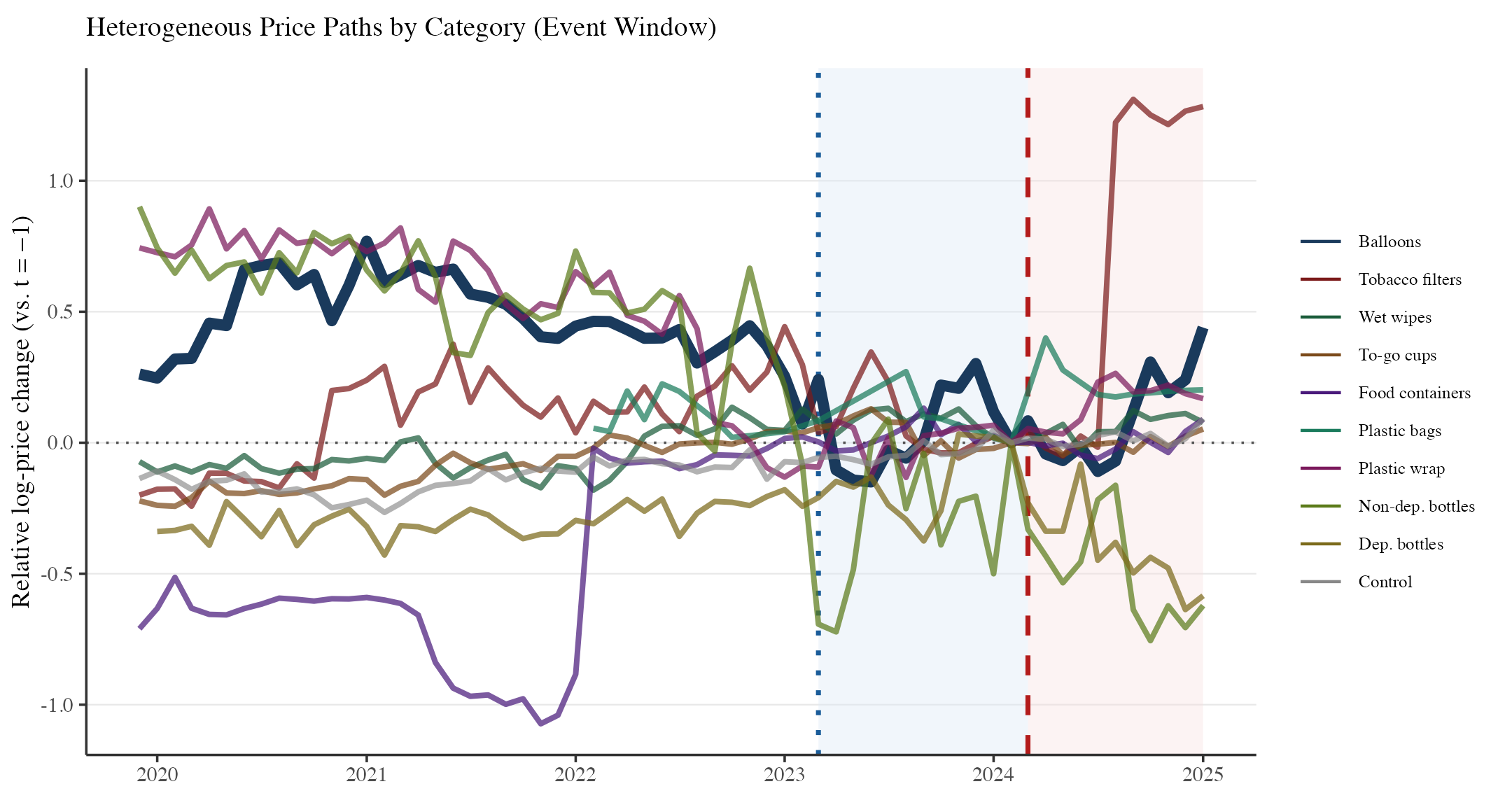}
  \caption{Relative 
log-price paths by treated category, normalized to zero at
  $t = -1$. The gray path denotes the control group. The blue dotted line and
  shaded region mark the reporting phase (from March 2023); the red dashed line
  and shaded region mark the payment phase (from March 2024). Category-level
  responses are heterogeneous: tobacco filters and balloons exhibit the largest
  upward deviations, while bottles and deposited bottles trend below the
  control path throughout the sample of raw offer spells.}
  \label{fig:results_category_paths_raw_text}
\end{figure}

Figure \ref{fig:results_category_paths_raw_text} plots relative log-price paths by
treated category in event time, normalized to zero at $t=-1$, with the control path
included for reference. The trajectories are heterogeneous: some categories remain
close to baseline around both policy thresholds, while others exhibit larger
deviations in the post-payment period. A modest average treatment effect may
therefore coexist with large category-specific responses if strongly affected
categories are offset by weakly affected ones. This motivates the sequential
specifications that separate the reporting phase from the payment phase and the
category-level analyses reported in subsequent sections.

\subsection{Parallel-Trend Diagnostics}
\label{subsec:pt}

\begin{figure}[H]
\hspace{-5pt}
  \includegraphics[width=\linewidth]{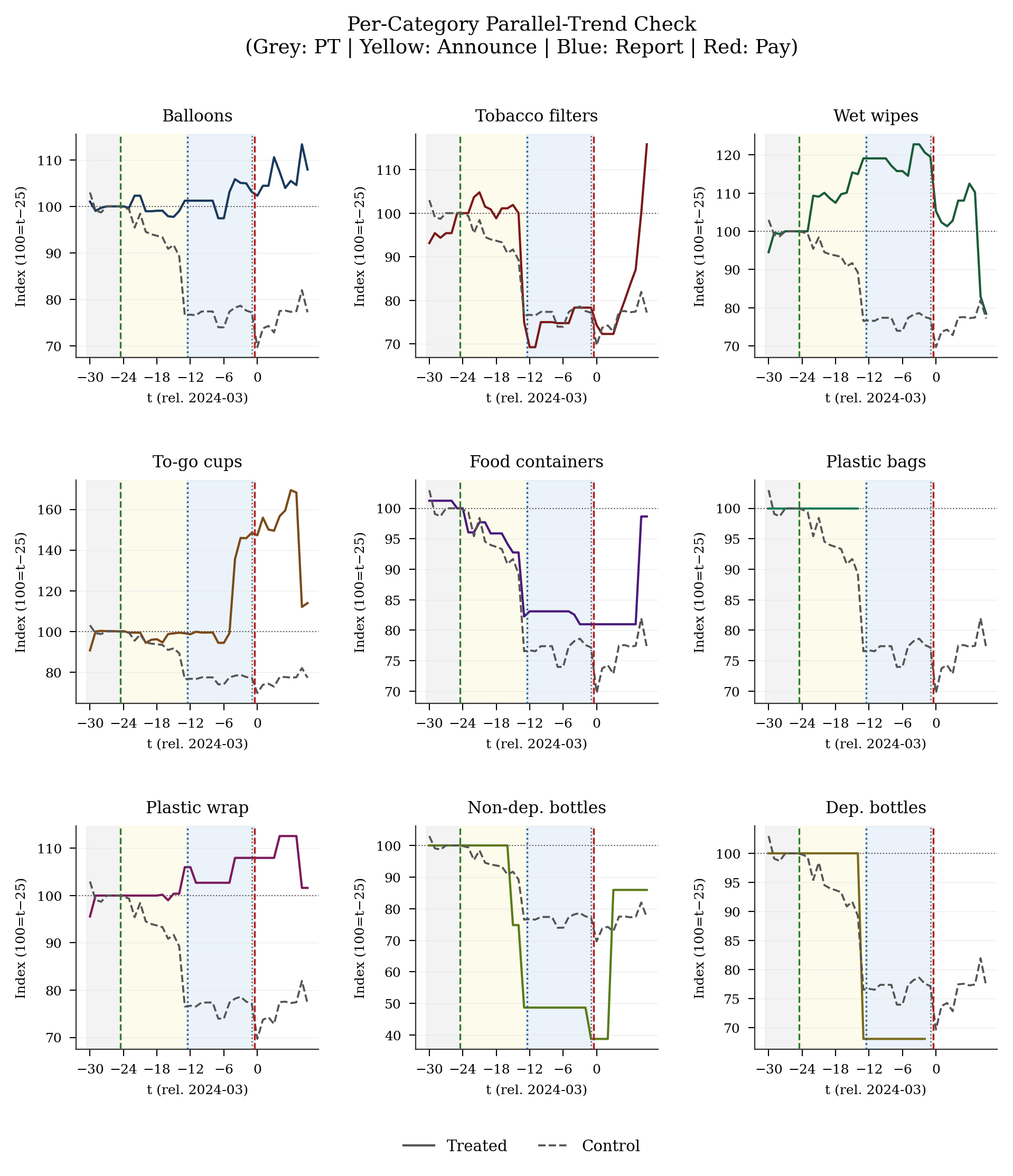}
  \caption{Per-category parallel-trend diagnostic using monthly aggregated and duration-weighted offer spells. Each panel plots the treated
  (solid) and matched-control (dashed) price index series, normalized
  to 100 at $t = -7$. The gray shaded 
 region marks the parallel-trend assessment
  window ($t = -30$ to $t = -25$); the yellow shaded region marks the first
  reporting year; the blue shaded region marks the payment anticipation phase
  leading to the payment onset at $t = 0$ (March 2024). The dashed/dotted horizontal lines mark the end of each time region, respectively. The horizontal axis
  measures months relative to the payment onset. Parallel-trend support is
  assessed over the gray window and alternatively the first six months of the
  reporting phase. Support is strongest for balloons, food containers, and plastic
  bags, where treated and control series co-move closely across both intervals.
  To-go cups, tobacco filters, and deposited bottles display reasonable pre-period
  tracking but exhibit moderate divergence entering the reporting phase. Wet wipes,
  plastic wrap, and non-deposited bottles show persistent and substantial gaps
  between treated and control series across both windows, providing little basis
  for a credible parallel-trends assumption in those categories.}
  \label{fig:parallel_trends_by_category_text}
\end{figure}

Figure \ref{fig:parallel_trends_by_category_text} provides a category-by-category
pre-trend diagnostic using a duration-weighted and aggregated panel of raw offer spells to remove the influence of extreme price outliers and provide some smoothing, indexing treated and matched-control series to 100 at $t=-7$.
Support for parallel trends is uneven: in some categories the two series track
closely over the pre-period; in others, divergence appears well before payment
onset. These diagnostics do not invalidate the design, but they imply that
credibility varies across product groups and that pooled estimates should be
interpreted as reduced-form averages over categories with different degrees of
pre-period comparability.

\vspace{-9pt}

\vspace{-12pt}

\subsection{Covariates and Energy Input-Price Controls}
\label{subsec:covariates}

The seller-characteristic controls include market location indicators,
sales channel indicators, payment method indicators, and shipping destination indicators, all coded as binary variables from the raw
platform fields documented in Appendix \ref{app:data_dictionary}.

Three monthly energy price series enter extended specifications as auxiliary
controls: Brent crude oil prices (USD per barrel, FRED St. Louis), the German energy
import index (2021\,=\,100, Destatis), and Austrian natural gas spot prices
(EUR per MWh, OEGPI). Their correlation structure is reported in
Appendix \ref{app:figures:energy_corr} Figure \ref{fig:energy_control_corr}.

Plastic packaging depends on petrochemical feedstocks---primarily naphtha,
ethylene, and propylene---so its input costs track broad fossil-energy prices
closely. The 2022--2023 European energy shock raised costs across plastics,
packaging, and logistics \citep{packagingeurope2022commodity,
plasticseurope2022energy}. If these shocks passed through more strongly to
treated supply chains than to the graphics card control, a treated--control
price divergence could arise before the SUP regime became binding and persist
into the policy window. Unit and month fixed effects capture time-invariant
unit differences and common monthly shocks but not differential trends arising
from heterogeneous input-price exposure. The lagged energy variables are
intended to absorb part of this cost channel; in the absence of product-level
input shares they remain partial controls rather than a structural solution.

\subsection{Outcome Variable}
The main outcome variable is the natural logarithm of the average price within each retailer–product–month; descriptive statistics for this variable appear in Appendix 
\ref{app:tables:descriptive}. Log prices reduce the influence of extreme values similar to the monthly duration-weighted aggregated panel analysis in Section 
\ref{sec:additional} and allow regression coefficients to be interpreted approximately as percentage changes for modest effect sizes. Baseline tables therefore use log average monthly prices as the primary outcome variable.

\section{Empirical Approach and Results}
\label{sec:empirical}

The analysis estimates the effect of the Austrian SUP regulatory regime on
posted e-commerce prices using high-frequency retail offer spells drawn from a
price-comparison platform covering stand-alone e-tailers and third-party
marketplace sellers, including those operating through Amazon and eBay. The observation window runs from January 2020 through end-2024 (till early 2025 for some spells). Two baseline
specifications are estimated on the Austria-only sample, both centered on
March 2024 as the compliance-payment date. This timing follows directly from
the Austrian packaging and SUP institutional sequence: firms report quantities
for the prior calendar year by 15 March and settle fees through participation
in collection and recycling systems, so March 2024 is the first date at which
the 2023 reporting year translates into an actual payment
obligation (for 
 the institutional background, including annual
reporting by 15 March, system-participation obligations from 1 January 2023,
and the compliance architecture of Austrian collection and recycling systems,
see \citet{eu2019sup,bmk2024,wko_packaging_2026,ara_licensing_2026,uba2024}).

\subsection{Panel Structure and Notation}
\label{subsec:notation}

Units $i=1,\dots,N$ are unique retailer--product pairs; $t=1,\dots,T$ indexes
calendar months. The outcome $Y_{it}$ is the log of the duration-weighted
average monthly price for unit $i$ in month $t$ (nominal 
price levels and median monthly prices are used in robustness checks; the main tables use
log average monthly price to ease interpretation and reduce the influence of
extreme values \citep{roth_santanna_2023}). Treatment status is
$D_i\in\{0,1\}$ and $Post_t=\mathbf{1}\{t\geq\text{2024:03}\}$. Relative
event time is $r_{it}=t-t_0$, where $t_0$ is March 2024; $r=-1$ as the omitted
reference period. Unit fixed effects are $\alpha_i$ and calendar-month fixed
effects are $\lambda_t$. Identification follows the standard
difference-in-differences logic: within-unit variation over time eliminates
permanent heterogeneity, while month effects absorb common calendar shocks
\citep{callaway2021multiple,roth_etal_2023}.

\subsection{Covariates}
\label{subsec:covariates_method}

The control vector $\mathbf{X}_{it}$ absorbs time-varying differences in
seller characteristics. Market location indicators cover Austria, Germany, the United Kingdom, the Netherlands, and Poland---the five countries that account
for the large majority of cross-border seller activity on the platform and that
registered the highest non-Austrian offer-spell volumes in the raw data.
Sales channel indicators distinguish online-only offers, collection, and in-store pickup. Payment method indicators cover Mastercard, Visa, American
Express, and Diners Club. Shipping destination indicators cover Austria,
Germany, the United Kingdom, Poland, the Netherlands, and Ireland. Additional
binary seller-attribute controls include country-specific sales and
availability markers.

Extended specifications add monthly input-price series---Brent crude from
FRED, the German energy import index (2021\,=\,100) from Destatis, and Austrian gas prices in \euro/MWh from OEGPI---plus their first three lags.
These series proxy petrochemical and packaging input costs. The 2022--2023
European energy shock provides the main motivation: it likely affected
SUP-intensive retail segments more severely than the graphics card control,
so a treated--control divergence may arise from differential cost exposure
rather than the regulation
\citep{packagingeurope2022commodity,plasticseurope2022energy}. Without product-level input shares these series remain reduced-form cost shifters
rather than structural indices, and they do not fully resolve the
differential trends concern---a limitation addressed further in
Section \ref{subsec:threats}.

\subsection{Estimating Equations}
\label{subsec:specs}

\noindent{Event study.}\vspace{3pt}

The dynamic specification is
\begin{equation}
Y_{it}
=
\sum_{\substack{r=-12\\r\neq -1}}^{12}
\beta_r
\bigl(\mathbf{1}\{r_{it}=r\}\times D_i\bigr)
+
\mathbf{X}_{it}'\boldsymbol{\gamma}
+
\alpha_i
+
\lambda_t
+
\varepsilon_{it}.
\label{eq:event_study}
\end{equation}
Pre-treatment 
coefficients ($r<0$) test for differential pre-trends;
post-treatment coefficients ($r\geq 0$) trace price adjustment as the policy
becomes binding. The design compares a single treated cohort against a
never-treated control group, so the multi-cohort contamination problems of
staggered settings do not apply
\citep{sun2021dynamic,callaway2021multiple}. Because all treated units share
a common treatment date, switching to the Sun--Abraham or Callaway--Sant'Anna
estimators would deliver the same comparison; those corrections address
cross-cohort contamination that is absent here.

\vspace{3pt}\noindent{Pooled TWFE.}\vspace{3pt}

Restricting \eqref{eq:event_study} to a single post-period indicator gives
the standard TWFE:
\begin{equation}
Y_{it}
=
\delta\bigl(D_i\times Post_t\bigr)
+
\mathbf{X}_{it}'\boldsymbol{\gamma}
+
\alpha_i
+
\lambda_t
+
u_{it}.
\label{eq:twfe}
\end{equation}
$\delta$ is the average post-March-2024 price gap between treated and control
units relative to the pre-period. With a single treated cohort against a
never-treated group, this coincides with the standard two-group
difference-in-differences estimand and under parallel trends.

\vspace{3pt}\noindent{Sequential TWFE with multiple periods.}\vspace{3pt}

The institutional timeline motivates separating two phases. Treated firms
first face a reporting requirement in March 2023 (quantities for 2022
declared) and then a financially binding payment in March 2024. Define
$Report_t=\mathbf{1}\{t\geq\text{2023:03}\}$ and
$Fee_t=\mathbf{1}\{t\geq\text{2024:03}\}$. The sequential model is
\begin{equation}
Y_{it}
=
\delta_R\bigl(D_i\times Report_t\bigr)
+
\delta_F\bigl(D_i\times Fee_t\bigr)
+
\mathbf{X}_{it}'\boldsymbol{\gamma}
+
\alpha_i
+
\lambda_t
+
\xi_{it}.
\label{eq:twfe_3x2}
\end{equation}
Because $Fee_t$ becomes active only after $Report_t$ is already in effect,
$\delta_R$ captures the reporting-only effect (March 2023 through
February 2024) relative to the pre-policy baseline, and $\delta_F$ captures
the incremental effect of the payment phase. The cumulative post-March-2024
effect is $\delta_R+\delta_F$. This is a multi-phase TWFE, not a
staggered-adoption design: all treated units move through both phases on the
same dates.

Category-level estimates run \eqref{eq:twfe} on sub-samples restricted to
treated units in category $c$ against the full control group, yielding a
category-specific coefficient $\delta_c$ for each $c=1,\dots,C$.

\subsection{Inference}
\label{subsec:inference}

Standard errors are reported under two clustering schemes: by unit
(retailer--product pair), which allows arbitrary serial correlation within
listings; and two-ways, by unit and calendar month, which additionally allows
cross-sectional dependence arising from platform-wide pricing movements
\citep{bertrand2004trust,cameron2011multiway}. The choice between these
schemes bears on inference in high-frequency retail panels; both are reported
as a robustness check.

\subsection{Pooled Baseline Estimates}
\label{subsec:pooled}

Table \ref{tab:twfe_main_text} reports the central pooled results. Column (1)
presents the standard TWFE \eqref{eq:twfe} with a single post-payment
indicator; column (2) presents the sequential TWFE \eqref{eq:twfe_3x2},
separating the report-only period from the payment-due period. Additional
estimates using aggregated panel data appear in Appendix \ref{app:tables:regression},
Table \ref{tab:app_twfe_3x2_main}.

\begin{table}[H]
\caption{Baseline TWFE
 estimates of SUP treatment effects on monthly prices.}
\label{tab:twfe_main_text}
\small
\setlength{\cellWidtha}{\textwidth/3-2\tabcolsep+0.5in}
\setlength{\cellWidthb}{\textwidth/3-2\tabcolsep-0.3in}
\setlength{\cellWidthc}{\textwidth/3-2\tabcolsep-0.2in}
\scalebox{1}[1]{\begin{tabularx}{\textwidth}{>{\raggedright\arraybackslash}m{\cellWidtha}>{\centering\arraybackslash}m{\cellWidthb}>{\centering\arraybackslash}m{\cellWidthc}}
\toprule
& \multicolumn{2}{c}{\textbf{Dependent Variable: ln(Average Monthly Price)}} \\
\cmidrule{2-3}
& \textbf{(1) Standard TWFE}& \textbf{(2) TWFE with Multiple Periods}\\
\midrule
Treated $\times$ Post-payment
  & 0.0398 *** &        \\
  & (0.0044)  &        \\

Treated $\times$ $\mathbf{1}\{\text{Report-only period}\}$ 
  &        & 0.0782 *** \\
  &        & (0.0082)  \\

Treated $\times$ $\mathbf{1}\{\text{Payment-due period}\}$
  &        & 0.0548 *** \\
  &        & (0.0052)  \\
\midrule
Unit fixed effects  & \checkmark 

 & \checkmark \\
Month fixed effects & \checkmark & \checkmark \\
Standard errors     & Clustered by unit & Clustered by unit \\
Observations        & 103{,}074 & 103{,}074 \\
Number of units     & 3213 & 3213 \\
$R^2$               & 0.9088 & 0.9097 \\
Within $R^2$        & 0.0023 & 0.0120 \\
RMSE                & 0.2382 & 0.2370 \\
\bottomrule
\end{tabularx}}
\noindent\footnotesize{{{Notes}}: 
Column (1) presents the standard TWFE specification
\eqref{eq:twfe} with a single post-payment interaction. Column (2) presents
the sequential TWFE specification \eqref{eq:twfe_3x2}, distinguishing the
report-only phase (March 2023--February 2024) from the payment-due phase
(from March 2024). All specifications include unit and month fixed effects.
Standard errors in parentheses are clustered at the unit
(retailer--product pair) level; see Section \ref{subsec:inference} and
Table \ref{tab:app_twfe_3x2_main} for retailer-level clustering. Significance:
*** $p<0.01$. 
}
\end{table}

The column (1) estimate of 0.0398 log points corresponds to an average post-treatment price increase of approximately 4.1 percent (a treated product priced at \euro10 before the policy rises to approximately \euro10.41; one priced at \euro25 reaches approximately \euro26.03).
The sequential TWFE in column (2) reveals a timing pattern obscured by the
pooled average. The report-only coefficient of 0.0782 log points
($\approx$8.1 percent) exceeds the payment-due coefficient of 0.0548 log
points ($\approx$5.6 percent). The larger coefficient in the reporting phase
relative to the payment phase is consistent with sellers adjusting prices in
anticipation of compliance obligations: a seller aware from March 2023 that
quantities must be tracked and ultimately settled has an incentive to adjust
prices in advance rather than absorb costs and reprice upon payment. The smaller payment-due coefficient indicates that the price response does not
constitute a discrete jump at the payment date but emerges earlier and
persists at a somewhat lower level once the system enters the payment phase.

\vspace{3pt}\noindent{Comparison to prior evidence.}\vspace{3pt}

A pooled effect of 4--8 percent is broadly consistent with the existing
literature on environmental-charge incidence. \citet{convery2007irish} find
near-full pass-through of the Irish plastic bag levy, which at \euro0.15 per
bag implied a retail price increase of approximately 4--6 percent depending on
the product. Fuel-tax pass-through estimates in \citet{marion2011fueltax}
cluster around 80--100 percent of the statutory rate in competitive retail
markets, translating into price responses of comparable magnitude. For Austrian and German SUP segments, \citet{reichel2025sup} reports a pooled
Austrian difference-in-differences contrast of approximately 13 index points
over a twelve-month post-policy window using a price-index event-study design;
a larger figure, attributable in part to the broader post-payment window and
the difference in normalization. The present 4 percent pooled estimate covers
a shorter window and is partly driven by lower-exposure categories; the
balloon sub-sample analyzed in Section \ref{subsec:balloons} demonstrates
that high-exposure products can exhibit effects exceeding 50 percent,
consistent with the upper range in \citet{reichel2025sup}. The \citet{homonoff2018bags} estimates for US bag charges are smaller (1--3
percent) and reflect lower statutory rates; \citet{walls2011drs} notes that
pass-through tends to rise with market concentration and product specificity,
both of which vary across the treated categories here. Taken together, the order of magnitude of the pooled estimate is plausible, though the phase
pattern---larger in the reporting phase than the payment phase---is less
directly comparable to the existing evidence, which does not generally
separate anticipatory from realized cost responses.

\vspace{3pt}\noindent{Energy confound.}\vspace{3pt}

A potential threat to a causal interpretation is that the 2022--2023 European
energy shock raised costs for plastic-intensive supply chains more than for
the graphics card control. If that differential persists into the 2023--24
window, part of the estimated treatment effect may reflect lagged cost
pass-through rather than regulatory incidence. The energy controls described
in Section \ref{subsec:covariates_method} do not fully resolve the
differential trends concern, as the lagged covariates are collinear with the
treatment timing. The concern therefore remains without product-level input
shares, and the estimates should be read with this caveat in mind.

\subsection{Sensitivity to the Level at Which Standard Errors Are Clustered}
\label{subsec:clustering}

Appendix \ref{app:tables:regression} Table \ref{tab:app_twfe_3x2_main} re-estimates the sequential TWFE
clustering standard errors at two levels: the unit (retailer--product pair)
and the retailer. Clustering at the unit level, which allows arbitrary serial
correlation within a listing, yields a report-only coefficient of 0.1377
($\approx$14.8 percent) and a payment-due coefficient of 0.0450 ($\approx$
4.6 percent). Clustering at the retailer level---which additionally allows
dependence across all listings from the same seller---leaves point estimates
unchanged but reduces precision substantially, particularly for the earlier
phase, pushing the payment-due coefficient below conventional significance
thresholds.

Two observations follow. The stability of point estimates across clustering
levels is reassuring: the estimated effects are not an artifact of the
variance estimator. The precision loss at the retailer level suggests that
residual dependence operates partly through seller-wide pricing decisions, so
inference should be interpreted cautiously where treatment variation is
concentrated within retailers.

\subsection{Balloon Prices: A High-Exposure Case Study}
\label{subsec:balloons}

Pooled estimates average over products that differ in regulatory exposure,
input-material intensity, and pass-through capacity. Balloons constitute a
useful high-exposure case study: a narrowly defined category with direct and
well-documented exposure to the SUP regime. Appendix \ref{app:tables:regression},
Table \ref{tab:app_balloon_prices_eventstudy} reports the event study for
Austrian balloon prices centered on March 2024 alongside the standard TWFE
\eqref{eq:twfe} and the sequential TWFE \eqref{eq:twfe_3x2} estimated on the
balloon sub-sample, with the full non-SUP control group retained in each case.

The TWFE estimate for balloons is 0.312 log points ($\approx$37 percent),
roughly eight times the pooled figure, and statistically robust across both
clustering levels. The sequential TWFE decomposition reveals the by-now-familiar pattern: the report-only coefficient (0.271, $\approx$31 percent)
exceeds the payment-due coefficient (0.198, $\approx$22 percent), and together they imply a cumulative post-March-2024 effect of 0.469 log points
($\approx$60 percent). The event study makes the timing concrete: the
payment-month coefficient is 0.446 log points ($\approx$55.9 percent), and subsequent months remain strongly elevated---0.442 at $t+1$ ($\approx$54.8
percent), 0.433 at $t+2$ ($\approx$53.2 percent), 0.516 at $t+3$ ($\approx$66.2 percent)---before gradually declining, with the coefficient falling to
approximately 4 percent by $t+9$ (December 2024).

Three observations are noteworthy. First, for highly exposed products the
regulatory impact is not merely detectable but substantial in magnitude.
Second, the persistence over several months points to gradual pass-through or
broad category repricing rather than a one-period spike. Third, the difference between a 55 percent balloon effect and the pooled 4 percent
illustrates why aggregation matters: mixing strongly and weakly affected
categories compresses the average. The balloon sub-sample comprises 178
observations across 15 products, so these estimates are indicative rather
than definitive, but they motivate the disaggregated heterogeneity analysis in
Section \ref{subsec:heterogeneity}.

\vspace{3pt}\noindent{Composition concern.}\vspace{3pt}

Part of the balloon price increase may reflect the exit of lower-priced
listings rather than repricing by continuing sellers. Figure 
\ref{fig:data_spell_counts_text} shows that treated observation counts
rise sharply from 2023 onward while the control series remains stable,
consistent with compositional drift. A balanced-panel robustness check
restricted to units surviving through both policy dates is reported in
Section \ref{subsec:balanced}; point estimates are somewhat smaller but the
qualitative pattern is preserved.

\subsection{Heterogeneity Across Product Categories}
\label{subsec:heterogeneity}

Category-level TWFE models \eqref{eq:twfe} estimated on category-$c$
sub-samples confirm that the pooled average obscures substantial underlying
heterogeneity. Appendix \ref{app:figures:coef_plots} Figure \ref{fig:category_coef_plot_raw} shows that
$\delta_c$ varies in both sign and magnitude across categories.
Figure \ref{fig:parallel_trends_by_category_text} indicates that
pre-treatment comparability with the control group is also uneven, so
category-level estimates should be assessed jointly by point estimate,
precision, and sample size rather than by coefficient magnitude alone.

A natural extension pools categories into exposure groups based on the fee
tiers visible in Figure \ref{fig:data_category_composition} and tests whether
estimated effects rise monotonically with ex ante regulatory intensity. A monotone pattern across fee-linked groups would strengthen the causal
interpretation of the pooled results.

\vspace{3pt}\noindent{Demand-side shifts and PFAS salience.}\vspace{3pt}

Consumer awareness of plastic-related harms---driven partly by EU-level SUP
communication and partly by media coverage of PFAS and microplastics---rose
over the sample period. If demand for affected categories shifted
independently of compliance costs, the estimated price response may partly
reflect market contraction rather than cost pass-through. No clean instrument
for separating regulatory-compliance salience from PFAS-driven consumer
responses is available; the estimates should be interpreted as equilibrium
price changes that may incorporate both cost and demand effects.

\subsection{Heterogeneity Across Sellers}
\label{subsec:seller_het}

The panel combines stand-alone e-tailers and marketplace-based sellers, who
differ in pricing technology, assortment breadth, and pass-through capacity
\citep{hackl2014marketstructure}. Splitting the sample by seller type and
re-estimating both \eqref{eq:twfe} and \eqref{eq:twfe_3x2} would clarify
whether the observed price response is concentrated in one segment of the
online market. Larger sellers may distribute fixed compliance costs across a
broader product range, accelerating pass-through; smaller sellers may face
proportionally heavier fixed burdens but weaker pricing power. These
mechanisms cannot be separated within the pooled estimates.

This distinction matters both economically and econometrically. Because retailer-level clustering in Section \ref{subsec:clustering} reduces
precision noticeably, seller heterogeneity constitutes one source of
within-group residual dependence. A seller-by-seller decomposition would
assist in interpreting both the mechanism and the clustering structure.

\subsection{Identification and Remaining Threats}
\label{subsec:threats}

Identification rests on treated and control prices following parallel trends
absent the policy, conditional on unit effects, month effects, and controls.
Pre-treatment event-study coefficients for $r<0$ provide a partial check.
Pre-trend tests have well-documented power limitations, and conditioning
inference on passing them can distort test size
\citep{roth_2022,rambachan_roth_2023}; clean pre-trends are therefore treated
as supportive evidence rather than conclusive validation.

\vspace{3pt}\noindent{Unobserved confounding.}\vspace{3pt}

If treated products differ from controls along unobserved dimensions---brand
positioning, consumer-segment exposure, or retailer strategic
priorities---parallel trends may fail even after conditioning on fixed effects
and controls.

\vspace{3pt}\noindent{Anticipation.}\vspace{3pt}

Firms may have begun adjusting prices before March 2023, the formal reporting
onset, through trade-association communications or interactions with licensing
systems such as ARA. If so, the pre-treatment event-study coefficients may
understate the true onset of the response and the reporting-phase coefficient
may underestimate the full adjustment. Extending the pre-period beyond the
current twelve-month window would help assess this possibility.

Because the control group is drawn from a separately sampled counterfactual
product set, interpretation also depends on the sample-construction choices
in Section \ref{subsec:control}---particularly the narrowing of the control
group and the baseline-survivor restriction. These push the estimand toward
units that remain observable throughout the window, so the estimates should
be read as internal to the matched surviving sample rather than as
representative of the full Austrian online retail market.

\subsection{Conceptual Mechanism}
\label{subsec:mechanism}

Figure \ref{fig:mechanism-sup-price-channels} summarizes the principal
channels through which posted prices in the treated SUP sample may change
over the observation window. The Austrian SUP regime can affect prices
directly through expected and realized compliance costs, with statutory
intensity differing across fee-schedule tiers (\euro225 and \euro450 per
ton). The 2022--2023 energy shock and its lagged pass-through may affect
treated categories differentially through heterogeneous input-mix exposure,
particularly where petrochemical, packaging, and logistics inputs constitute
a larger share of total costs. These cost channels may interact with
price-adjustment costs, administrative salience, seller exit, and resulting
market-share reallocation within the treated segment.

Observed price changes cannot be attributed exclusively to regulatory
pass-through. Posted-price responses may combine regulation-induced cost
changes, exposure-weighted input-cost shocks, seller-side repricing
frictions, composition effects from exit and survival, and non-policy demand
shifts. Because the data contain prices but not quantities, and because
product-level input shares are not observed, these channels cannot be
separated. The estimates are therefore reduced-form equilibrium price
responses in the treated online segment rather than a structural estimate of
pass-through of the statutory fee alone.

\begin{figure}[H]

\centering 
\resizebox{0.73\textheight}{!}{%
\begin{tikzpicture}[
    node distance=9mm and 11mm,
    >=Latex,
    box/.style={
        draw, rounded corners=2pt, align=center,
        minimum height=10mm, text width=4.0cm,
        font=\small, fill=gray!6
    },
    mech/.style={
        draw, rounded corners=2pt, align=center,
        minimum height=10mm, text width=4.2cm,
        font=\small, fill=blue!6
    },
    resultbox/.style={
        draw, rounded corners=2pt, align=center,
        minimum height=10mm, text width=4.0cm,
        font=\small, fill=red!6
    },
    note/.style={
        draw, rounded corners=2pt, align=left,
        text width=8.4cm, font=\footnotesize, fill=yellow!10
    },
    solidarrow/.style={->, thick},
    dashedarrow/.style={->, thick, dashed}
]
\node[box] (energyshock)
    {Energy price shock\\(2022--2023, upward cost pressure)};
\node[box, below=of energyshock] (lagenergy)
    {Lagged energy-price pass-through\\(waning but possibly relevant through 2026)};
\node[box, below=of lagenergy] (supreg)
    {Austrian SUP regulation\\reporting onset: Mar 2023\\payment due: Mar 2024\\
     fee tiers: \euro225 and \euro450 per ton};
\node[box, below=of supreg] (news)
    {Other exogenous shocks:\\PFAS, plastic salience, demand shifts};
\node[mech, right=20mm of energyshock] (inputmix)
    {Input-mix exposure\\(petrochemical, packaging, logistics intensity)};
\node[mech, below=of inputmix] (supexposure)
    {Category-specific SUP exposure\\(direct vs.\ indirect and fee-schedule intensity)};
\node[mech, below=of supexposure] (pac)
    {Price-adjustment costs\\relabeling, repricing, admin burden};
\node[mech, below=of pac] (exit)
    {Firm exit and survival selection};
\node[mech, right=22mm of supexposure, text width=4.8cm, fill=green!8] (supproducts)
    {Observed treated SUP product prices\\in retail offer spells};
\node[resultbox, right=20mm of supproducts] (marketshare)
    {Market-share reallocation\\within the treated segment};
\node[resultbox, above=of marketshare] (competition)
    {Composition and competition effects};
\node[resultbox, below=of marketshare] (pricechange)
    {Observed posted price changes\\in the estimation sample};
\node[note, below=26mm of supproducts, xshift=36mm] (note1) {
\textbf{Interpretation:} The observed price response is reduced-form. It may
combine:\\
(i) exposure-weighted input-cost shocks;\\
(ii) SUP pass-through (\euro225/\euro450 tiers);\\
(iii) price-adjustment costs and administrative salience;\\
(iv) firm exit and composition changes;\\
(v) non-policy demand or media shocks.\\[1mm]
With prices but no product-level input shares or quantities, these
channels cannot be cleanly disentangled.
};
\draw[solidarrow] (energyshock) -- (inputmix);
\draw[solidarrow] (lagenergy)   -- (inputmix);
\draw[solidarrow] (supreg)      -- (supexposure);
\draw[solidarrow] (supreg)      -- (pac);
\draw[dashedarrow](news)        -- (supproducts);
\draw[solidarrow] (inputmix)    -- (supproducts);
\draw[solidarrow] (supexposure) -- (supproducts);
\draw[solidarrow] (pac)         -- (supproducts);
\draw[solidarrow] (pac)         -- (exit);
\draw[solidarrow] (exit)        -- (marketshare);
\draw[solidarrow] (marketshare) -- (competition);
\draw[solidarrow] (competition) -- (pricechange);
\draw[solidarrow] (supproducts) -- (pricechange);
\node[
    draw, rounded corners=4pt, dashed, inner sep=7pt,
    fit=(energyshock)(lagenergy)(supreg)(news)(inputmix)(supexposure)
        (pac)(exit)(supproducts)(competition)(marketshare)(pricechange)(note1),
    label={[font=\small]above:Likely mechanism linking SUP policy, input shocks,
           and observed price changes}
] {};
\end{tikzpicture}%
}

\caption{Conceptual mechanism for observed price changes in treated SUP offer
spells. Posted-price changes may reflect overlapping channels:
exposure-weighted energy and input-cost shocks, category-specific SUP
fee-schedule intensity, price-adjustment costs, seller exit and composition
effects, and non-policy demand shocks. The empirical estimates are
reduced-form equilibrium price responses rather than a structural estimate of
pass-through of the statutory fee.}
\label{fig:mechanism-sup-price-channels}
\end{figure}

\subsection{Summary of Findings}
\label{subsec:summary}

The evidence yields three principal conclusions. First, the Austrian SUP
compliance regime is associated with a positive price response in the treated
online segment: the pooled effect is economically modest but robust, on the
order of 4--6 percent depending on specification, and falls within the range
of prior pass-through estimates for comparable environmental charges
\citep{convery2007irish,marion2011fueltax,reichel2025sup}.

Second, timing matters. Prices begin rising during the reporting phase,
before any fee is due. The report-only coefficient of 0.0782 implies that a
\euro15 item reaches approximately \euro16.22 before the first payment
date---a larger adjustment than the \euro15.84 implied by the payment-due
coefficient alone. Sellers appear to respond to the legal and administrative
onset of the regime rather than exclusively to realized monetary obligations.

Third, aggregation conceals substantial heterogeneity. The balloon case
demonstrates that high-exposure products can exhibit effects exceeding 50
percent in the immediate post-treatment period, sustained over several months
before fading. A \euro5 balloon item rising to \euro7.80 in the payment
month represents a qualitatively different phenomenon from the pooled 4
percent average. Disaggregated category analysis, seller-type splits, and composition checks are the natural next steps; the pooled estimate provides a
useful benchmark but is unlikely to be the most informative object in a
setting where regulatory exposure, packaging intensity, and pass-through
capacity vary substantially across the sample.

\section{Additional Analyses}
\label{sec:additional}

The pooled TWFE results in Section \ref{sec:empirical} establish a positive
average price response but leave four questions unresolved: whether effects
differ across categories in a way that maps onto the SUP fee schedule; whether
the baseline estimates reflect within-product repricing or compositional
turnover; whether the response is concentrated in one segment of the seller
population; and whether the price increase coincides with demand contraction
or demand expansion. This section addresses each in turn using the aggregated
panel of retailer--product--month offer spells described in
Section \ref{sec:data}.

\subsection{Category-Level Heterogeneity and Fee-Tier Test}
\label{subsec:cat_het}

The nine treated keyword categories differ in their position in the Austrian
fee schedule: balloons and tobacco filters face \euro450/ton, while the
remaining seven categories face \euro225/ton. If the estimated price response
reflects SUP pass-through, effects should be systematically larger in the
higher-tier group. Table \ref{tab:A1_cat_het} reports category-specific
coefficients $\hat{\beta}_c$ from the joint interaction model
\begin{equation}
Y_{it}
=
\sum_{c=1}^{C}
\beta_c
\bigl(D_i^c \times \mathbf{1}\{\tau \geq -24\}\bigr)
+
\alpha_i + \lambda_t + \varepsilon_{it},
\label{eq:joint_cat}
\end{equation}
where $D_i^c = \mathbf{1}\{D_i=1,\, C_{ic}=1\}$ and $\tau$ is months
relative to the payment date. Standard errors are clustered at the retailer
level (231 
 clusters).

\begin{table}[H]
\caption{Category-level pass-through heterogeneity.}
\label{tab:A1_cat_het}

\setlength{\cellWidtha}{\fulllength/7-2\tabcolsep-0.4in}
\setlength{\cellWidthb}{\fulllength/7-2\tabcolsep+0.2in}
\setlength{\cellWidthc}{\fulllength/7-2\tabcolsep-0.2in}
\setlength{\cellWidthd}{\fulllength/7-2\tabcolsep-0.4in}
\setlength{\cellWidthe}{\fulllength/7-2\tabcolsep-0.5in}
\setlength{\cellWidthf}{\fulllength/7-2\tabcolsep-0.5in}
\setlength{\cellWidthg}{\fulllength/7-2\tabcolsep+1.8in}
\scalebox{1}[1]{\begin{tabularx}{\fulllength}{>{\raggedright\arraybackslash}m{\cellWidtha}>{\raggedright\arraybackslash}m{\cellWidthb}>{\raggedleft\arraybackslash}m{\cellWidthc}>{\raggedleft\arraybackslash}m{\cellWidthd}>{\raggedleft\arraybackslash}m{\cellWidthe}>{\raggedleft\arraybackslash}m{\cellWidthf}>{\raggedright\arraybackslash}m{\cellWidthg}}
\toprule
\textbf{Tier} & \textbf{Category} & $\boldsymbol{\hat{\beta}}_\textbf{\textit{c}}$ & \textbf{SE} & \textbf{\textit{p}} & \textbf{\textit{N}} & \textbf{Most Expensive Keyword-Based Selected Treated Sample Product} \\
\midrule
\multicolumn{7}{l}{\textit{\euro{}450 per ton 
}} \\
& Balloons        & $0.447$ {***}& $(0.101)$ & $0.000$ & $150$  &
  \textit{Konstsmide led motiv szenerie heissluftba\ldots} (\euro{}74)  \\
& Tobacco filters & $0.342$ {***}& $(0.102)$ & $0.001$ & $122$  &
  \textit{Zebra ladeger\"{a}t f\"{u}r rw\,420, zigarettenanz\ldots} (\euro{}272) \\
\addlinespace
\multicolumn{7}{l}{\textit{\euro{}225 per ton}} \\
& Plastic bags    & $2.696$ *** & $(0.103)$ & $0.000$ & $4$    &
  \textit{Litepanels leichte tragetasche f\"{u}r astra\ldots} (\euro{}250) \\
& Food containers & $0.765$ ***& $(0.110)$ & $0.000$ & $60$   &
  \textit{Blanco sitybox einh\"{a}ngbare kunststoffscha\ldots}(\euro{}43)  \\
& Dep.\ bottles   & $0.221$ & $(0.244)$ & $0.366$ & $2$ &
  \textit{Dennerle co2-adapter nano mehrwegflasche\ldots}(\euro{}23)  \\
& To-go cups      & $0.001$ & $(0.202)$ & $0.997$ & $1545$ &
  \textit{Villeroy \& boch anmut platinum no.\,1 kaffe\ldots}(\euro{}207) \\
& Non-dep.\ bottles & $-0.130$ & $(0.271)$ & $0.633$ & $6$  &
  \textit{Schott zwiesel basic bar selection wasser\ldots}(\euro{}41)  \\
& Plastic wrap    & $-0.372$ ***           & $(0.116)$ & $0.002$ & $60$   &
  \textit{Qeridoo fu\ss{}s\"{a}ckchen f\"{u}r fahrradanh\"{a}nger\ldots} (\euro{}123) \\
& Wet wipes       & $-0.435$   ***         & $(0.130)$ & $0.001$ & $104$  &
  \textit{B + w photo-clear 18 $\times$ 18\,cm mikrofaser-reinig\ldots} (\euro{}44) \\
\midrule
\multicolumn{7}{l}{\textit{Tier pooled test
  ($H_0$: $\beta_{\text{\euro{}450}} = \beta_{\text{\euro{}225}}$)}} \\
& \euro{}225/ton avg  & $-0.023$ & $(0.179)$ & $0.896$ &---&---\\
& \euro{}450/ton avg  & $0.403$ {***} & $(0.092)$ & $0.000$ &---&---\\
& Difference          & $0.427$ {**} & $(0.201)$ & $0.035$ &---&---\\
\bottomrule
\end{tabularx}}

\noindent\footnotesize{{Fixed effects}: month (within-month demeaning).
  \textit{Standard errors}: retailer-clustered (\textbf{231} clusters), shown in
  parentheses. $\hat{\beta}_c$: coefficient on
  $D_i^c \times \mathbf{1}\{\tau \geq -24\}$ from the joint model
  \eqref{eq:joint_cat}. Tier-test standard error is conservative.
  $N$: treated units in category. Most expensive product: highest mean
  price across all spells in the estimation window.
  Plastic bags ($N=4$) should be interpreted with caution.
  {***} \textit{p} < 0.01, {**} \textit{p} < 0.05. 
}
\end{table}

\begin{figure}[H]
\hspace{-5pt}
  \includegraphics[width=\linewidth]{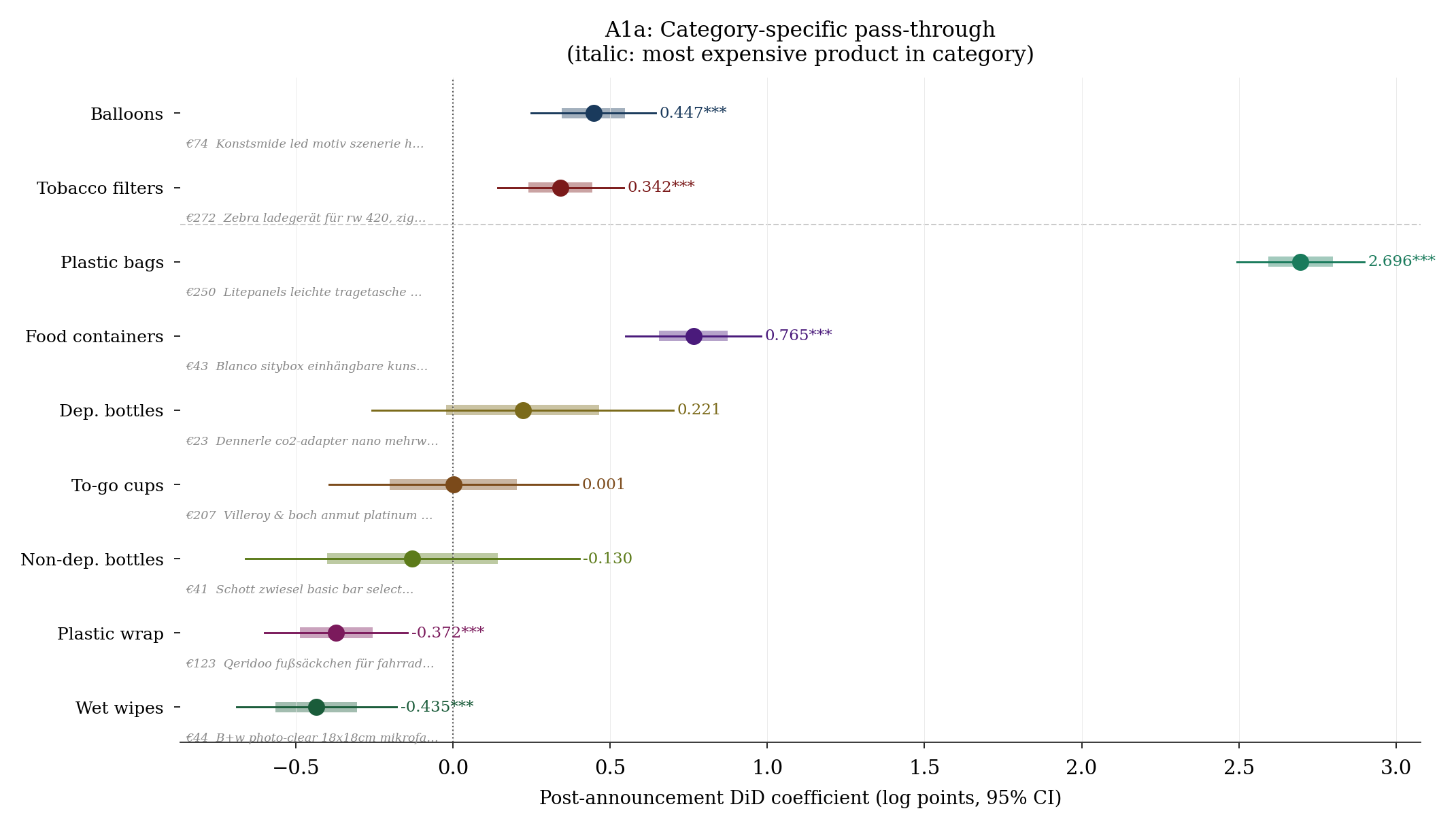}
  \caption{{Category-specific} 
pass-through coefficients $\hat{\beta}_c$ from
  the joint model \eqref{eq:joint_cat}. Thick bars: $\pm 1\hat{\sigma}$; thin
  lines: 95\% CI. Retailer-clustered standard errors; month fixed effects.
  Italic labels in grey show the most expensive product in each category.  Significance: *** $p < 0.01$.}
  \label{fig:A1a}
\end{figure}

The estimates are heterogeneous in both sign and magnitude.
Figure \ref{fig:A1a} plots the coefficients with 95\% confidence intervals
and labels the most expensive product in each category.

\begin{figure}[H]
\hspace{-3pt}
  \includegraphics[width=\linewidth]{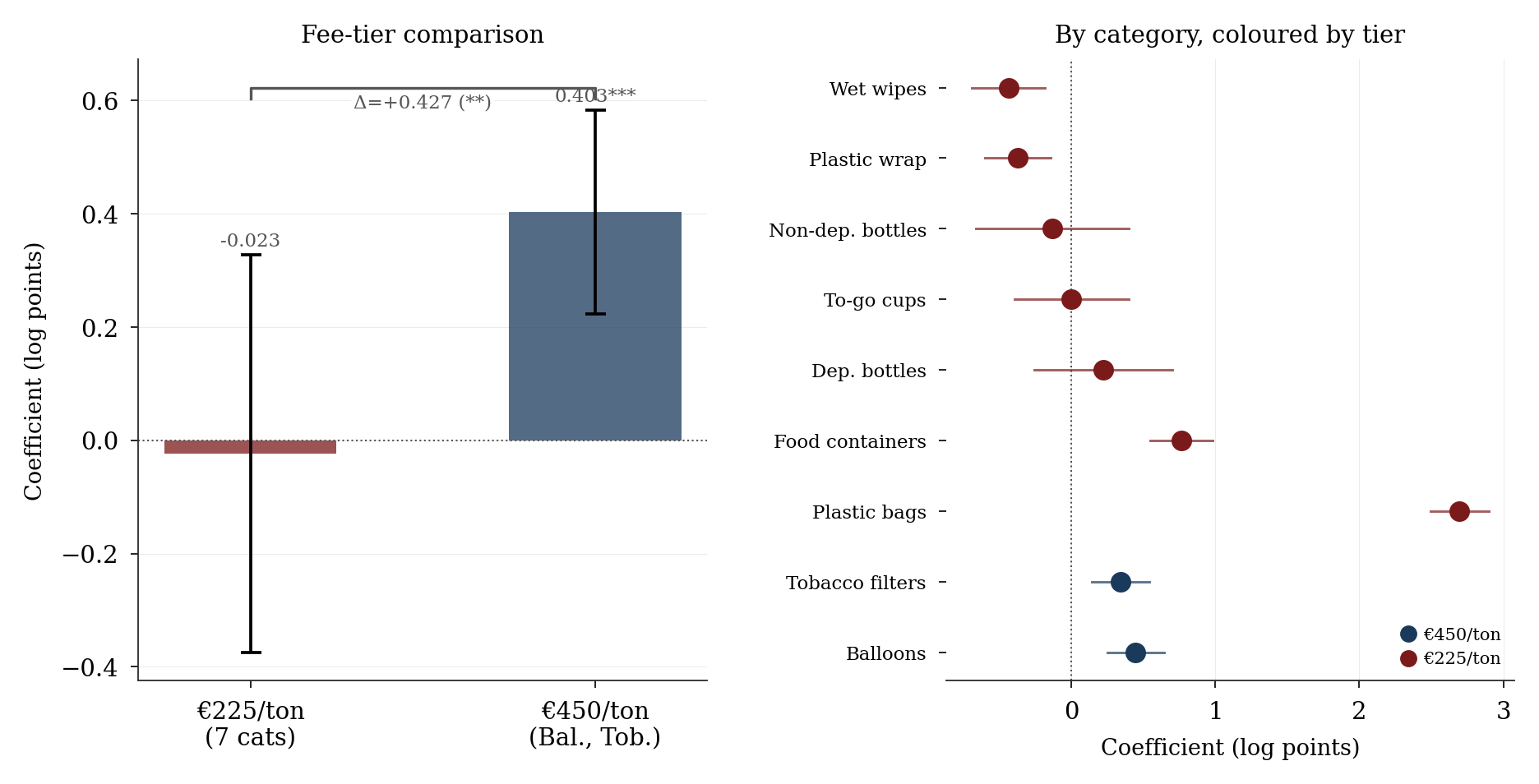}
  \caption{\textbf{Left panel}: 
 Tier-average DiD coefficients for the \euro225/ton
  and \euro450/ton groups; bracket shows the tier difference
  $\Delta=+0.427$ (**). \textbf{Right panel}: Individual category estimates colored
  by fee tier. Both \euro450/ton categories (balloons, tobacco filters) sit
  clearly above zero; the \euro225/ton group is dispersed around zero with
  several negative estimates. Retailer-clustered standard errors; month fixed
  effects. Significance: ** $p < 0.05$, *** $p < 0.01$.}
  \label{fig:A1b}
\end{figure}

Among the \euro450/ton categories, balloons show a coefficient of 0.447 log
points ($\approx$56 percent, $p<0.001$) and tobacco filters 0.342 log points
($\approx$41 percent, $p=0.001$). The \euro225/ton group is heterogeneous:
plastic bags (2.696, driven by a very thin sample of four units) and food
containers (0.765) show large positive effects, while to-go cups (0.001) and
deposit bottles (0.221) are near zero, and plastic wrap ($-0.372$) and wet
wipes ($-0.435$) are significantly negative. The tier-pooled test in the lower
panel of Table \ref{tab:A1_cat_het} finds an average \euro450/ton coefficient
of 0.403 ($p<0.001$) against an average \euro225/ton coefficient of $-0.023$
($p=0.896$), with a tier difference of 0.427 ($p=0.035$).
Figure \ref{fig:A1b} presents this comparison graphically.

\vspace{-6pt}

Three observations are noteworthy. First, the fee-tier difference is
statistically significant and economically large: a 0.43 log-point gap between
the two groups is consistent with the view that higher statutory rates generate
larger price responses. Second, the negative coefficients for plastic wrap and
wet wipes indicate that certain \euro225/ton categories experienced price
declines over the same period, possibly reflecting competitive pressure,
product-mix shifts, or the natural-material substitution dynamic described in
Section \ref{subsec:control}. Third, the plastic bag estimate of 2.696
warrants caution given a sample of four units; it is excluded from the tier
average in the robustness checks below.

\begin{figure}[H]
\hspace{-6pt}
  \includegraphics[width=\linewidth]{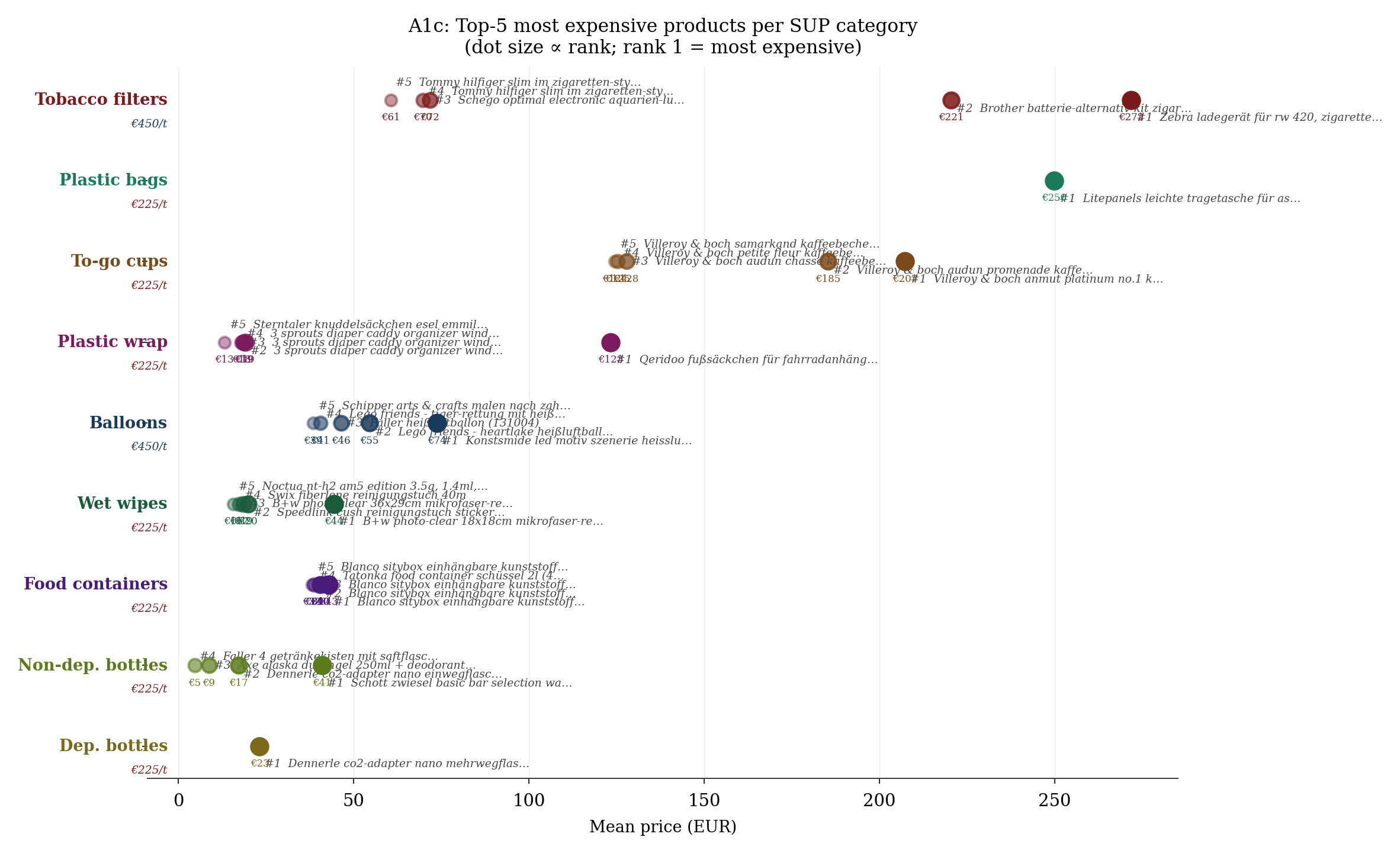}
  \caption{Top-five 
most expensive products per SUP category by mean price
  over the estimation window. Dot size is proportional to price rank
  (rank 1 = most expensive). Categories are ordered by fee tier; fee rates
  shown below category labels.}
  \label{fig:A1c}
\end{figure}

Figure \ref{fig:A1c} plots the five most expensive products per category by
mean price, providing a sense of the product landscape underlying each
coefficient.

\vspace{-6pt}
\unskip

\subsection{Balanced-Panel Robustness}
\label{subsec:balanced}

The baseline panel is unbalanced: treated observation counts rise sharply from
2023 onward while the control series remains stable
(Section \ref{subsec:pooled}). If lower-priced listings exit
disproportionately following the policy, the average treated price rises
mechanically without any within-product repricing. To separate these margins,
the sample is restricted to units observed in at least 25 of the estimation
months---a balanced sub-panel that excludes the most intermittently observed
listings, representing a conservative alternative to a strict survival
restriction.

\begin{figure}[H]
\hspace{-5pt}
  \includegraphics[width=\linewidth]{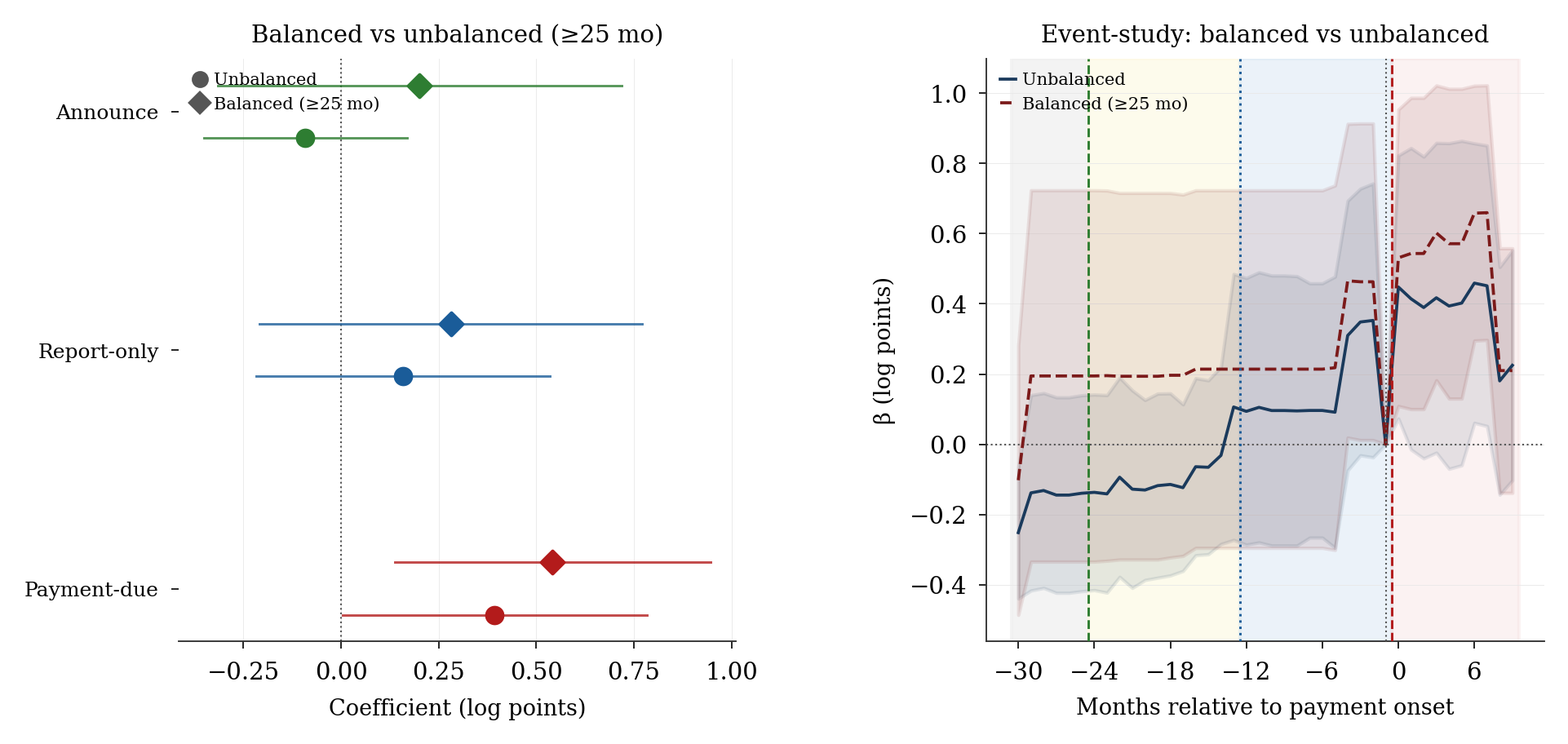}
  \caption{\textbf{Left panel}: Three-period TWFE coefficients for the balanced
  ($\geq $25 months, diamonds) and unbalanced (circles) panels. \textbf{Right panel}: The gray shaded region marks the regulatory assessment phase from $t=-30$ to $t=-25$; the yellow shaded region marks the reporting phase from $t=-24$ to $t=-13$; and the blue shaded region marks the payment implementation phase from $t=-12$ to $t=0$. 
Event-study paths for both samples with 95\% confidence bands.
  Retailer-clustered standard errors; month fixed effects.}
  \label{fig:A2}
\end{figure}

Figure \ref{fig:A2} compares the three-period TWFE coefficients and the full
event-study path between the balanced and unbalanced samples.

The payment-due coefficient rises from 0.394 in the unbalanced panel to 0.541
in the balanced sub-panel; the report-only coefficient increases from 0.158 to
0.219. Point estimates are larger for the longer-lived listings, not smaller,
which is inconsistent with the compositional exit hypothesis: if exit of
lower-priced listings were driving the baseline result, restricting the sample
to survivors would be expected to compress rather than amplify the coefficient.
The event-study paths track each other reasonably through the pre-period before
diverging in the payment phase, consistent with the balanced sub-panel
capturing a set of products whose sellers respond more persistently to the
compliance regime. Wider confidence bands in the balanced sample reflect the
smaller unit count. Taken together, these results indicate that the baseline
estimates are not primarily artefacts of changing sample composition.

\subsection{Seller-Type Heterogeneity}
\label{subsec:seller_type}

\begin{figure}[H]
\hspace{-5pt}
  \includegraphics[width=\linewidth]{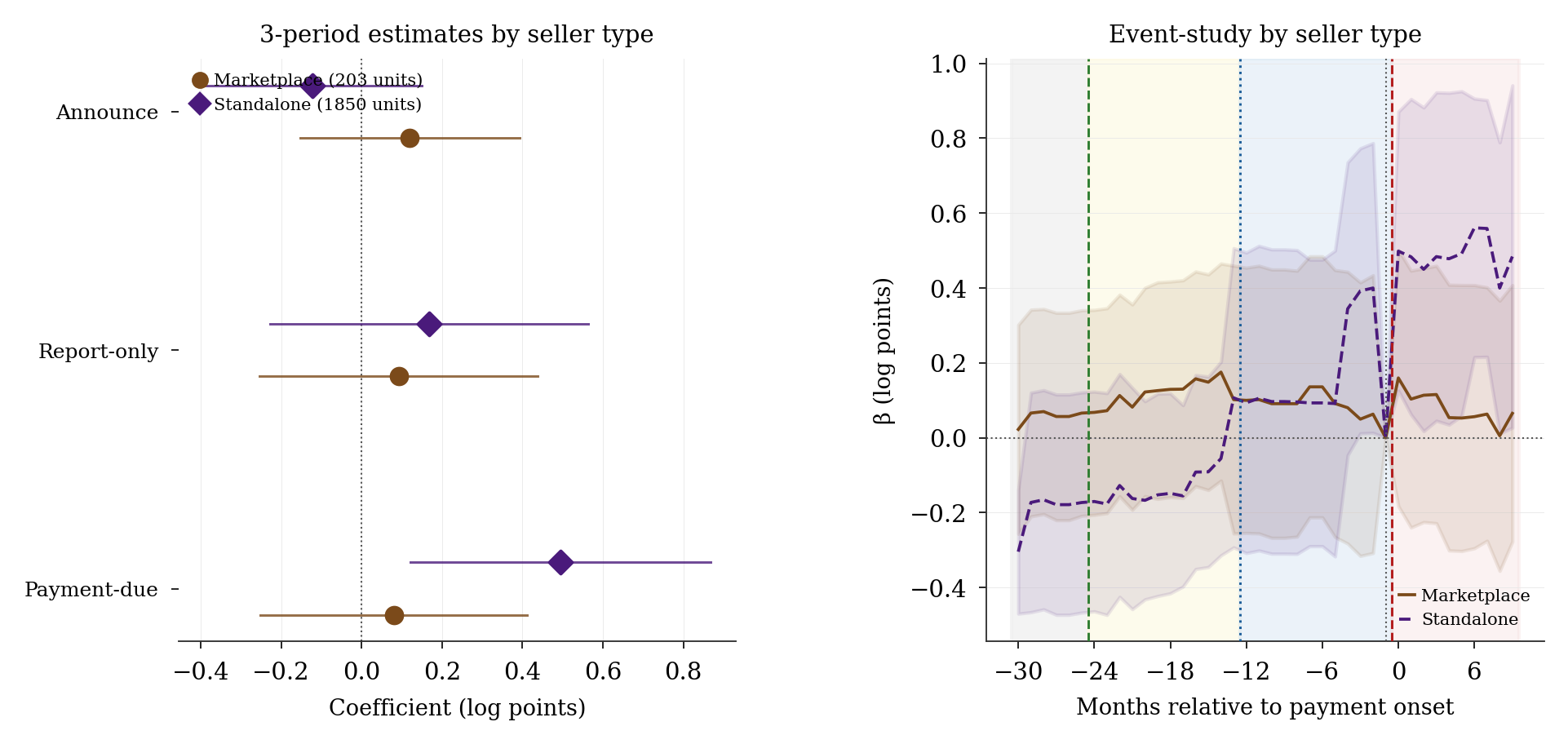}
  \caption{\textbf{Left panel}: Three-period TWFE coefficients by seller type---
  marketplace (203 treated units, circles) and standalone (1850 treated
  units, diamonds). \textbf{Right panel}: The gray shaded region marks the regulatory assessment phase from $t=-30$ to $t=-25$; the yellow shaded region marks the reporting phase from $t=-24$ to $t=-13$; and the blue shaded region marks the payment implementation phase from $t=-12$ to $t=0$. 
Event-study paths by seller type with 95\%
  confidence bands. Retailer-clustered standard errors; month fixed effects.}
  \label{fig:A3}
\end{figure}

The panel combines stand-alone e-tailers and marketplace-based sellers
(Amazon, eBay, and similar platforms, identified by retailer slug patterns).
The two groups differ in pricing technology, assortment breadth, and compliance cost structure: marketplace sellers operate within platform-level
pricing tools and may face different fixed compliance costs per listing compared to
stand-alone shops. Figure \ref{fig:A3} reports the three-period TWFE
coefficients and event-study paths separately for the 203 treated marketplace
units and 1850 treated standalone units.

The payment-due coefficient for standalone sellers is 0.494 log points
($\approx$64 percent), roughly six times the marketplace estimate of 0.080
($\approx$8 percent). The report-only phase shows a similar pattern:
standalone sellers register a larger anticipatory response than marketplace
sellers in both the point estimate and the event-study path. The marketplace
path drifts slightly upward from the reporting onset but remains close to zero
through most of the post-payment window, while the standalone path rises more
clearly and remains elevated.

Two interpretations are consistent with this pattern. First, standalone
sellers may face higher per-unit compliance and repricing costs that are more
fully passed on, while marketplace sellers---who route compliance through
platform infrastructure---may absorb a greater share of the burden or
distribute it across larger volumes. Second, the marketplace sub-sample is
approximately ten times smaller (203 versus 1850 units), so the wider
confidence bands for that group do not preclude effects of similar magnitude
to the standalone estimate. The seller-type split should therefore be
interpreted as indicative evidence on the concentration of the pricing
response rather than as a definitive decomposition.

\subsection{Alternative Outcome Variables and Economic Magnitude}
\label{subsec:outcomes}

\begin{figure}[H]
\hspace{-3pt}
  \includegraphics[width=\linewidth]{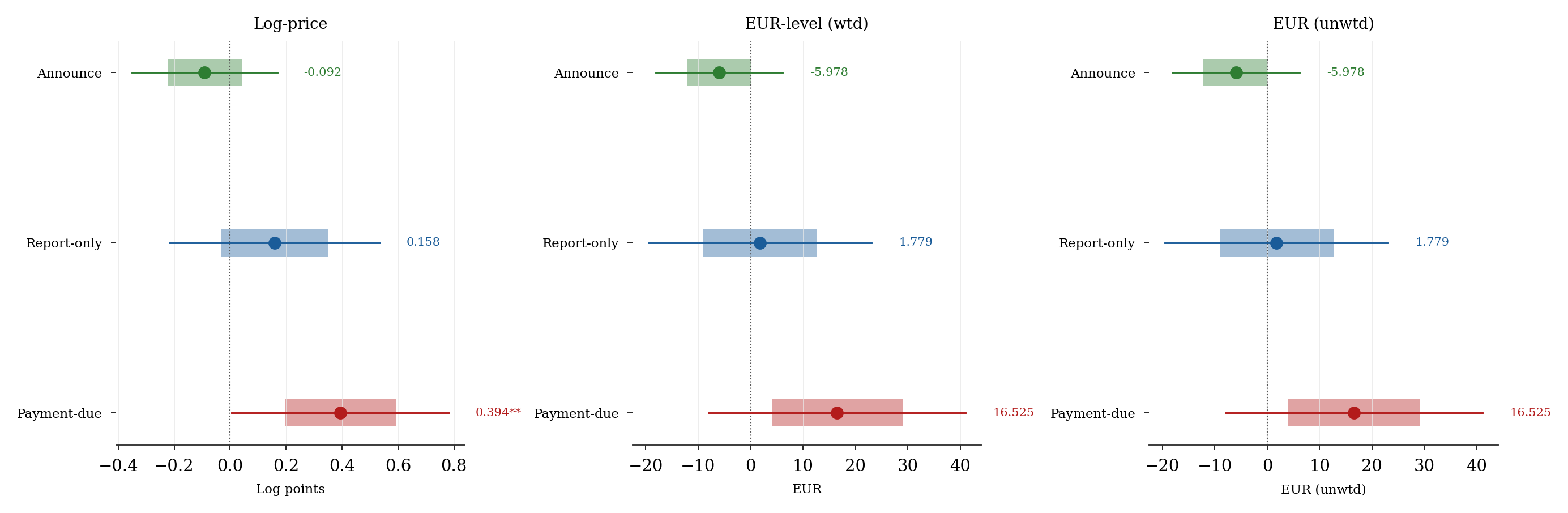}
  \caption{Multi-period 
  TWFE estimates under three outcome definitions: log
  price (\textbf{left}), EUR-level duration-weighted price (\textbf{center}), and EUR-level
  unweighted price (\textbf{right}). Point estimates and 95\% confidence intervals.
  Retailer-clustered standard errors; month fixed effects.  Significance: ** p < 0.05.}
  \label{fig:A4a}
\end{figure}

The baseline outcome is $\ln(\text{average monthly price})$. Two
considerations motivate alternative specifications: log-point effects are
harder to assess economically without a reference price level, and duration-weighted averages may assign disproportionate weight to long-lived
spells. Figure \ref{fig:A4a} reports three-period estimates for three
outcomes side by side: log price, EUR-level duration-weighted price, and EUR-level unweighted price.

The announcement and report-only coefficients are near zero across all three
outcomes, while the payment-due coefficient is positive and statistically
significant in log points (0.394, $p<0.05$) and positive but less precisely
estimated in EUR levels (\euro16.5 in both weighted and unweighted
specifications). The qualitative timing pattern---payment phase dominant---is
robust across all three outcome definitions. The EUR-level estimates are large
in absolute terms, reflecting the presence of high-priced products in the
sample (to-go cups at \euro207, tobacco filters at \euro272) that generate
substantial EUR changes even for modest log-point effects.

\begin{figure}[H]
\hspace{-6pt}
  \includegraphics[width=\linewidth]{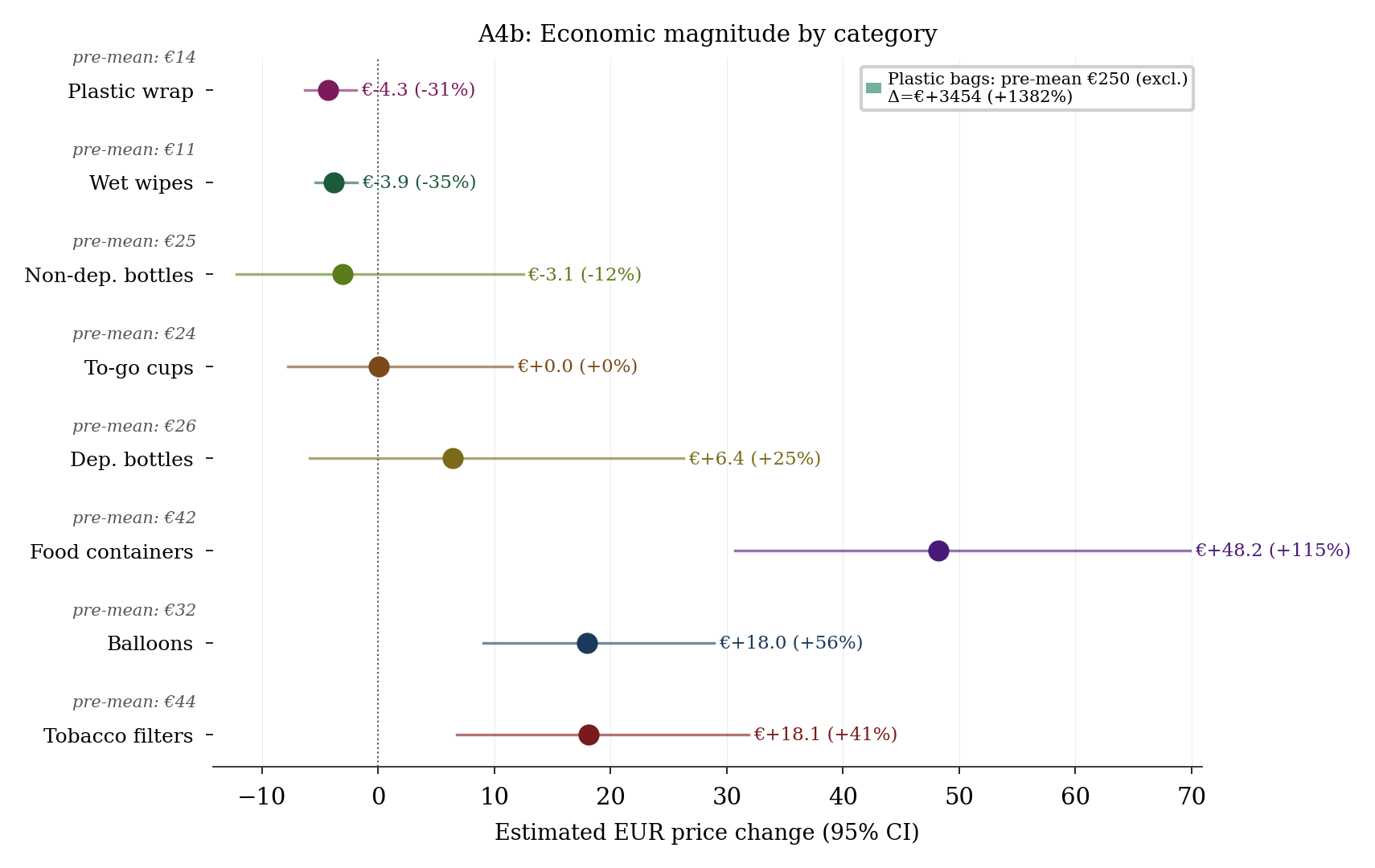}
  \caption{Estimated 
EUR price change by category, computed as
  $\Delta\hat{p}_c =
  \bar{p}_c^{\,\text{pre}}\times(\exp(\hat{\delta}_c)-1)$. Pre-treatment
  category means shown above each row; whiskers are 95\% CI. Plastic bags
  excluded (pre-mean: \euro250, $\Delta=+$\euro3454, sample of four
  units).}
  \label{fig:A4b}
\end{figure}

Figure \ref{fig:A4b} translates the category-specific log-point coefficients
from Table \ref{tab:A1_cat_het} into estimated EUR price changes using
pre-treatment category means, $\Delta\hat{p}_c = \bar{p}_c^{\,\text{pre}}
\times (\exp(\hat{\delta}_c)-1)$.

The EUR magnitudes vary substantially across categories. Food containers
exhibit the largest positive effect (\euro+48.2, $+115$ percent on a
pre-mean of \euro42), followed by balloons (\euro+18.0, $+56$ percent) and
tobacco filters (\euro+18.1, $+41$ percent). To-go cups are near zero in EUR
terms (\euro+0.0) despite constituting the largest category by unit count.
Plastic wrap and wet wipes show negative EUR changes ($-$\euro4.3 and
$-$\euro3.9, respectively), with food containers pulling the pooled estimate
upward. This decomposition reveals that the pooled log-point average
constitutes a unit-count-weighted mixture of economically diverse outcomes:
three categories account for most of the positive signal while two sizeable
categories move in the opposite direction.

\vspace{-6pt}

\subsection{Demand-Side Proxy: Listing Frequency and Duration}
\label{subsec:demand}

Price increases that reflect cost pass-through need not coincide with demand
contraction in a platform setting where sellers can adjust listing intensity
in response to market conditions. The panel contains two demand-adjacent
variables available at the unit--month level: the number of distinct price
spells in a month ($\ln n_{\text{spells},it}$) and the total active-price
days covered ($\ln \text{days}_{it}$). Neither constitutes a direct measure
of consumer demand---higher spell counts may reflect more frequent repricing
rather than greater sales volumes---but both provide an initial indication of
whether the policy coincides with retraction or expansion of market presence.

\begin{figure}[H]
\hspace{-6pt}
  \includegraphics[width=\linewidth]{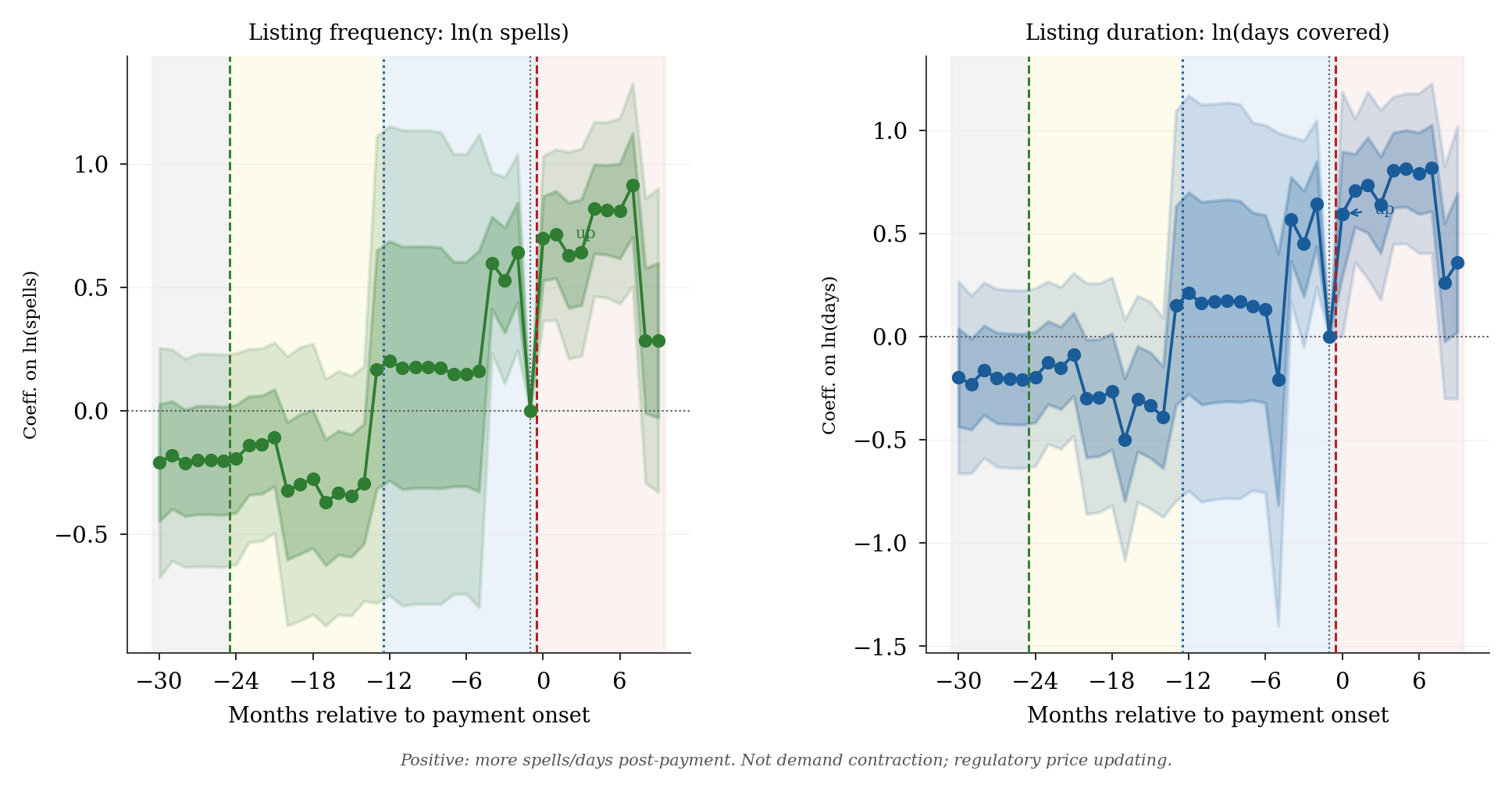}
  \caption{Event-study. The gray shaded region marks the regulatory assessment phase from $t=-30$ to $t=-25$; the yellow shaded region marks the reporting phase from $t=-24$ to $t=-13$; and the blue shaded region marks the payment implementation phase from $t=-12$ to $t=0$. 
 coefficients for $\ln(\textit{n\_spells})$
 (\textbf{left}) and
  $\ln(\texttt{days\_covered})$ (\textbf{right}) relative to the payment date. Green
  dashed line: reporting onset ($\tau=-24$); blue dotted line: announcement
  ($\tau=-12$); red dashed line: payment onset ($\tau=0$). Both series rise
  post-payment rather than fall. Retailer-clustered standard errors; month
  fixed effects.}
  \label{fig:A5a}
\end{figure}

Figure \ref{fig:A5a} plots event-study coefficients for these two outcomes
using the same TWFE specification as the main results with $\tau$ relative to
the payment date.

Both $\ln n_{\text{spells}}$ and $\ln \text{days}$ rise after the payment
date relative to the control group. Spell counts increase by approximately
0.5--0.9 log points in the post-payment window; days covered rises by a
similar magnitude. This pattern is inconsistent with demand contraction as
the primary driver of the price increase: if consumers were substituting away
from treated products in response to higher prices, market presence would be
expected to contract. The evidence is instead consistent with sellers
increasing their activity on the platform---raising listing frequency and
coverage---around and following the compliance onset, possibly as they update
price lists and refresh product pages in response to the administrative burden
of the regime.

This finding does not preclude demand contraction at the transaction level,
as spell counts and days measure listing activity rather than realized sales.
It does indicate, however, that the price increase documented in the main
results is not accompanied by visible retraction from the platform on the
treated side.

\begin{figure}[H]
  \includegraphics[width=0.75\linewidth]{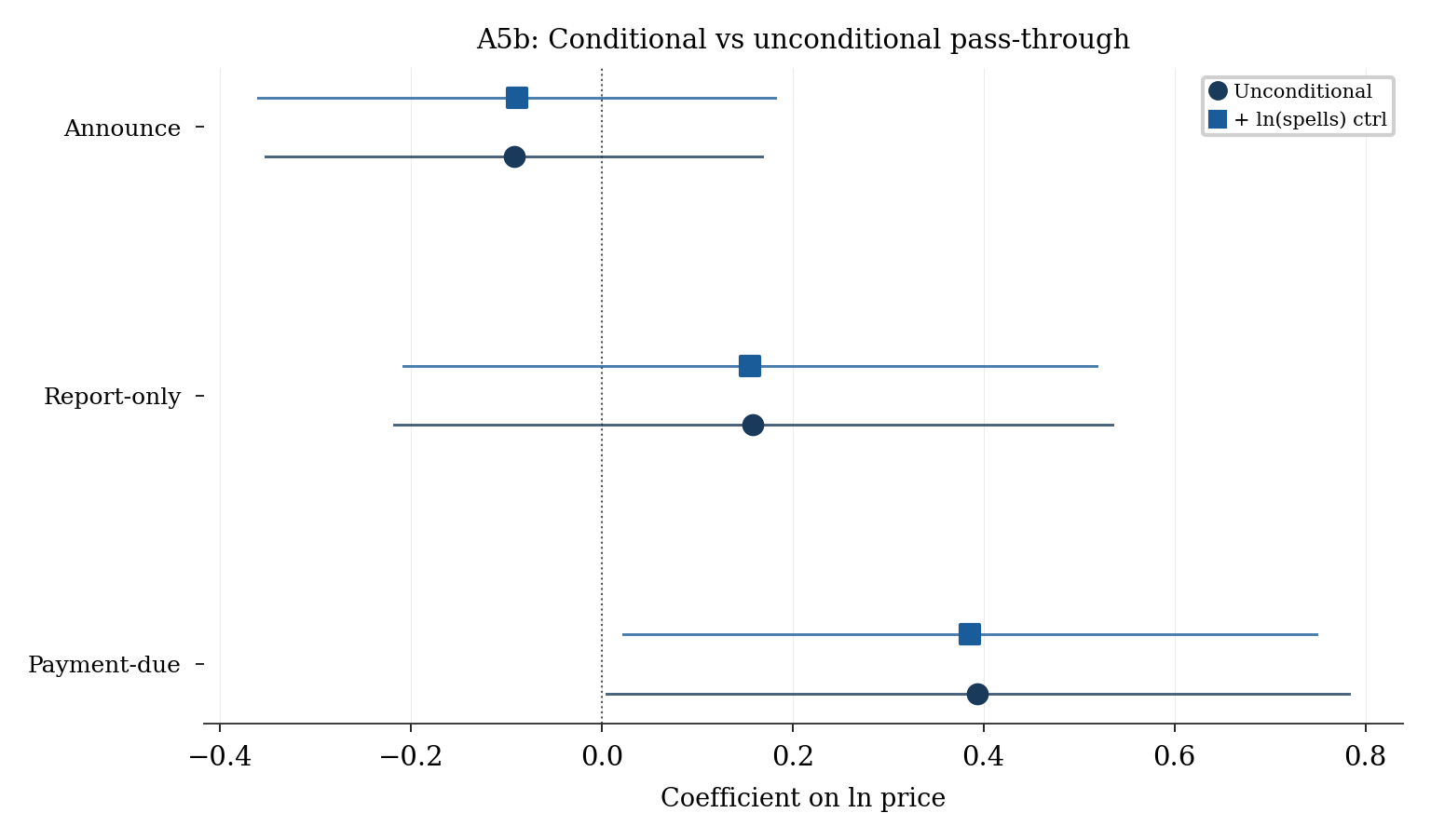}
  \caption{Unconditional 
 (circles) vs.\ conditional on
  $\ln(\texttt{n\_spells})$ (squares) three-period price coefficients. Adding
  the spell-count control leaves the payment-due coefficient effectively
  unchanged (0.394 vs.\ 0.397), indicating that the price effect is not
  driven by compositional shifts in listing frequency within unit--months.
  Retailer-clustered standard errors; month fixed effects.}
  \label{fig:A5b}
\end{figure}

Figure \ref{fig:A5b} examines whether the payment-due price coefficient is
sensitive to controlling for $\ln n_{\text{spells}}$ directly---a test of
whether the price effect operates through repricing conditional on listing
behavior or partly through the composition of listings.

Adding $\ln n_{\text{spells}}$ as a control leaves the payment-due
coefficient essentially unchanged (0.397 versus 0.394, unconditional). The price effect is not driven by shifts in listing frequency within unit--months;
it reflects genuine within-unit price adjustment, conditional on continued
market presence. The report-only and announcement coefficients are also stable
across the two specifications.

\vspace{-6pt}

\vspace{-12pt}

\subsection{Summary of Additional Analyses}
\label{subsec:add_summary}

Four findings emerge from this section. First, category-level pass-through is
tied to the fee schedule: the \euro450/ton tier (balloons, tobacco filters)
shows a tier-average effect of 0.403 log points against $-0.023$ for the
\euro225/ton group, a statistically significant difference of 0.427
($p=0.035$). The monotone fee-tier pattern supports a causal interpretation
of the pooled estimate. Second, the baseline estimates do not primarily
reflect a composition artifact: the balanced-panel sub-sample of long-lived
listings produces larger coefficients, not smaller, suggesting that the
pooled effect is, if anything, attenuated in the full unbalanced panel by the
presence of short-lived, intermittently observed listings that may not have
adjusted prices. Third, the pricing response is concentrated among standalone
e-tailers (0.494) rather than marketplace sellers (0.080), pointing to
differences in compliance cost structure or pricing technology between the two
segments. Fourth, the price increase coincides with rising rather than falling
platform presence among treated products and is robust to controlling for
listing frequency, which together rule out the most direct form of the
demand-contraction hypothesis and confirm that the estimated effect reflects
within-unit repricing rather than compositional shifts.

\section{Conclusions} 
\label{sec:conclusion}

This paper asked whether the Austrian SUP compliance regime---reporting
obligations, system-participation requirements, and per-quantity fee settlement
introduced under Directive (EU) 2019/904---passes through to posted online
retail prices. The pooled answer is yes: treated products are on average about
$4.1\%$ more expensive after the first payment date relative to the control
group. The more revealing finding is in the timing. The sequential TWFE
model puts the larger price adjustment in the reporting phase---around $8.1\%$---
with an incremental payment-phase effect of $5.6\%$. Sellers began repricing
before any fee was due, consistent with anticipatory pass-through of expected
compliance costs rather than a contemporaneous reaction to realized payments.
For balloons---a category with direct and transparent SUP exposure---event-study
coefficients exceed $50\%$ in the months immediately following the first payment
date and remain elevated through most of the post-treatment window before fading
around December 2024.

The additional analyses in Section \ref{sec:additional} sharpen this picture
along four dimensions. Category-level pass-through is tied to the statutory fee
schedule: the \euro450/ton tier (balloons and tobacco filters) averages 0.403
log points against $-0.023$ for the \euro225/ton group, a difference of 0.427
($p=0.035$) that supports a causal reading of the pooled estimate. Restricting
to a balanced sub-panel of long-lived listings raises the payment-due coefficient
from 0.394 to 0.541, arguing against compositional exit as the primary driver
of the baseline result. The pricing response is concentrated among standalone
e-tailers (payment-due coefficient 0.494) rather than marketplace sellers (0.080),
pointing to differences in compliance cost structure or pricing technology between
the two segments. And listing frequency and days covered both rise after the
payment date rather than fall, which is inconsistent with demand contraction and
consistent with sellers actively updating price lists and refreshing listings in
response to the compliance burden.

Four conclusions follow. First, SUP compliance costs in Austrian online retail
are not fully absorbed by producers: a meaningful share reaches posted prices,
and the pass-through is larger for the more heavily taxed categories. Second,
the reporting obligation alone---before any fee is due---is sufficient to trigger
price adjustment, consistent with administrative salience rather than only the
monetary charge driving firm behavior \citep{homonoff2018bags}. Third, the pooled estimate is a lower bound for the most exposed categories: the \euro450/ton
tier, balloons in particular, shows effects many times larger, and the parallel
analysis in \citet{reichel2025sup} finds full-period contrasts of around
20 index points for Austrian balloons in a price-index design. Fourth, the demand
proxies available in the platform data---spell counts and days covered---point
toward repricing and listing expansion rather than market retraction on the treated
side, suggesting that the price increase is not primarily a composition artifact.

\subsection{Limitations}
\label{subsec:limitations}

The main limitations are discussed where they arise---the control-group distance
and baseline-survivor restriction in Section \ref{subsec:control}, the identification threats including lagged energy confounds, composition effects,
and demand-side shifts in Section \ref{subsec:threats}. Two points are worth
restating. The estimand is internal to the treated-versus-non-SUP comparison and
should not be read as representative of the full Austrian online retail market.
The data contain prices but not quantities, so pass-through can be measured but
consumption responses, substitution toward reusable alternatives, and downstream
environmental effects cannot.

\subsection{Directions for Future Research}
\label{subsec:future}

Several of the extensions flagged in the body of the paper remain open. On the
demand side, transaction-level click data, which would allow a
cleaner test of whether the price increases documented here translate into
substitution toward non-SUP alternatives---the mechanism the regulation is
designed to activate. The listing-frequency results in Section \ref{subsec:demand}
are suggestive but cannot settle this, since spell counts measure market presence
rather than sales. Scanner data, platform transaction records, or linked
administrative data on reported quantities would convert the price-incidence
evidence into a welfare and environmental assessment.

On the supply side, the seller-type split in Section \ref{subsec:seller_type}
shows that standalone e-tailers drive the result while marketplace sellers show
much smaller effects. Understanding whether this reflects differences in
compliance cost structure, pricing technology, or the platform's own role in
absorbing or transmitting regulatory costs is a natural next step. A structural
pass-through model that maps the Austrian fee schedule onto observed price changes
by seller type and category would also allow a sharper test of whether the
magnitude of the response is consistent with full, partial, or more-than-full
pass-through given the statutory rate.

Finally, adding 2025 data would allow cleaner separation of permanent price-level
changes from temporary spikes, and reveal whether competitive pressure eventually
erodes the pass-through visible in the post-payment window.

\subsection{An Ex Post Identification Challenge: The 2026 Petrochemical Shock}
\label{subsec:hormuz}

Any extension of this analysis into 2026 data will face a severe new
identification problem. The geopolitical disruption of early 2026, and the
associated near-closure of a key transit route for seaborne oil and liquefied
natural gas, generated a sharp simultaneous increase in petrochemical feedstock
costs---naphtha, polyethylene, polypropylene, PET---of a magnitude with no close
historical precedent \citep{cnbc_hormuz_blockade2026,packaginggw_oilshock2026}.
Because SUP products are manufactured primarily from these feedstocks, any study
extending an Austrian panel into 2026 will observe a cost shock entirely unrelated
to SUP compliance obligations that will affect treated categories---food containers,
beverage cups, bottles, wraps---far more than a non-SUP control group. This is
structurally identical to the 2022--2023 energy confound discussed in
Section \ref{subsec:threats}, but larger in magnitude and more concentrated in
the specific inputs that go into SUP products. The category-level heterogeneity
results in Section \ref{subsec:cat_het} make this especially relevant: the same
categories that show the largest pass-through---food containers, balloons,
tobacco filters---are also among the most petrochemical-intensive, so a feedstock
shock would generate exactly the treated--control divergence a DiD design would
otherwise attribute to the regulation.

Four design adaptations would be needed for any evaluation covering 2026 data.
First, the control group should be selected with explicit attention to
petrochemical input intensity, not only platform-level comparability, so the
feedstock shock affects treated and control categories as symmetrically as
possible. Second, input-price controls should include product-specific resin
series---polypropylene, PET, polyethylene---rather than only broad energy indices.
Third, any event window covering 2026 should treat the shock onset as a named
identification threat rather than absorbing it silently into time fixed effects.
Fourth, a triple-difference design exploiting cross-category variation in resin
intensity alongside the SUP treatment variable could partially separate the
regulatory effect from the feedstock shock, given a credible measure of
category-level input intensity.

The 2022--2023 Ukrainian energy shock and the 2026 petrochemical disruption are two
episodes from the same underlying source of identification risk. Research designs
in petrochemical-intensive sectors should treat input-price heterogeneity as a
first-order threat rather than a robustness footnote.

\subsection{Broader Context}
\label{subsec:context}

The finding that Austrian online retailers pass SUP compliance costs through to
posted prices extends a body of evidence showing that producer-facing
environmental charges are not silently absorbed in the supply chain
\citep{fullerton2001environmental,weyl2013passthrough,marion2011fueltax,
besley1999salestax,convery2007irish,homonoff2018bags}. The fee-tier monotonicity
result---larger pass-through for \euro450/ton than \euro225/ton
products---provides within-sample evidence that the magnitude of the regulatory
burden matters, complementing the cross-study comparisons in the broader
literature. The paper also complements \citet{reichel2025sup}, who reaches
similar qualitative conclusions using a price-index event-study design and finds
pooled DiD contrasts of around 13 index points over twelve months and 19 index
points over the full period. Two studies using different outcome metrics,
aggregation approaches, and control group definitions consistently find positive
price incidence for Austrian SUP products in online retail, and the category and
seller heterogeneity documented here suggests that the aggregate figure masks
substantial variation in who bears the cost and how quickly it is passed on.

If the goal of SUP charges is to shift relative prices toward non-regulated
alternatives, the evidence suggests the mechanism is at least partially operative
in Austrian online retail. Whether the response is large enough to materially
reduce SUP consumption or generate environmentally meaningful reductions in
litter cannot be assessed without a greater quantity of data---
that remains the binding
constraint for any full welfare evaluation of the regime.

\vspace{6pt}

\section*{Funding}
This research was supported by the Johannes Kepler University Open Access Publishing Fund and the federal state of Upper Austria.

\section*{Data availability}
The data used in this article is subject to third-party licensing restrictions and therefore cannot be shared publicly. A small subsample for replication purposes may be made publicly available upon request via the authors' GitHub page.

\section*{Acknowledgments}
I thank my supervisor for helpful comments on the initial version of the manuscript and its revision. I also thank the three anonymous reviewers for their valuable comments and evaluations of the manuscript. Finally, I thank the editor for handling the manuscript and for allowing sufficient time to prepare the second revision while I was engaged in graduate studies.

\section*{Conflicts of interest}
The author declares no
conflicts of interest.



\begin{thebibliography}{999}

\bibitem[{European Union}(2019)]{eu2019sup}
{European Union}.
\newblock Directive (EU) 2019/904 of the European Parliament and of the Council
  of 5 June 2019 on the reduction of the impact of certain plastic products on
  the environment,  2019.
\newblock \textit{Off. J. Eur. Union} \textit{L 155}, 1--19.

\bibitem[Fullerton and Metcalf(2001)]{fullerton2001environmental}
Fullerton, D.; Metcalf, G.E.
\newblock Environmental controls, scarcity rents, and pre-existing distortions.
\newblock {\em J. Public Econ.} {\bf 2001}, {\em 80}, 249--267.
\newblock {\url{https://doi.org/10.1016/S0047-2727(00)00087-6}}.

\bibitem[Weyl and Fabinger(2013)]{weyl2013passthrough}
Weyl, E.G.; Fabinger, M.
\newblock Pass-through as an economic tool: Principles of incidence under
  imperfect competition.
\newblock {\em J. Political Econ.} {\bf 2013}, {\em 121}, 528--583.
\newblock {\url{https://doi.org/10.1086/670401}}.

\bibitem[Marion and Muehlegger(2011)]{marion2011fueltax}
Marion, J.; Muehlegger, E.
\newblock Fuel tax incidence and supply conditions.
\newblock {\em J. Public Econ.} {\bf 2011}, {\em 95}, 1202--1212.
\newblock {\url{https://doi.org/10.1016/j.jpubeco.2011.04.003}}.

\bibitem[Besley and Rosen(1999)]{besley1999salestax}
Besley, T.J.; Rosen, H.S.
\newblock Sales taxes and prices: An empirical analysis.
\newblock {\em Natl. Tax J.} {\bf 1999}, {\em 52}, 157--178.
\newblock {\url{https://doi.org/10.1086/NTJ41789387}}.

\bibitem[Convery et al.(2007)Convery, McDonnell, and
  Ferreira]{convery2007irish}
Convery, F.; McDonnell, S.; Ferreira, S.
\newblock The most popular tax in Europe? Lessons from the Irish plastic bags
  levy.
\newblock {\em Environ. Resour. Econ.} {\bf 2007}, {\em
  38}, 1--11.
\newblock {\url{https://doi.org/10.1007/s10640-006-9059-2}}.

\bibitem[Homonoff(2018)]{homonoff2018bags}
Homonoff, T.A.
\newblock Can small incentives have large effects? The impact of taxes versus
  bonuses on disposable bag use.
\newblock {\em Am. Econ. Journal: Econ. Policy} {\bf 2018}, {\em
  10}, 177--210.
\newblock {\url{https://doi.org/10.1257/pol.20150261}}.

\bibitem[Walls(2011)]{walls2011drs}
Walls, M.
\newblock \textit{Deposit--Refund Systems in Practice and Theory};
\newblock Discussion Paper 11-47; Resources for the Future: Washington, DC, USA, 2011.

\bibitem[{European Parliament and Council}(2025)]{bmk2024}
European Parliament and Council.
\newblock Regulation (EU) 2025/40 of the European Parliament and of the Council of 19 December 2024 on packaging and packaging waste, amending Regulation (EU) 2019/1020 and Directive (EU) 2019/904, and repealing Directive 94/62/EC.
\newblock \emph{Off. J. Eur. Union} \textbf{2025}.
\newblock Available online: \url{http://data.europa.eu/eli/reg/2025/40/oj} (accessed on 18 May 2026).


\bibitem[Reichel(2025)]{reichel2025sup}
Reichel, F.
\newblock Political Interventions to Reduce Single-Use Plastics ({SUPs}) and
  Price Effects: An Event Study for {Austria} and {Germany}.
\newblock \textit{arXiv} \textbf{2025}, arXiv:2510.15617v2.

\bibitem[Roth et al.(2023)Roth, Sant'Anna, Bilinski, and Poe]{roth_etal_2023}
Roth, J.; Sant'Anna, P.H.C.; Bilinski, A.; Poe, J.
\newblock What's Trending in Difference-in-Differences? A Synthesis of the
  Recent Econometrics Literature.
\newblock {\em J. Econom.} {\bf 2023}, {\em 235}, 2218--2244.
\newblock {\url{https://doi.org/10.1016/j.jeconom.2023.03.008}}.
\bibitem[{Altstoff Recycling Austria AG}(2026)]{ara_licensing_2026}
{Altstoff Recycling Austria AG}.
\newblock Lizenzierungsservices f{\"u}r Verpackungen. 2026.
\newblock Available online: \url{https://www.ara.at/ara-lizenzierung-verpackungen} (accessed on 14 April 2026).

\bibitem[{Wirtschaftskammer {\"O}sterreich}(2026)]{wko_packaging_2026}
{Wirtschaftskammer {\"O}sterreich}.
\newblock Information on the Austrian Packaging Ordinance 2014. 2026.
\newblock Available online: \url{https://www.wko.at/en/information-on-the-austrian-packaging-ordinance-2014} (accessed on 14 April 2026).

\bibitem[{Umweltbundesamt (UBA)}(2024)]{uba2024}
{Umweltbundesamt (UBA)}.
\newblock Onlineplattform DIVID des Einwegkunststofffonds Gestartet. 2024.
\newblock Available online: \url{https://www.umweltbundesamt.de/presse/pressemitteilungen/onlineplattform-divid-des-einwegkunststofffonds} (accessed on 14 April 2026).

\bibitem[{Packaging Europe}(2022)]{packagingeurope2022commodity}
{Packaging Europe}.
\newblock What Do Rising Commodity Prices Linked with the War in Ukraine Mean
  for Glass, Metal and Plastic Packaging? 2022.
\newblock Available online: \url{https://packagingeurope.com/news/what-do-rising-commodity-prices-linked-with-the-war-in-ukraine-mean-for-glass-metal-and-plastic-packaging/8266.article} (accessed on 14 April 2026).

\bibitem[{Plastics Europe}(2022)]{plasticseurope2022energy}
{Plastics Europe}.
\newblock The Impact of the Energy Crisis on the European Plastics
  Manufacturers. 2022.
\newblock Available online: \url{https://plasticseurope.org/media/the-impact-of-the-energy-crisis-on-the-european-plastics-manufacturers/} (accessed on 14 April 2026).

\bibitem[Roth and Sant'Anna(2023)]{roth_santanna_2023}
Roth, J.; Sant'Anna, P.H.C.
\newblock When Is Parallel Trends Sensitive to Functional Form?
\newblock {\em Econometrica} {\bf 2023}, {\em 91}, 737--747.
\newblock {\url{https://doi.org/10.3982/ECTA19402}}.

\bibitem[Callaway and Sant'Anna(2021)]{callaway2021multiple}
Callaway, B.; Sant'Anna, P.H.C.
\newblock Difference-in-differences with multiple time periods.
\newblock {\em J. Econom.} {\bf 2021}, {\em 225}, 200--230.

\bibitem[Sun and Abraham(2021)]{sun2021dynamic}
Sun, L.; Abraham, S.
\newblock Estimating dynamic treatment effects in event studies with
  heterogeneous treatment effects.
\newblock {\em arXiv} {\bf 2021}, arXiv:1804.05785.

\bibitem[Bertrand et al.(2004)Bertrand, Duflo, and
  Mullainathan]{bertrand2004trust}
Bertrand, M.; Duflo, E.; Mullainathan, S.
\newblock How much should we trust differences-in-differences estimates?
\newblock {\em Q. J. Econ.} {\bf 2004}, {\em 119}, 249--275.

\bibitem[Cameron et al.(2011)Cameron, Gelbach, and Miller]{cameron2011multiway}
Cameron, A.C.; Gelbach, J.B.; Miller, D.L.
\newblock Robust inference with multiway clustering.
\newblock {\em J. Bus. Econ. Stat.} {\bf 2011}, {\em 29}, 238--249.
\newblock {\url{https://doi.org/10.1198/jbes.2010.07136}}.

\bibitem[Hackl et al.(2014)Hackl, Kummer, Winter-Ebmer, and
  Zulehner]{hackl2014marketstructure}
Hackl, F.; Kummer, M.E.; Winter-Ebmer, R.; Zulehner, C.
\newblock Market Structure and Market Performance in E-Commerce.
\newblock {\em Eur. Econ. Rev.} {\bf 2014}, {\em 68}, 199--218.
\newblock {\url{https://doi.org/10.1016/j.euroecorev.2014.03.007}}.

\bibitem[Roth(2022)]{roth_2022}
Roth, J.
\newblock Pretest with Caution: Event-Study Estimates after Testing for
  Parallel Trends.
\newblock {\em Am. Econ. Rev. Insights} {\bf 2022}, {\em 4}, 305--322.
\newblock {\url{https://doi.org/10.1257/aeri.20210236}}.

\bibitem[Rambachan and Roth(2023)]{rambachan_roth_2023}
Rambachan, A.; Roth, J.
\newblock A More Credible Approach to Parallel Trends.
\newblock {\em Rev. Econ. Stud.} {\bf 2023}, {\em 90}, 2555--2591.
\newblock {\url{https://doi.org/10.1093/restud/rdad018}}.

\bibitem[{CNBC}(2026)]{cnbc_hormuz_blockade2026}
{CNBC}.
\newblock \textit{Hormuz Blockade Could Deepen World's Worst Energy Crisis---And Risk a Dangerous Misstep};
  CNBC: Englewood Cliffs, NJ, USA, 2026.
\newblock Available online: \url{https://www.cnbc.com/} (accessed on 14 April 2026).

\bibitem[{Packaging Gateway}(2026)]{packaginggw_oilshock2026}
{Packaging Gateway}.
\newblock \textit{Oil Shock Sends Packaging Costs Soaring};
  Packaging Gateway: London, UK, 2026.
\newblock Available online: \url{https://www.packaging-gateway.com/} (accessed on 14 April 2026).

\end{thebibliography}

\appendix
\appendix
\setcounter{table}{0}
\setcounter{figure}{0}
\section[\appendixname \thesection]{Guide to the Online Appendix
}

This online appendix contains material that supports but does not repeat the
main text.  Appendix \ref{app:sample_construction} documents the construction of treated and
 control samples, including the full keyword-matching rules and
the baseline-survival restriction.  Appendix \ref{app:data_dictionary}
provides a variable-level description of the raw offer-spell files and the
derived unit--month panel.  Appendix \ref{app:pipeline} sets out the
step-by-step panel construction pipeline and reports sample counts.
Appendix \ref{app:figures} contains figures that do not appear in the main
text: the energy-control correlation matrix, coefficient plots based on the
disaggregated (raw offer spell-level) data that complement the aggregated monthly duration-weighted main
results, and the energy input-price series.
Appendix \ref{app:tables} collects the descriptive and regression tables
referenced in the main text that are not reproduced there.
Appendix \ref{app:additional} provides brief summaries in prose and
cross-references for additional analyses in Section \ref{sec:additional}; the corresponding
figures appear in the main text as Figures \ref{fig:A1a}--\ref{fig:A5b} and are not repeated here. A much longer more illuminating appendix including even more figures will be made available online by the author in the near future upon request.
\section[\appendixname \thesection]{Sample Construction and Counterfactual Design}
\label{app:sample_construction}

This appendix documents the construction of the treated and counterfactual
samples from the underlying retail offer-spell data and clarifies the gap
between the counterfactual design initially envisaged and the control group
ultimately implemented in the estimation code.  Because the credibility of
the empirical design turns in part on the comparability of treated and
untreated observations, these sample-construction choices warrant explicit
documentation.

The raw data comprise offer-level observations spanning 2020:01--2024:12 (till early 2025 for some spells).
Let $s$ index offer spells, where each spell corresponds to a
product--retailer listing observed over a finite interval defined by start
and end timestamps.  Two raw input files underlie the analysis sample:
\emph{plastics\_regulation\_obs.csv}, which contains the treated-side
observations, and \emph{plastics\_regulation\_counterfactual\_obs.csv},
which contains the counterfactual observations.
Appendix \ref{app:data_dictionary} provides a variable-level description of
these files.

\subsection[\appendixname \thesubsection]{Construction of the Treated Sample}
\label{app:treated_sample}

The treated sample is defined through keyword matches in product titles,
recorded in the raw variable $\mathit{produkt\_bez}$.  Matching is
implemented through case-insensitive string patterns designed to identify
product groups plausibly exposed to the Austrian SUP-related regulatory regime.  Let $M_i^{T} \in \{0,1\}$ denote an indicator equal to
one if product $i$ matches at least one treated keyword pattern, and zero
otherwise.  A product is assigned to the treated sample whenever
$M_i^{T}=1$, subject to the subsequent cleaning, deduplication, and panel construction steps of the empirical pipeline.

Substantively, the aim is to capture product classes that either fall
directly within the regulatory scope or are sufficiently close to the
affected domain that their prices may respond to the altered compliance
environment.  Table \ref{tab:treated_keywords_appendix} reports the keyword
categories used to construct the treated sample.  The first column gives the
internal category label used in the data pipeline, the second provides a
natural English translation, and the third lists the underlying keyword
patterns.  These categories should be read as operational search rules
rather than as a legal classification in the strict statutory sense.

\begin{table}[H]
\caption{Keyword categories used to construct the treated sample.}
\label{tab:treated_keywords_appendix}
\footnotesize

\setlength{\cellWidtha}{\fulllength/3-2\tabcolsep-1in}
\setlength{\cellWidthb}{\fulllength/3-2\tabcolsep-00in}
\setlength{\cellWidthc}{\fulllength/3-2\tabcolsep+1in}
\scalebox{1}[1]{\begin{tabularx}{\fulllength}{>{\raggedright\arraybackslash}m{\cellWidtha}>{\raggedright\arraybackslash}m{\cellWidthb}>{\raggedright\arraybackslash}m{\cellWidthc}}
\toprule
\textbf{SUP Category} & \textbf{Translation} &
  \textbf{Keyword Patterns} \\
\midrule
tabak &
Tobacco-related products and filters &
tabakfilter; zigarettenfilter; zigarette; rauchwaren; nikotinprodukt;
tabakprodukt; e-zigarette; rauchger\"{a}t; tabakr\"{o}hre;
filterzigarette \\
becher &
Single-use cups and to-go drink cups &
to-go becher; einwegbecher; kaffeebecher; kunststoffbecher; trinkbecher;
coffee to go; takeaway becher; getr\"{a}nkebecher; wegwerfbecher;
plastikbecher \\
lebensmittelbehaelter &
Single-use food containers &
lebensmittelbeh\"{a}lter; takeaway box; to-go beh\"{a}lter; einwegbox;
essensbox; men\"{u}schale; mittagsschale; essensbeh\"{a}lter; food
container; kunststoffschale \\
tueten\_folien &
Plastic wraps, films, and small packaging &
folienverpackung; verpackungsfolie; plastikfolie; s\"{a}ckchen;
t\"{u}tchen; verpackungseinheit; folie verpackung; beutelverpackung;
kunststoffverpackung; kleinverpackung \\
flaschen\_ohne\_pfand &
Non-deposit beverage bottles, especially disposable plastic bottles &
getr\"{a}nkeflasche; pet flasche; einwegflasche; kunststoffflasche;
saftflasche; wasserflasche; getr\"{a}nkebeh\"{a}lter; softdrinkflasche;
limonadenflasche; to-go flasche \\
flaschen\_mit\_pfand &
Deposit and reusable beverage bottles &
mehrwegflasche; pfandflasche; getr\"{a}nkeflasche pfand;
getr\"{a}nkebeh\"{a}lter pfand; r\"{u}ckgabeflasche; flasche mit pfand;
getr\"{a}nkeflasche mehrweg \\
plastiktueten &
Plastic carrier bags and shopping bags &
kunststofftragetasche; plastikt\"{u}te; einkaufst\"{u}te; leichte
tragetasche; d\"{u}nne plastikt\"{u}te; tragetasche einweg;
einwegtragetasche; t\"{u}te plastik; kleine plastiktasche \\
%
%
%
%
feuchttuecher &
Wet wipes and disposable cleaning or hygiene wipes &
feuchttuch; reinigungstuch; hygienetuch; babyfeuchttuch; pflegetuch;
intimtuch; kosmetiktuch; einwegtuch; nassreinigungstuch;
desinfektionstuch \\
luftballons &
Balloons and balloon decoration products &
luftballon; ballon latex; partyballon; heliumballon; einwegballon;
deko ballon; ballonset; kinderballon; ballon dekoration \\
\bottomrule
\end{tabularx}}

\end{table}

Products matching one or more of these patterns are assigned to the treated
sample.  The treated sample is therefore title-defined.  Its coverage
consequently depends on the informativeness and consistency of product naming
conventions in the marketplace data.

\subsection[\appendixname \thesubsection]{Intended Counterfactual Design}
\label{app:intended_counterfactual}

The original counterfactual design was meant to select products similar in
retail context and use environment, yet not directly exposed to the
SUP regime.  Let $M_i^{C,0} \in \{0,1\}$ denote an indicator for
membership in this originally intended counterfactual design.  The guiding
idea was to construct a comparison group from reusable, natural-material,
refillable, or otherwise non-regulated alternatives to the treated product
groups.  Such a design would have produced a conceptually closer control
group by comparing treated products with adjacent goods that occupy similar
retail environments while plausibly remaining outside the direct scope of
the regulation.

Table \ref{tab:counterfactual_keywords_appendix} lists these intended
counterfactual categories.  As with the treated keywords, the first column
gives the internal design label, the second provides a natural English
translation, and the third reports the underlying keyword patterns.  These
categories are best understood as an empirical approximation to a conceptual
design rather than as the final implemented sample rule.

\begin{table}[H]
\caption{Intended keyword categories for the counterfactual sample.}
\label{tab:counterfactual_keywords_appendix}
\small

\setlength{\cellWidtha}{\fulllength/3-2\tabcolsep-0.5in}
\setlength{\cellWidthb}{\fulllength/3-2\tabcolsep-0.3in}
\setlength{\cellWidthc}{\fulllength/3-2\tabcolsep+0.8in}
\scalebox{1}[1]{\begin{tabularx}{\fulllength}{>{\raggedright\arraybackslash}m{\cellWidtha}>{\raggedright\arraybackslash}m{\cellWidthb}>{\raggedright\arraybackslash}m{\cellWidthc}}
\toprule
\textbf{Category} & \textbf{Translation} &
  \textbf{Keyword Patterns} \\
\midrule
Tabakprodukte---rare substitutes &
Tobacco alternatives and non-standard smoking substitutes &
\%kr\"{a}uterzigarette\%; \%pfeife\%; \%zigarre\%; \%snus\%;
\%verdampfer ohne nikotin\%; \%rauchfreies nikotin\% \\
To-go cups---reusable or natural materials &
Reusable or natural-material drink cups &
\%keramikbecher\%; \%emaillebecher\%; \%kupferbecher\%;
\%kokusnussbecher\%; \%mehrweg goblet\%; \%glas tumbler\% \\
Food containers---non-plastic materials &
Food containers of glass, ceramic, wood, or similar materials &
\%tiffin box\%; \%glasdose\%; \%keramikbeh\"{a}lter\%;
\%holzbox\%; \%wachstuchbox\% \\
Bags and wraps---natural or biodegradable materials &
Natural-material bags, wraps, and biodegradable packaging &
\%wachstuch\%; \%stoffverpackung\%; \%juteS\"{a}ckchen\%;
\%leinenbeutel\%; \%reispapierverpackung\% \\
Bottles without deposit---alternative materials &
Non-deposit bottles made from alternative materials &
\%glasflasche klein\%; \%keramikflasche\%; \%steinzeugflasche\%;
\%kupferflasche\%; \%emailleflasche\%; \%bambusflasche\% \\
Bottles with deposit---reusable alternatives &
Reusable and refillable bottle alternatives &
\%milchflasche glas\%; \%bierflasche glas\%; \%nachf\"{u}llflasche\%;
\%tee flasche glas\%; \%flasche aus holz\% \\
Carrier bags---textile-based alternatives &
Textile and reusable carrier bags &
\%baumwolltasche handgefertigt\%; \%jute beutel bio\%;
\%papiert\"{u}te deluxe\%; \%leinentasche\%; \%upcycling tasche\%;
\%netztasche\% \\
Wet-wipe alternatives---textile products &
Reusable cloth-based wipe alternatives &
\%baumwolltuch\%; \%waschlappen bio\%; \%leinen tuch\%;
\%textiltuch\%; \%nachhaltiges pflegetuch\% \\
Balloon substitutes---decorative alternatives &
Non-balloon decorative substitutes &
\%stoffgirlande\%; \%papierrosette\%; \%wimpelkette\%;
\%papierlaterne\%; \%naturdeko\%; \%holzdeko\%; \%stoffblume\% \\
\bottomrule
\end{tabularx}}

\end{table}

Had this broader design been implemented in full, the counterfactual sample
would have resembled a material- and use-case-adjacent comparison group.  In that sense, the intended design was conceptually stronger than a generic
untreated control because it sought to hold fixed at least part of the
retail environment in which treated products were sold.

\subsection[\appendixname \thesubsection]{Implemented Counterfactual Sample}
\label{app:implemented_counterfactual}

In the final implementation, the counterfactual product list was not the
broader set of adjacent alternatives summarized in
Table \ref{tab:counterfactual_keywords_appendix}.  Instead, the implemented
control sample was narrowed to a single keyword, \emph{grafikkarte}
(graphics card).  Let $M_i^{C} \in \{0,1\}$ denote the implemented
counterfactual-match indicator.  In the final code, $M_i^{C}=1$ if and only
if the product title matches this keyword rule and satisfies the additional
life-cycle restrictions discussed below.

The implemented counterfactual sample is therefore not a broad
taxonomy-matched control group based on reusable or non-regulated
substitutes.  Rather, it is a selected untreated sample defined by one
specific product keyword.  This narrowing matters for the economic
interpretation of the estimated treatment effects because the untreated
comparison group is now drawn from a more specialized product domain that is
unlikely to reproduce the full retail context of the treated goods.  Graphics
cards share the same platform infrastructure and price-reporting conventions
as the treated products but are driven by entirely different cost
factors---semiconductor cycles and, during part of the sample,
cryptocurrency-related demand.  If an effect is detectable against this
economically distant counterfactual, it is unlikely to reflect correlated
cost shocks between treated and control products.  Results under the broader
non-SUP control pool are reported in the main text.

\subsection[\appendixname \thesubsection]{Implemented Counterfactual Filter and Baseline-Survivor
  Restriction}
\label{app:survivor_restriction}

The control sample is further restricted by an explicit baseline-survivor
filter.  Let $b_i$ denote the product birth timestamp and $d_i$ the product
death timestamp.  Let $t_0^{\text{base}}$ denote the baseline timestamp
corresponding to 1 January 2020 00:00:00 UTC; in the data pipeline,
$t_0^{\text{base}} = 1{,}577{,}836{,}800$.  The implemented counterfactual
sample can then be written as
\[
  \mathcal{C}
  =
  \Bigl\{
    i : M_i^{C}=1,\;
    b_i \le t_0^{\text{base}},\;
    \bigl(d_i=\varnothing \;\text{or}\; d_i > t_0^{\text{base}}\bigr)
  \Bigr\}.
\]
Operationally, this means that the implemented control sample is restricted
to products that were already active at the start of the sample period.  All
products entering after 2020:01 are excluded from the control group by
construction, as are products that had already disappeared before the
initial sample month.  The same filter applies in the graphics card
stress-test specification.

\subsection{Implications for Panel Composition and Identification}
\label{app:panel_composition}

The baseline-survivor restriction has direct implications for panel
composition and, in turn, for the interpretation of the identifying
comparison.  Because the control sample is conditioned on being alive at the
beginning of the sample window, it mechanically over-represents older, more
persistent, and potentially more established products.  Conversely, it
under-represents short-lived products, later entrants, and products with more
intermittent market presence.

Formally, the observed control sample is not drawn from the full untreated
population $\mathcal{U}$, but from the selected subset
$\mathcal{C} \subset \mathcal{U}$ satisfying the survival condition at
$t_0^{\text{base}}$.  Three consequences follow: the panel is tilted toward
incumbent products; if turnover patterns differ between treated and untreated
goods, the restriction induces differential sample selection unrelated to the
regulation; and the final panel represents a survivor cohort defined at
baseline rather than the full market over 2020--2025.

This is not random attrition but systematic conditioning on survival at the
start of the window.  If survival correlates with price levels, retailer
persistence, assortment quality, or product type, treatment--control
comparability may be affected.  The design remains valid, but the estimand
is narrower: the treatment effect should be read relative to a specific
untreated survivor cohort rather than the full population of available
untreated goods.  The parallel-trend diagnostics in the main text (Figure \ref{fig:parallel_trends_by_category_text})
should be read with both the composition of the control pool and this
survival conditioning in mind.

\subsection[\appendixname \thesubsection]{Summary of Interpretation}
\label{app:design_summary}

Four features of the sample-construction process are central for
interpretation.  First, the treated sample is defined through keyword matches
in product titles.  Second, the intended counterfactual design originally
relied on adjacent but plausibly non-regulated product alternatives.  Third,
the implemented counterfactual design was narrowed to the single keyword
\emph{grafikkarte} (graphics card).  Fourth, the implemented control sample
excludes post-2020 entrants through a baseline-survivor restriction.

Taken together, these choices imply that the control group should not be
interpreted as a broad taxonomy-matched market comparison.  Instead, it is a
selected counterfactual sample defined by a specific keyword and conditioned
on survival at the start of the observation window.  The empirical estimates
should be interpreted against that design choice throughout the paper.

\section[\appendixname \thesection]{Data Dictionary}
\label{app:data_dictionary}

This appendix documents the variables contained in the raw retail
offer-spell files used to construct the treated and counterfactual samples
described in Appendix \ref{app:sample_construction}.  The underlying files
are \emph{plastics\_regulation\_obs.csv} and
\emph{plastics\_regulation\_counterfactual\_obs.csv}.  Both contain retail
offer spells observed over 2020:01--2025.  The unit of observation is an
individual retail offer spell---a listing for a product at a particular
retailer that remains active over an interval defined by start and end
timestamps.

\subsection[\appendixname \thesubsection]{Notes on Variable Encoding}
\label{app:dict_notes}

Several coding details warrant explicit mention.  Timestamp variables such
as $\mathit{dtimebegin}$, $\mathit{dtimeend}$, $\mathit{dtime\_birth}$, and $\mathit{dtime\_death}$ are stored as Unix epoch time in seconds; conversion
uses \texttt{as.POSIXct(., origin = ``1970-01-01'', tz = ``UTC'')}.  Variables
prefixed by country codes ($\mathit{oe\_}$, $\mathit{de\_}$,
$\mathit{is\_}$, $\mathit{liefert\_}$) encode market-specific conditions,
seller origin, or delivery availability.  Several fields appear to have been
duplicated during joins or merge operations ($\mathit{produkt\_id\_1}$,
$\mathit{haendler\_bez\_1}$, $\mathit{dtimebegin\_1}$,
$\mathit{dtimeend\_1}$); these are noted as such in
Table \ref{tab:data_dictionary_main}.
The empirical analysis relies primarily on five groups of variables: (i)
identifiers ($\mathit{angebot\_id}$, $\mathit{produkt\_id}$,
$\mathit{haendler\_bez}$); (ii) timing variables ($\mathit{dtimebegin}$,
$\mathit{dtimeend}$, $\mathit{week}$); (iii) core price measures
($\mathit{preis\_min}$, $\mathit{preis\_avg}$, $\mathit{preis\_max}$);
(iv) sample-assignment variables ($\mathit{produkt\_bez}$,
$\mathit{subsubkat}$); and (v) market-coverage and seller-characteristic
variables ($\mathit{is\_at}$, $\mathit{is\_de}$, $\mathit{is\_uk}$,
$\mathit{is\_nl}$, $\mathit{is\_pl}$, and the delivery indicators
$\mathit{liefert}_{\ell}$).  Together, these variables are sufficient to
reconstruct panel timing, define treated and control observations, and derive the main price and availability outcomes.

\subsection[\appendixname \thesubsection]{Raw Offer-Spell Files}
\label{app:dict_raw}

\vspace{-6pt}
\begin{table}[H]
\caption{Variable-level data dictionary for the raw retail
  offer-spell files.}
  \label{tab:data_dictionary_main}

\footnotesize
\setlength{\cellWidtha}{\fulllength/4-2\tabcolsep-0.8in}
\setlength{\cellWidthb}{\fulllength/4-2\tabcolsep-1.1in}
\setlength{\cellWidthc}{\fulllength/4-2\tabcolsep+1.2in}
\setlength{\cellWidthd}{\fulllength/4-2\tabcolsep+0.7in}
\scalebox{1}[1]{\begin{tabularx}{\fulllength}{>{\raggedright\arraybackslash}m{\cellWidtha}>{\raggedright\arraybackslash}m{\cellWidthb}>{\raggedright\arraybackslash}m{\cellWidthc}>{\raggedright\arraybackslash}m{\cellWidthd}}
\toprule
\textbf{Raw Variable} & \textbf{Type} & \textbf{Description} &
  \textbf{Notes/Example} \\
\midrule
angebot\_id        & integer    & Unique identifier for the offer-spell observation.
  & Offer-level primary key. \\
produkt\_id        & integer    & Product identifier.
  & Links multiple offers to the same product. \\
haendler\_bez      & string     & Retailer or seller name/identifier.
  & Example: amazon-de. \\
preis\_min         & numeric    & Minimum observed price during the offer spell.
  & Measured in observed currency units. \\
preis\_avg         & numeric    & Average observed price during the offer spell.
  & In some rows equal to preis\_min and preis\_max. \\
preis\_max         & numeric    & Maximum observed price during the offer spell.
  & Captures within-spell price variation. \\
avail              & integer    & Availability status code.
  &  \\
oe\_vk             & num./ind.  & Austria-specific shipping or sales condition. 
  &  \\
oe\_nn             & num./ind.  & Austria-specific condition variable. 
  &  \\
de\_vk             & num./ind.  & Germany-specific shipping or sales condition. 
  &  \\
de\_nn             & num./ind.  & Germany-specific condition variable. 
  &  \\
oe\_kr             & numeric    & Austria-specific cost measure, plausibly shipping cost. 
  & Missing in some rows. \\
de\_kr             & numeric    & Germany-specific cost measure, plausibly shipping cost. 
  & Example values include 3.99. \\
anz\_angebote      & integer    & Number of offers associated with the product or spell.
  & Likely contemporaneous offer count. \\
dtimebegin         & Unix time  & Start timestamp of the offer spell.
  & Integer; Unix epoch seconds. \\
dtimeend           & Unix time  & End timestamp of the offer spell.
  & Integer; Unix epoch seconds. \\
produkt\_id\_1     & integer    & Duplicate or joined product identifier.
  & Matches produkt\_id in example rows. \\
dtime\_birth       & Unix time  & Product birth or first-seen timestamp.
  & Product-level life-cycle marker. \\
dtime\_death       & Unix time  & Product death or last-seen timestamp.
  & Product-level life-cycle marker. \\
produkt\_bez       & string     & Product title or description.
  & Used in keyword-based sample assignment. \\
subsubkat          & string     & Fine product category.
  & Examples: spzgfig, blufscifi. \\
week               & int./str.  & Calendar week identifier.
  & Example: 202045. \\
clicks\_ijt        & numeric    & Clicks for the product--retailer--time cell. 
  & Naming suggests item $i$, retailer $j$, time $t$. \\
haendler\_bez\_1   & string     & Duplicate or joined retailer identifier.
  & Mirrors haendler\_bez. \\
is\_at             & binary     & Retailer associated with Austria.
  & Equals 1 for Austria-specific sellers. \\
is\_de             & binary     & Retailer associated with Germany.
  & Equals 1 for Germany-specific sellers. \\
is\_uk             & binary     & Retailer associated with the United Kingdom.
  & Equals 1 for UK-specific sellers. \\
is\_nl             & binary     & Retailer associated with the Netherlands.
  & Equals 1 for Netherlands-specific sellers. \\
laden              & binary     & In-store purchase option. 
  & German term suggests a physical-store channel. \\
abhol              & binary     & Pick-up or click-and-collect option.
  & Often empty in example rows. \\
online             & binary     & Online purchase availability.
  & Frequently equals 1. \\
lon                & numeric    & Longitude coordinate of retailer location.
  & Example around 11.59. \\
lat                & numeric    & Latitude coordinate of retailer location.
  & Example around 48.18. \\
versandk\_default  & str./num.  & Default shipping-cost field.
  & Mixed content possible. \\
kunden\_id         & integer    & Customer or seller account identifier. 
  & Appears retailer-account specific. \\
mastercard         & binary     & Mastercard accepted.
  & Equals 1 if accepted. \\
visa               & binary     & Visa accepted.
  & Equals 1 if accepted. \\
amex               & binary     & American Express accepted.
  & Equals 1 if accepted. \\
dinersclub         & binary     & Diners Club accepted.
  & Often missing or zero. \\
vk\_at             & binary     & Austrian $vk$ condition. 
  &  \\
vk\_de             & binary     & German $vk$ condition. 
  &  \\
nn\_at             & binary     & Austrian $nn$ condition. 
  &  \\
nn\_de             & binary     & German $nn$ condition.
  &  \\
liefert\_at        & binary     & Seller delivers to Austria.
  & Equals 1 if delivery to Austria offered. \\
liefert\_de        & binary     & Seller delivers to Germany.
  & Equals 1 if delivery to Germany offered. \\
liefert\_uk        & binary     & Seller delivers to the United Kingdom.
  & Equals 1 if delivery to the UK offered. \\
liefert\_pl        & binary     & Seller delivers to Poland.
  & Equals 1 if delivery to Poland offered. \\
liefert\_nl        & binary     & Seller delivers to the Netherlands.
  & Equals 1 if delivery to Netherlands offered. \\
liefert\_ie        & binary     & Seller delivers to Ireland.
  & Equals 1 if delivery to Ireland offered. \\
is\_pl             & binary     & Retailer associated with Poland.
  & Equals 1 for Poland-specific sellers. \\
dtimebegin\_1      & Unix time  & Duplicate or joined start timestamp.
  &  \\
dtimeend\_1        & Unix time  & Duplicate or joined end timestamp.
  &  \\
row\_num           & integer    & Row sequence number within deduplication procedure.
  & Equals 1 in the sample excerpt. \\
\bottomrule
\end{tabularx}}

\end{table}
\unskip
\subsection[\appendixname \thesubsection]{Derived Unit--Month Panel
  (\emph{unit\_month\_weighted\_prices.csv})}
\label{app:dict_panel}

After aggregation the unit--month panel contains one row per unit--month
pair: $102{,}627$ rows in full (September 2012 to January 2025) and $43{,}371$ rows in
the estimation window (September 2021 to December 2024).  Table \ref{tab:dict:panel}
lists the variables in the derived file.

\begin{table}[H]
\caption{Derived unit--month panel data dictionary
  (\emph{unit\_month\_weighted\_prices.csv}).}
  \label{tab:dict:panel}

\setlength{\cellWidtha}{\fulllength/3-2\tabcolsep-0.7in}
\setlength{\cellWidthb}{\fulllength/3-2\tabcolsep-1.7in}
\setlength{\cellWidthc}{\fulllength/3-2\tabcolsep+2.4in}
\scalebox{1}[1]{\begin{tabularx}{\fulllength}{>{\raggedright\arraybackslash}m{\cellWidtha}>{\raggedright\arraybackslash}m{\cellWidthb}>{\raggedright\arraybackslash}m{\cellWidthc}}
\toprule
\textbf{Column Name} & \textbf{Type} & \textbf{Description} \\
\midrule
sample\_flag               & string  & \emph{treated} or \emph{control}; see Table \ref{tab:data_dictionary_main}. \\
unit\_id                   & string  & Retailer--product identifier (\emph{produkt\_id\_\_haendler\_bez}). 3219 unique units: 2580 treated, 639 controls. \\
product\_id                & integer & Numeric product identifier; 675 unique products. \\
retailer\_id               & string  & Retailer slug; 314 unique retailers. \\
month\_date                & Date    & First calendar day of the observation month. \\
price                      & numeric & Duration-weighted geometric mean price (EUR) for unit $i$ in month $t$. \\
price\_unweighted          & numeric & Unweighted arithmetic mean of \emph{preis\_avg} across spells; retained for specification checks. \\
total\_days\_covered       & integer & Total spell-days with positive overlap in month $t$. \\
n\_spells\_in\_month       & integer & Number of distinct spells contributing to the unit-month observation. \\
dur\_days                  & numeric & Mean total spell duration in days across contributing spells. \\
product\_title             & string  & First non-empty product description carried over from \emph{produkt\_bez}. \\
treated                    & binary  & Treatment indicator $D_i$; equals 1 for treated units. \\
post\_treat\tnote{a} (a)      & binary  & Equals 1 if \emph{month\_date} is on or after the payment date as recorded in the CSV; the stored anchor differs from the paper. \\
rel\_month\tnote{a} (a)       & integer & Months relative to the CSV anchor date (1 December 2023). \textit{Not used in regressions.} \\
ln\_price                  & numeric & $\ln({price})$ when price $>0$; primary regression outcome. \\
cat\_tabak                 & boolean & Tobacco-filter units. \\
cat\_becher                & boolean & To-go cup units. \\
cat\_lebensmittelbehaelter & boolean & Food-container units. \\
cat\_tueten\_folien        & boolean & Plastic-wrap or small-bag units. \\
cat\_flaschen\_ohne\_pfand & boolean & Non-deposit bottle units. \\
cat\_flaschen\_mit\_pfand  & boolean & Deposit bottle units. \\
cat\_plastiktueten         & boolean & Plastic bag units. \\
cat\_feuchttuecher         & boolean & Wet-wipe units. \\
cat\_luftballons           & boolean & Balloon units. \\
\bottomrule
\end{tabularx}}

\noindent\footnotesize{$^\text{a}$ Reconstructed in the analysis scripts from \emph{month\_date} using
  the treatment dates adopted in the paper.  The stored values of
  \emph{post\_treat} and \emph{rel\_month} in the CSV use a different anchor
  date (1 December 2023) and are therefore \emph{not} used in estimation; event
  time is rebuilt from \emph{month\_date}. Data is subject to third-party platform licensing and cannot be made available in raw form by the author.}
\end{table}

\section[\appendixname \thesection]{Panel Construction Pipeline}
\label{app:pipeline}

This appendix records the step-by-step transformation from raw offer spells
to the unit--month panel used in all regressions.  All steps are implemented
in \textsf{R} using \emph{data.table}.

\vspace{3pt}\noindent{Step 1: Ingestion and Austria filter.}\vspace{3pt}

Read both CSV files with \texttt{data.table::fread()}.  Coerce variables to
the target types listed in Table \ref{tab:data_dictionary_main}.  Restrict
to $\mathit{is\_at}_r=1$; rows with missing $\mathit{is\_at}$ are treated
as zero and removed.

\vspace{3pt}\noindent{Step 2: Timestamp resolution.}\vspace{3pt}

The preferred pair is $(\mathit{dtimebegin\_1},\;\mathit{dtimeend\_1})$ with
fallback to $(\mathit{dtime\_birth},\;\mathit{dtime\_death})$.  Resolved
timestamps are converted from epoch seconds to UTC calendar dates.  If $e_r < s_r$, the end date is reset to the start date.  Rows without a
resolved start are removed.  Spell duration:
$\mathit{dur\_days}_r = \max(1,\; e_r - s_r + 1)$.

\vspace{3pt}\noindent{Step 3: Unit identifier.}\vspace{3pt}

The unit is the retailer--product pair, identified by concatenating
$\mathit{produkt\_id}$, a double underscore, and $\mathit{haendler\_bez}$.

\vspace{3pt}\noindent{Step 4: Month expansion.}\vspace{3pt}

Each spell $r$ is expanded into one record per calendar month it overlaps.

\vspace{3pt}\noindent{Step 5: Days of overlap.}\vspace{3pt}

$d_{r,m} = \max\!\bigl(0,\;\min(e_r, m_{\mathrm{end}}) -
\max(s_r, m_{\mathrm{start}}) + 1\bigr)$.  Pairs with zero overlap are
dropped.

\vspace{3pt}\noindent{Step 6: Duration-weighted aggregation.}\vspace{3pt}

For unit $i$ in month $t$ with active spells $\mathcal{S}_{it}$, the duration-weighted price is
\[
  p_{it}
  \;=\;
  \exp\!\left(
    \frac{\displaystyle\sum_{r \in \mathcal{S}_{it}} d_{r,t}\ln p_r}
         {\displaystyle\sum_{r \in \mathcal{S}_{it}} d_{r,t}}
  \right).
\]
The regression outcome is $\ln p_{it}$, stored as \emph{ln\_price}.

\vspace{3pt}\noindent{Step 7: Treatment indicators and event time.}\vspace{3pt}

Payment onset is 1 March 2024; event time $\tau_{it}$ is the signed month
distance from March 2024.  The three phases are: announcement
($\tau \in [-24,-13]$), reporting ($\tau \in [-12,-1]$), payment
($\tau \ge 0$).

\vspace{3pt}\noindent{Step 8: SUP category assignment.}\vspace{3pt}

Treated units are assigned by first-match keyword search on
$\mathit{produkt\_bez}$ in the order: balloons, tobacco filters, wet wipes,
to-go cups, food containers, plastic bags, plastic wrap, non-deposit bottles,
deposit bottles.  Balloons and tobacco filters carry the
\texteuro{}450-per-tonne fee; the remainder carry \texteuro{}225 per tonne.

\vspace{3pt}\noindent{Step 9: Selected estimation window.}\vspace{3pt}

The main sample retains $\tau_{it} \in [-30,+9]$ (September 2021 to December 2024).

\vspace{3pt}\noindent{Identification note.}\vspace{3pt}

Because $\mathit{ln\_price}$ is constant within units across months
(spell-average prices broadcast per row), unit fixed effects absorb all
within-unit variation.  The paper uses pooled OLS with calendar-month fixed
effects:
\begin{equation*}
  \ln p_{it}
  =
  \gamma_m
  + \sum_{k \in K} \beta_k
    \bigl(D_i \cdot \mathbf{1}\{k\}_{it}\bigr)
  + \varepsilon_{it},
\end{equation*}
where $\gamma_m$ is removed by within-month demeaning,
$K = \{\text{Announce},\;\text{Report},\;\text{Payment}\}$, and standard
errors are clustered at the retailer level (314 clusters).

Table \ref{tab:sample:counts} reports the resulting unit and observation
counts.

\begin{table}[H]
\caption{Sample composition: unit counts by group and category.}
\label{tab:sample:counts}
\footnotesize
\setlength{\cellWidtha}{\textwidth/3-2\tabcolsep+0.4in}
\setlength{\cellWidthb}{\textwidth/3-2\tabcolsep-0.2in}
\setlength{\cellWidthc}{\textwidth/3-2\tabcolsep-0.2in}
\scalebox{1}[1]{\begin{tabularx}{\textwidth}{>{\raggedright\arraybackslash}m{\cellWidtha}>{\raggedleft\arraybackslash}m{\cellWidthb}>{\raggedleft\arraybackslash}m{\cellWidthc}}
\toprule
\textbf{Group} & \textbf{Full Panel} & \textbf{Est. Window} \\
\midrule
Treated (SUP)   & 2580 & 2053 \\
\quad Balloons            &    212  &         \\
\quad Tobacco filters     &    170  &         \\
\quad Wet wipes           &    131  &         \\
\quad To-go cups          & 1903 &         \\
\quad Food containers     &     80  &         \\
\quad Plastic bags        &      4  &         \\
\quad Plastic wrap        &     69  &         \\
\quad Non-dep.\ bottles   &      8  &         \\
\quad Dep.\ bottles       &      3  &         \\
Control                   &    639  &     500 \\
\midrule
Total                     & 3219 & 2553 \\
Full panel rows           & \multicolumn{2}{l}{102,627 (September 2012 to January 2025)} \\
Selected estimation window rows    & \multicolumn{2}{l}{43{,}371 (September 2021 to December 2024)} \\
\bottomrule
\end{tabularx}}
\noindent\footnotesize{Notes: Selected estimation window: $\tau \in [-30,+9]$ relative to 1 March 2024. Category counts follow the first-match rule with \euro450-per-tonne categories evaluated first.}
\end{table}
\unskip

\section[\appendixname \thesection]{Additional Figures}
\label{app:figures}

This appendix contains figures that do not appear elsewhere in the paper.
Figures \ref{fig:energy_control_corr}--\ref{fig:energy_full_sample} document
the energy input-price controls.  Figures \ref{fig:twfe_coef_plot_raw}--\ref{fig:balloon_eventstudy_raw} are coefficient plots based on the original
\emph{disaggregated} offer-spell data and thus complement---rather than
repeat---the corresponding aggregated results reported in the main text.

\subsection{Correlation Among Energy Control Variables}
\label{app:figures:energy_corr}

\begin{figure}[H]
  \includegraphics[width=0.45\textwidth]%
    {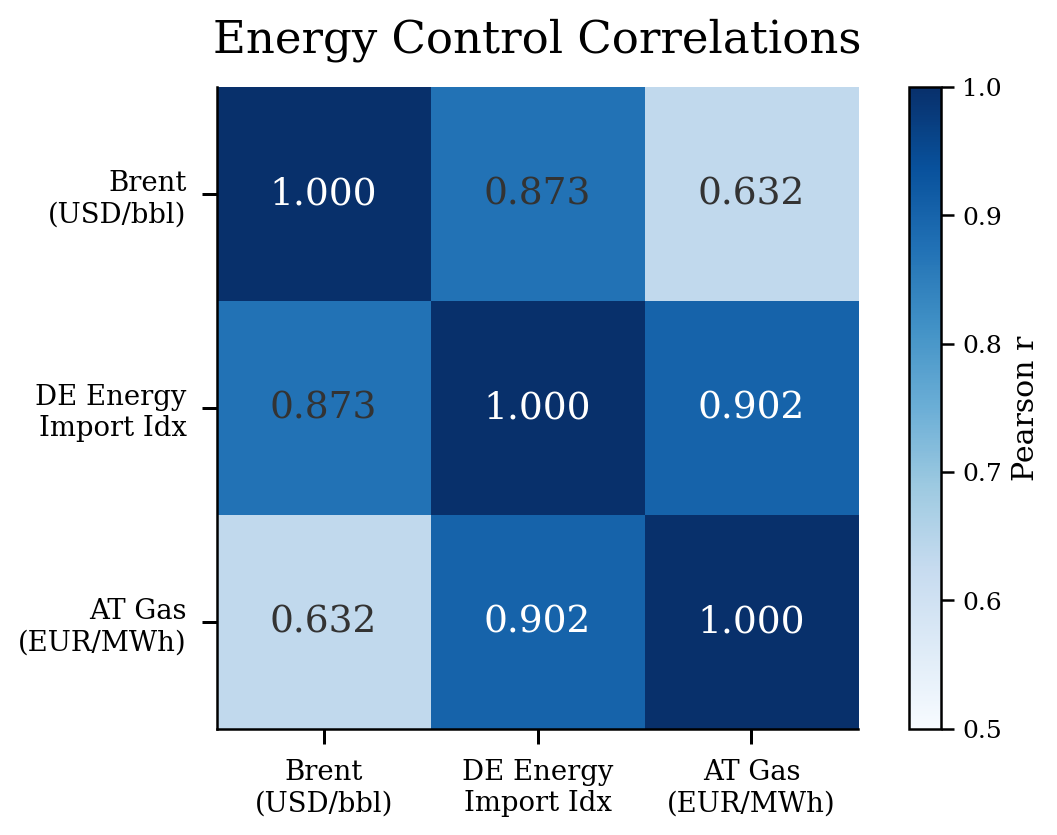}
  \caption[Pairwise correlations among energy control variables]{%
    Pairwise Pearson correlations among Brent crude prices (USD/bbl,
    FRED), the German energy import price index (2021\,=\,100, Destatis),
    and Austrian natural gas spot prices (\euro/MWh, OEGPI).}
  \label{fig:energy_control_corr}
\end{figure}

Figure \ref{fig:energy_control_corr} reports pairwise Pearson correlations
among the three energy control variables.  The high pairwise correlations
(Brent--DE import index 0.87; DE import index--AT gas 0.90) motivate caution
about including all three simultaneously; this collinearity is discussed in
Section \ref{subsec:covariates_method} and is the reason the energy controls enter as partial rather
than fully structural cost indices.

\vspace{-3pt}
\unskip

\subsection[\appendixname \thesubsection]{Regression Coefficient Plots: Disaggregated Data}
\label{app:figures:coef_plots}

The main text reports results for the aggregated duration-weighted monthly
panel.  The figures in this subsection replicate the key regression outputs
using the original, disaggregated offer-spell data.  Because each raw spell
enters as a separate observation, the estimating environment differs from the
aggregated panel, but the qualitative patterns are consistent throughout.

\begin{figure}[H]
\hspace{-6pt}
  \includegraphics[width=\textwidth]{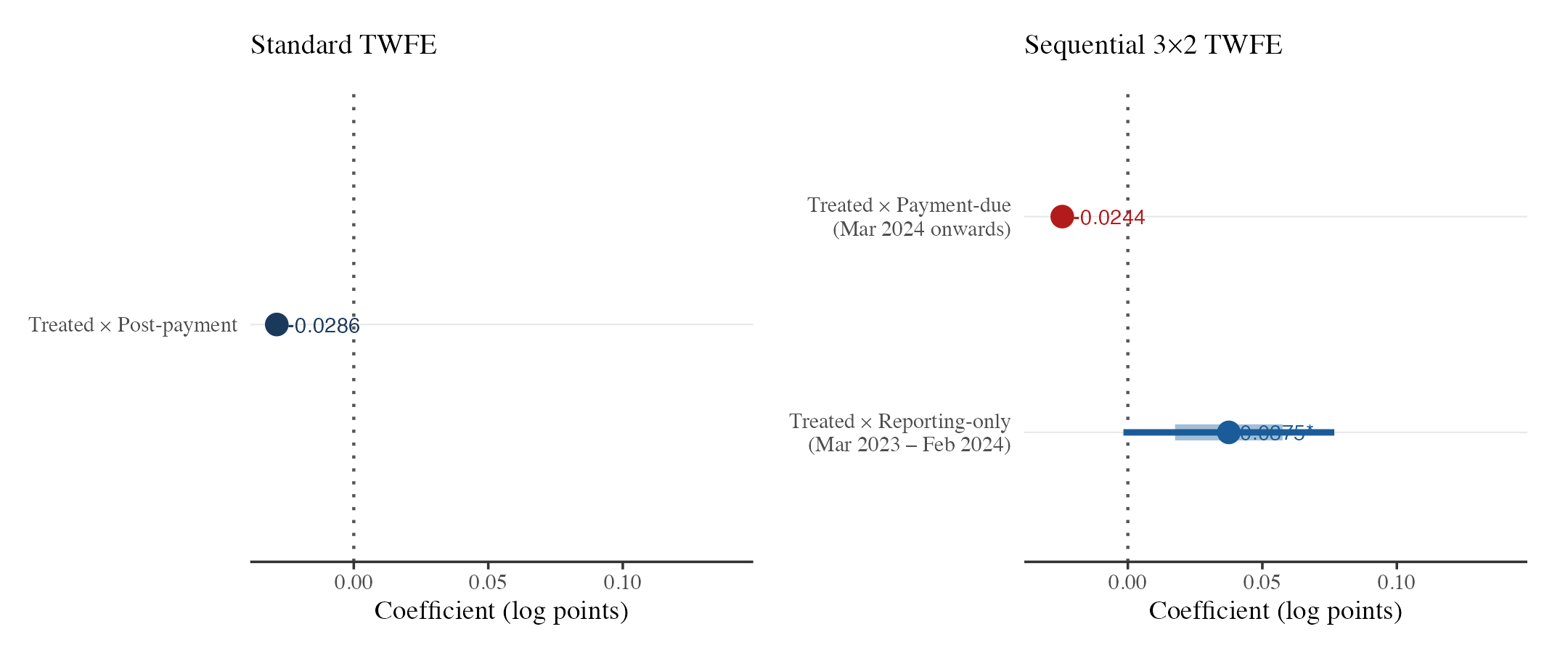}
  \caption{Coefficient plots 
for the pooled TWFE specifications using the
    original disaggregated offer-spell sample.  \textbf{Left}: Standard
    TWFE with a single post-payment interaction.  \textbf{Right}:
    Sequential multiperiod TWFE ($3\times2$) separating the reporting-only and
    payment-due phases.  Points show coefficient estimates; horizontal
    bars indicate 95\% confidence intervals. Significance: * $p < 0.1$.}
  \label{fig:twfe_coef_plot_raw}
\end{figure}

Figure \ref{fig:twfe_coef_plot_raw} shows the pooled TWFE estimates on the
disaggregated data.  The left panel is the standard post-payment TWFE; the
right panel is the sequential $3\times2$ specification.  The larger
adjustment during the reporting-only phase is visible here as in the
aggregated counterpart.

\begin{figure}[H]
\hspace{-6pt}
  \includegraphics[width=0.9\textwidth]{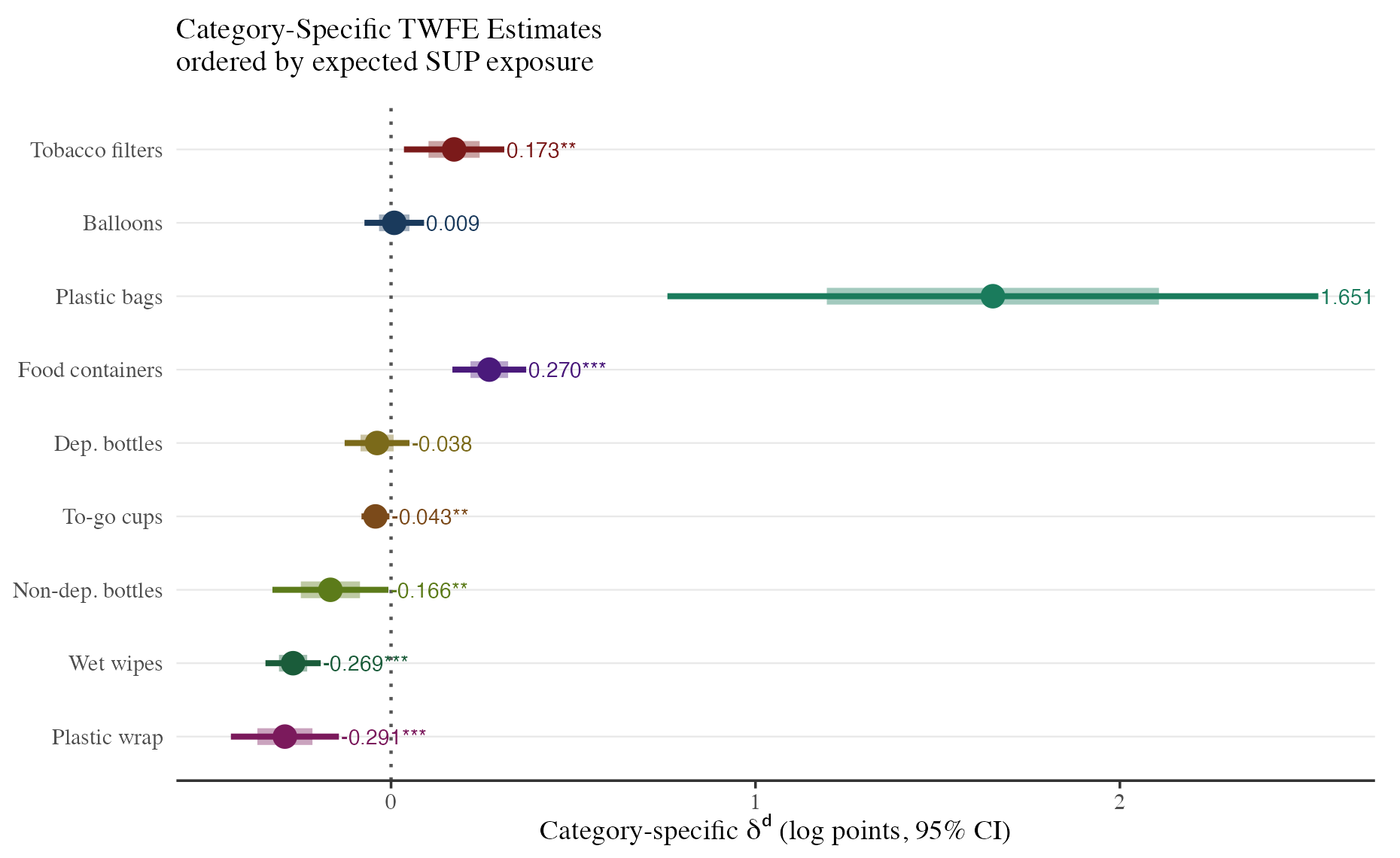}
  \caption{Category-specific 
 TWFE estimates for the original disaggregated
    offer-spell sample, ordered by expected SUP exposure.
    Points show category-level treatment coefficients (log points);
    horizontal bars indicate 95\% confidence intervals. Significance: ** $p < 0.05$, *** $p < 0.01$.}
  \label{fig:category_coef_plot_raw}
\end{figure}

Figure \ref{fig:category_coef_plot_raw} reports category-specific TWFE
coefficients on the disaggregated data, ordered by expected SUP
exposure.  The pooled average masks substantial heterogeneity.  Tobacco
filters and food containers display positive and comparatively precise
estimates; wet wipes, plastic wrap, and non-deposit bottles show negative
coefficients.  To-go cups, the largest treated category, exhibits only a
small negative estimate.  Plastic bags has a large positive point estimate
but very wide confidence intervals given only four treated units in the raw
data.

\begin{figure}[H]
  \includegraphics[width=0.6\textwidth]{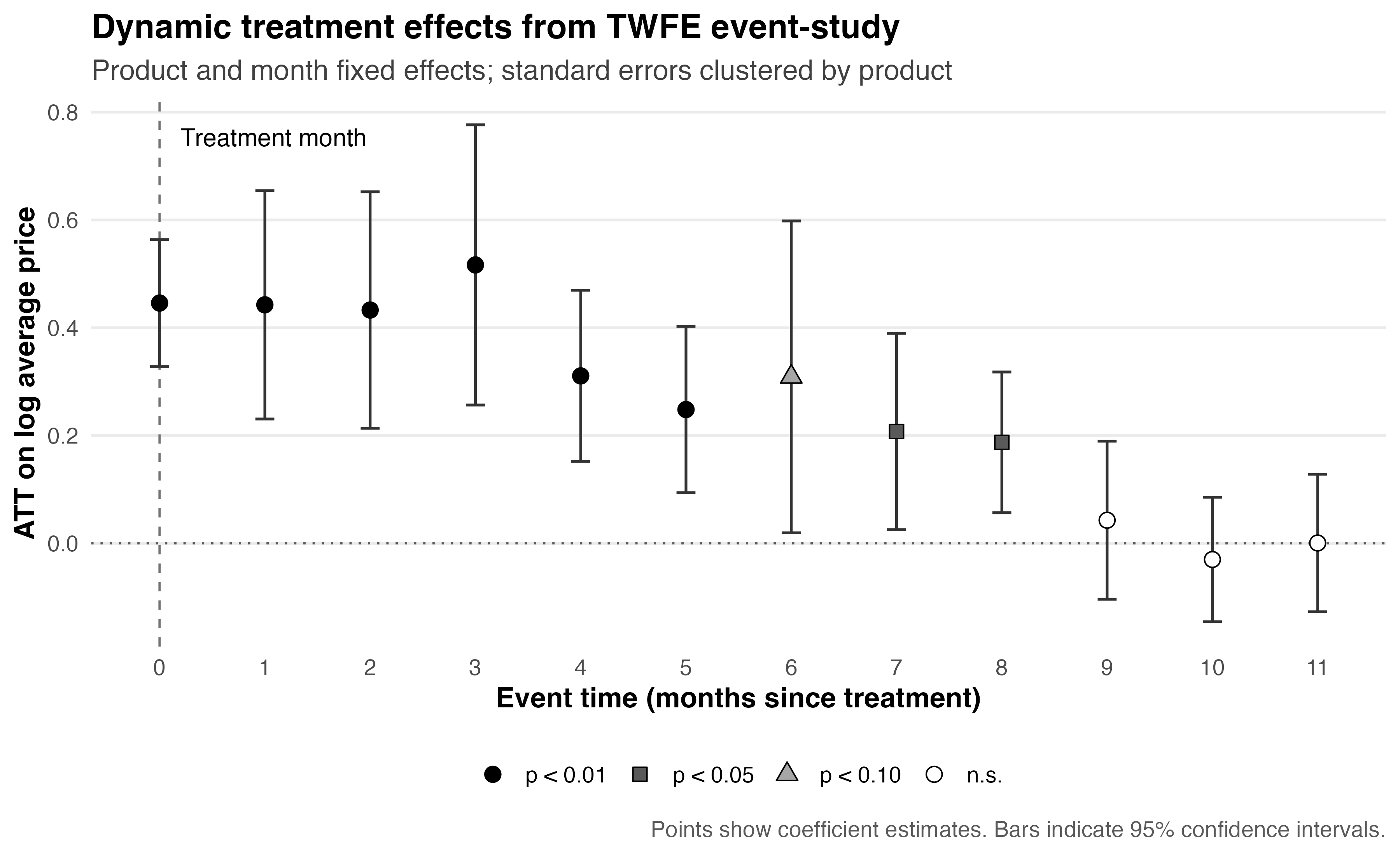}
  \caption{Event-study 
coefficients for Austrian balloon prices in the original
    disaggregated offer-spell data.  Points show month-specific
    coefficient estimates relative to the omitted pre-payment reference
    period; vertical bars indicate 95\% confidence intervals.  The vertical dashed line marks the first payment month (March 2024).}
  \label{fig:balloon_eventstudy_raw}
\end{figure}

Figure \ref{fig:balloon_eventstudy_raw} presents the balloon event study on
the disaggregated data.  The timing and magnitude of the price response are
consistent with the aggregated results reported in Section \ref{subsec:balloons} and
Table \ref{tab:app_balloon_prices_eventstudy}: a large, immediate increase
at the first payment date that remains elevated for several months before
fading.

\vspace{-6pt}

\vspace{-12pt}

\vspace{-12pt}
\unskip

\subsection[\appendixname \thesubsection]{Energy Input-Price Series}
\label{app:energy_plots}

\begin{figure}[H]
\hspace{-2pt}
  \includegraphics[width=\textwidth]{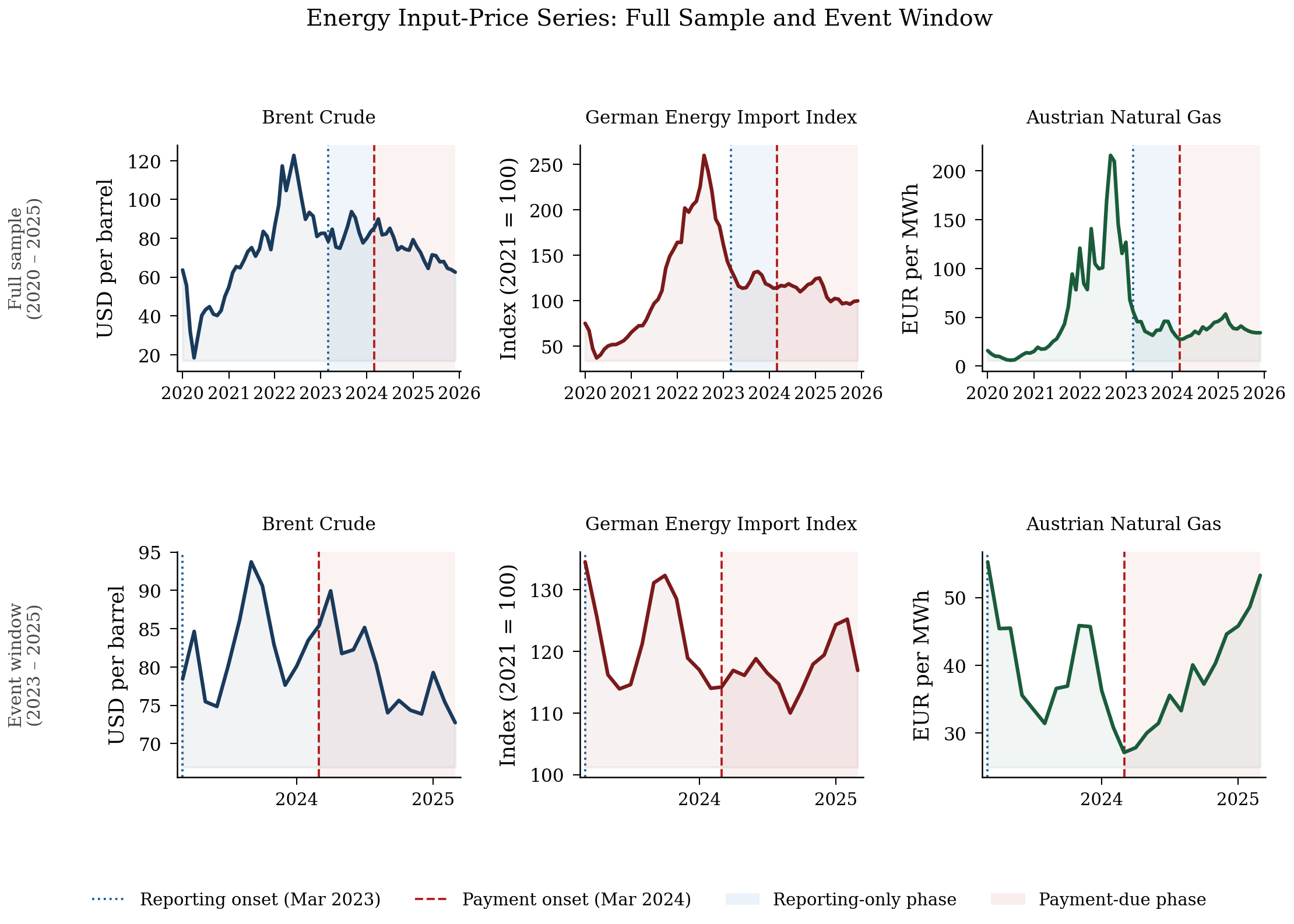}
  \caption{Energy input-price 
controls: full sample (top row, 2020--2025) and
    event window (bottom row, March 2023 to March 2025).  \textit{Left:}
    Brent crude oil (USD/bbl, FRED).  \textit{Center:} German energy
    import price index (2021\,=\,100, Destatis).  \textit{Right:}
    Austrian natural gas spot price (\euro/MWh, OEGPI).  Vertical dotted
    line: reporting-phase onset (March 2023).  Vertical dashed line:
    Payment-phase onset (March 2024).  Blue shading: Reporting-only
    period.  Red shading: Payment-due period.}
  \label{fig:energy_2x3}
\end{figure}

Figure \ref{fig:energy_2x3} presents the three energy controls in a
$2\times3$ grid.  The top row covers the full sample (January 2020 to
December 2025); the bottom row restricts attention to the estimation event
window (March 2023 to March 2025), which brackets treatment onset by twelve
months on either side.

All three series peak sharply in 2022, reflecting the European energy crisis
following Russia's invasion of Ukraine in February of that year.  The German
energy import index reaches approximately 260 (August 2022), Austrian gas
prices peak at approximately \euro216/MWh (September 2022), and Brent crude
reaches roughly \$122/bbl (June 2022).  By reporting onset in March 2023, all
three had declined substantially but remained materially above pre-2022
levels, creating the potential for lagged input-cost confounding during the
reporting phase.  By payment onset in March 2024 energy prices had fallen
further; this declining backdrop is, if anything, favorable for identifying
the regulatory effect in the payment-due period.

\begin{figure}[H]
\hspace{-2pt}
  \includegraphics[width=\textwidth]{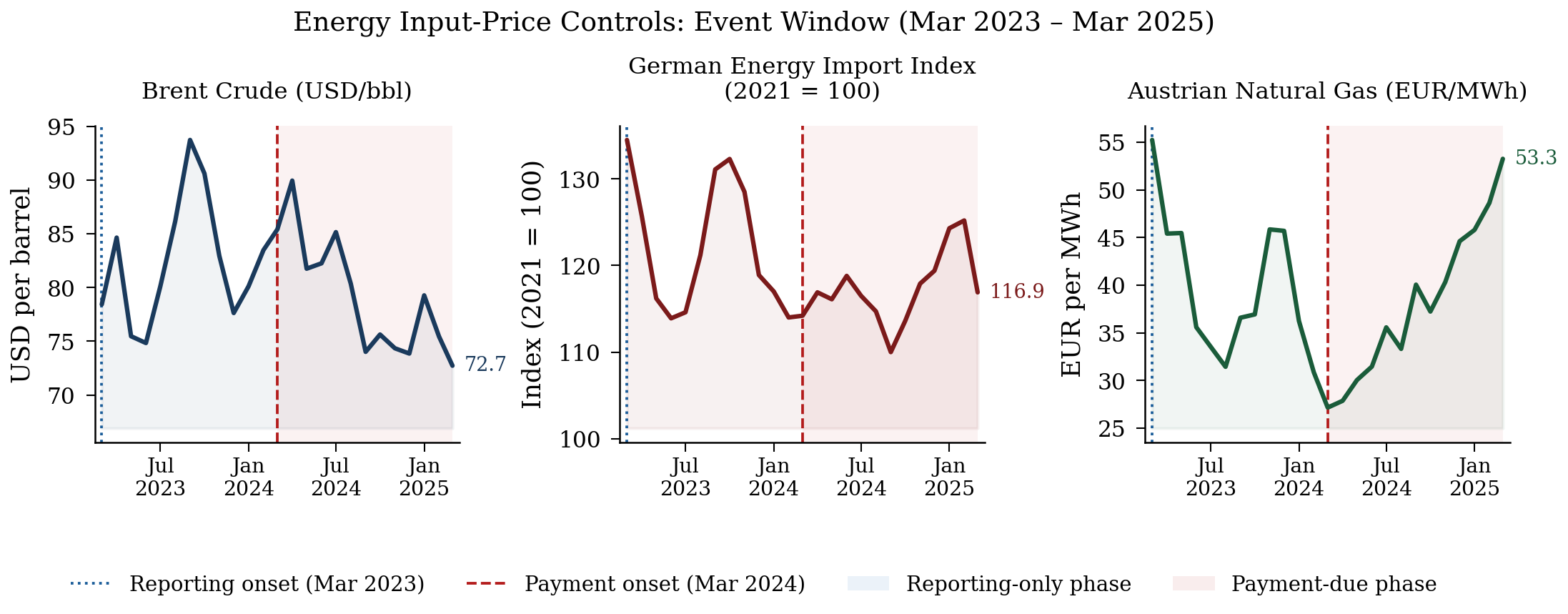}
  \caption[Energy input-price controls: event window]{%
    Energy input-price controls, event window only (March 2023 to
    March 2025).  \textbf{Left}: Brent crude.  \textbf{Center}: German
    energy import index.  \textbf{Right}: Austrian gas price.  Annotated
    values indicate the end-of-window observation.  Shading and vertical
    lines follow Figure \ref{fig:energy_2x3}.}
  \label{fig:energy_event_window}
\end{figure}

\begin{figure}[H]
\hspace{-2pt}
  \includegraphics[width=\textwidth]{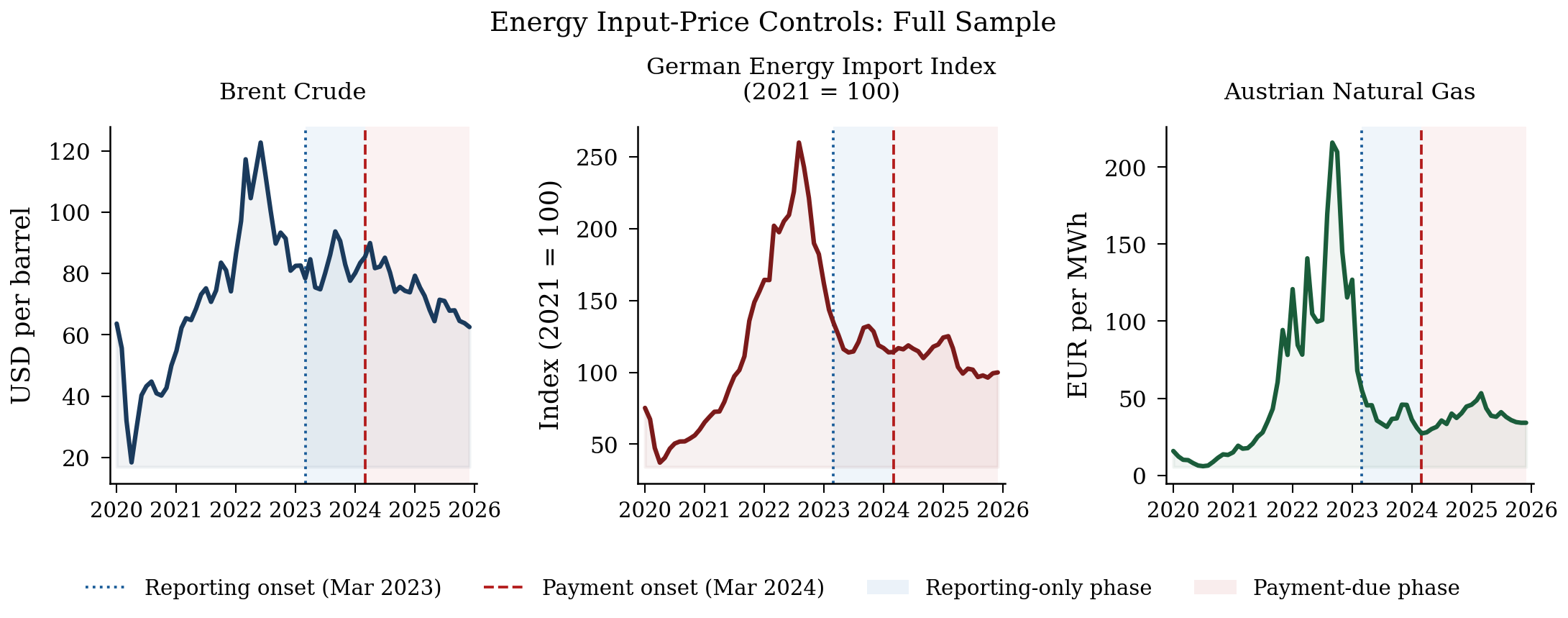}
  \caption[Energy input-price controls: full sample]{%
    Energy input-price controls, full sample (January 2020 to
    December 2025).  {\textbf{Left}:} Brent crude. {\textbf{Center}:}
    German energy import index.  {\textbf{Right}:} Austrian gas price.
    Shading and vertical lines follow Figure \ref{fig:energy_2x3}.  The 2022 price spike motivates the inclusion of lagged energy controls in
    extended specifications.}
  \label{fig:energy_full_sample}
\end{figure}

Figures \ref{fig:energy_event_window} and \ref{fig:energy_full_sample}
reproduce the event-window and full-sample panels at a larger scale for ease
of inspection, with end-of-window values annotated.

\unskip

\unskip

\section[\appendixname \thesection]{Tables}
\label{app:tables}

\subsection[\appendixname \thesubsection]{Descriptive Sample Statistics}
\label{app:tables:descriptive}

The treated offer-spell panel is larger than the control panel both in
observations and in active units.  The treated sample contains
1{,}738{,}870 spell-month observations from 20{,}721 units, while the
control sample contains 1{,}039{,}833 observations from 7054 units.
In levels, mean monthly prices are similar across groups, although the
treated sample exhibits substantially greater dispersion.  Median prices are
somewhat higher among the treated listings than among the controls.

Within the treated sample, exposure is concentrated in a few categories,
specifically, to-go cups (15{,}933 units; 1{,}312{,}574 observations).
balloons, tobacco filters, wet wipes, and plastic wrap contribute meaningful
shares; plastic bags and deposit bottles remain very small.  Price levels
vary markedly across categories, with the lowest means for wet wipes and
plastic wrap and the highest for tobacco filters and plastic bags.

The monthly energy controls vary substantially across the policy phases.  Brent
crude prices are highest on average during the reporting-only period, while
the German energy import index is elevated both before the policy and during
the reporting-only phase.  Austrian gas prices are most volatile in the
pre-policy period (SD\,=\,60.05; maximum above EUR 215/MWh) before
stabilizing at lower levels in later phases.

\begin{table}[H]
\caption{Descriptive statistics: 
offer-spell panel by treatment status.}
\label{tab:spell_sumstats_appendix}
\small
\setlength{\cellWidtha}{\textwidth/7-2\tabcolsep-0.1in}
\setlength{\cellWidthb}{\textwidth/7-2\tabcolsep-0.15in}
\setlength{\cellWidthc}{\textwidth/7-2\tabcolsep-00in}
\setlength{\cellWidthd}{\textwidth/7-2\tabcolsep+0.2in}
\setlength{\cellWidthe}{\textwidth/7-2\tabcolsep-0.15in}
\setlength{\cellWidthf}{\textwidth/7-2\tabcolsep-00in}
\setlength{\cellWidthg}{\textwidth/7-2\tabcolsep+0.2in}
\scalebox{1}[1]{\begin{tabularx}{\textwidth}{>{\raggedright\arraybackslash}m{\cellWidtha}>{\raggedleft\arraybackslash}m{\cellWidthb}>{\raggedleft\arraybackslash}m{\cellWidthc}>{\raggedleft\arraybackslash}m{\cellWidthd}>{\raggedleft\arraybackslash}m{\cellWidthe}>{\raggedleft\arraybackslash}m{\cellWidthf}>{\raggedleft\arraybackslash}m{\cellWidthg}}
\toprule
 & \multicolumn{3}{c}{\textbf{Treated (SUP Products)}}
 & \multicolumn{3}{c}{\textbf{Control (Non-SUP Products)}} \\
\cmidrule{2-7}
 & \textbf{Full} & \textbf{Pre-Policy} & \textbf{Post-Payment}
 & \textbf{Full} & \textbf{Pre-Policy} & \textbf{Post-Payment} \\
\midrule
Obs.\  & 1{,}738{,}870 & 870{,}500 & 468{,}450
       & 1{,}039{,}833 & 690{,}900 & 163{,}051 \\
Units  & 20{,}721 &---&---& 7054 &---&---\\
\addlinespace
\multicolumn{7}{l}{\textit{Average monthly price (EUR)}} \\
\quad Mean   & 24.99 & 23.95 & 26.86 & 24.38 & 24.65 & 24.71 \\
\quad SD     & 373.42 & 526.32 & 38.49 & 70.75 & 83.67 & 23.92 \\
\quad Median & 18.95 & 18.09 & 19.29 & 15.64 & 14.99 & 16.95 \\
\quad p10    &  8.90 &  9.36 &  8.51 &  6.13 &  5.96 &  7.05 \\
\quad p90    & 38.19 & 36.38 & 40.94 & 53.90 & 53.90 & 59.47 \\
\addlinespace
\multicolumn{7}{l}{\textit{Log average monthly price}} \\
\quad Mean   & 2.94 & 2.92 & 2.97 & 2.81 & 2.79 & 2.89 \\
\quad SD     & 0.61 & 0.54 & 0.70 & 0.80 & 0.82 & 0.76 \\
\bottomrule
\end{tabularx}}
\noindent\footnotesize{{Notes:} Post-payment refers to months from March 2024 onward.
  Price statistics are computed on spell-level average prices.}
\end{table}

\vspace{-12pt}
\begin{table}[H]
\caption{Treated 
sample composition by keyword category.}
\label{tab:category_composition_appendix}
\small
\setlength{\cellWidtha}{\textwidth/7-2\tabcolsep+0.25in}
\setlength{\cellWidthb}{\textwidth/7-2\tabcolsep+0.15in}
\setlength{\cellWidthc}{\textwidth/7-2\tabcolsep-0.25in}
\setlength{\cellWidthd}{\textwidth/7-2\tabcolsep-0.1in}
\setlength{\cellWidthe}{\textwidth/7-2\tabcolsep+0.05in}
\setlength{\cellWidthf}{\textwidth/7-2\tabcolsep-0in}
\setlength{\cellWidthg}{\textwidth/7-2\tabcolsep-0.1in}
\scalebox{1}[1]{\begin{tabularx}{\textwidth}{>{\raggedright\arraybackslash}m{\cellWidtha}>{\raggedright\arraybackslash}m{\cellWidthb}>{\raggedleft\arraybackslash}m{\cellWidthc}>{\raggedleft\arraybackslash}m{\cellWidthd}>{\raggedleft\arraybackslash}m{\cellWidthe}>{\raggedleft\arraybackslash}m{\cellWidthf}>{\raggedleft\arraybackslash}m{\cellWidthg}}
\toprule
\textbf{Category} & \textbf{SUP Fee (\euro/t)} & \textbf{Units} &\textbf{ Obs.}
  & \textbf{Mean Price} & \textbf{Mean ln(\textit{p})} & \textbf{SD ln(\textit{p})} \\
\midrule
Tobacco filters   & 450 &    935 &    83{,}394 &  67.94 & 3.291 & 0.866 \\
Balloons          & 450 & 2078 &  110{,}254 &  29.60 & 3.159 & 0.735 \\
Plastic bags      & 225 &     11 &     1168 & 232.35 & 5.434 & 0.171 \\
Food containers   & 225 &    319 &    39{,}027 &  37.74 & 3.609 & 0.212 \\
Dep.\ bottles     & 225 &     29 &    11{,}372 &  27.62 & 3.303 & 0.175 \\
To-go cups        & 225 & 15{,}933 & 1{,}312{,}574 & 22.83 & 2.944 & 0.542 \\
Non-dep.\ bottles & 225 &    127 &    20{,}480 &  21.75 & 2.763 & 0.907 \\
Plastic wrap      & 225 &    562 &    57{,}559 &  14.14 & 2.458 & 0.614 \\
Wet wipes         & 225 &    727 &   103{,}042 &  12.02 & 2.370 & 0.413 \\
\addlinespace
Control (non-SUP) &---& 7054 & 1{,}039{,}833
  & 24.38 & 2.81 & 0.80 \\
\bottomrule
\end{tabularx}}
\noindent\footnotesize{{Notes:} Categories ordered by expected SUP exposure,
  first by fee tier then by mean log price.  Control denotes
  baseline-survivor graphics card listings.}
\end{table}

\vspace{-12pt}
\begin{table}[H]
\caption{Descriptive 
statistics: monthly energy input-price controls by
  policy phase.}
\label{tab:energy_sumstats_appendix}
\small

\setlength{\cellWidtha}{\fulllength/10-2\tabcolsep+0.75in}
\setlength{\cellWidthb}{\fulllength/10-2\tabcolsep+0.3in}
\setlength{\cellWidthc}{\fulllength/10-2\tabcolsep-0.15in}
\setlength{\cellWidthd}{\fulllength/10-2\tabcolsep-0.15in}
\setlength{\cellWidthe}{\fulllength/10-2\tabcolsep-0.15in}
\setlength{\cellWidthf}{\fulllength/10-2\tabcolsep-0.15in}
\setlength{\cellWidthg}{\fulllength/10-2\tabcolsep-0.15in}
\setlength{\cellWidthh}{\fulllength/10-2\tabcolsep+0in}
\setlength{\cellWidthi}{\fulllength/10-2\tabcolsep-0.15in}
\setlength{\cellWidthj}{\fulllength/10-2\tabcolsep-0.15in}
\scalebox{1}[1]{\begin{tabularx}{\fulllength}{>{\raggedright\arraybackslash}m{\cellWidtha}>{\raggedright\arraybackslash}m{\cellWidthb}>{\raggedleft\arraybackslash}m{\cellWidthc}>{\raggedleft\arraybackslash}m{\cellWidthd}>{\raggedleft\arraybackslash}m{\cellWidthe}>{\raggedleft\arraybackslash}m{\cellWidthf}>{\raggedleft\arraybackslash}m{\cellWidthg}>{\raggedleft\arraybackslash}m{\cellWidthh}>{\raggedleft\arraybackslash}m{\cellWidthi}>{\raggedleft\arraybackslash}m{\cellWidthj}}
\toprule
\textbf{Series} & \textbf{Phase} & \textbf{\textit{N}} & \textbf{Mean} & \textbf{SD} & \textbf{Min} & \textbf{p25} & \textbf{Median} & \textbf{p75} & \textbf{Max} \\
\midrule
Brent crude (USD/bbl)
  & Pre-policy     & 38 &  71.68 & 26.17 &  18.38 &  51.19 &  73.66 &  88.95 & 122.71 \\
  & Reporting-only & 12 &  82.34 &  5.80 &  74.84 &  78.23 &  81.53 &  85.02 &  93.72 \\
  & Payment-due    & 22 &  74.18 &  7.72 &  62.54 &  68.02 &  73.94 &  80.09 &  89.94 \\
DE energy import index
  & Pre-policy     & 38 & 121.33 & 67.75 &  37.20 &  61.50 &  99.45 & 177.83 & 260.00 \\
  & Reporting-only & 12 & 122.33 &  7.71 & 113.90 & 115.80 & 120.05 & 129.15 & 134.50 \\
  & Payment-due    & 22 & 110.11 &  9.47 &  96.40 & 100.47 & 113.90 & 116.90 & 125.20 \\
AT gas (EUR/MWh)
  & Pre-policy     & 38 &  61.99 & 60.05 &   5.93 &  13.33 &  31.38 & 100.50 & 215.93 \\
  & Reporting-only & 12 &  39.91 &  7.46 &  30.86 &  35.09 &  36.76 &  45.54 &  55.25 \\
  & Payment-due    & 22 &  37.86 &  6.55 &  27.16 &  34.09 &  37.56 &  40.82 &  53.27 \\
\bottomrule
\end{tabularx}}

\noindent\footnotesize{{Notes:} Monthly observations January 2020 through
  December 2025.  Pre-policy: before March 2023.  Reporting-only:
  March 2023--February 2024.  Payment-due: from March 2024.}
\end{table}
\unskip

\subsection{Regression Results}
\label{app:tables:regression}

Table \ref{tab:app_twfe_3x2_main} reports the multiperiod TWFE ($3\times2$) estimates
under two clustering schemes: unit-level (column 1) and retailer-level
(column 2).  The point estimates are identical across both columns because
clustering affects only the standard errors.  Clustering at the retailer
level reduces precision sharply for the earlier phase---the report-only
coefficient falls below conventional significance thresholds---while the
payment-due coefficient remains qualitatively stable.  This result, discussed
in Section \ref{subsec:clustering} of the main text, reflects residual dependence operating
partly through seller-wide pricing decisions.

\begin{table}[H]
\caption{Multiperiod TWFE estimates: sensitivity to clustering level.}
\label{tab:app_twfe_3x2_main}
\small
\setlength{\cellWidtha}{\textwidth/3-2\tabcolsep+0.5in}
\setlength{\cellWidthb}{\textwidth/3-2\tabcolsep-0.3in}
\setlength{\cellWidthc}{\textwidth/3-2\tabcolsep-0.2in}
\scalebox{1}[1]{\begin{tabularx}{\textwidth}{>{\raggedright\arraybackslash}m{\cellWidtha}>{\centering\arraybackslash}m{\cellWidthb}>{\centering\arraybackslash}m{\cellWidthc}}
\toprule
 & \multicolumn{2}{c}{\textbf{Dependent Variable: ln(Average Monthly
   Price)}} \\
\cmidrule{2-3}
 & \textbf{(1) Multiperiod TWFE Clustered by Unit}& \textbf{(2) Multiperiod TWFE Clustered by Retailer}\\
\midrule
Treated $\times$ $\mathbf{1}\{\text{Legal-to-payment period}\}$
  & 0.1377 *** & 0.1377 \\
  & (0.0240)  & (0.0945) \\[4pt]
Treated $\times$ $\mathbf{1}\{\text{Post-payment period}\}$
  & 0.0450    & 0.0450 \\
  & (0.0291)  & (0.0756) \\
\midrule
Unit fixed effects  & Yes & Yes \\
Month fixed effects & Yes & Yes \\
Standard errors     & Unit & Retailer \\
Observations        & 51{,}537 & 51{,}537 \\
$R^2$               & 0.9197 & 0.9197 \\
Within $R^2$        & 0.0075 & 0.0075 \\
\bottomrule
\end{tabularx}}
\noindent\footnotesize{Notes:} The omitted category is the pre-legal period; both
  coefficients are interpreted relative to that baseline.  Column (1)
  clusters at the unit (retailer--product pair) level.  Column (2) clusters
  at the retailer level.  All specifications include unit and month fixed
  effects.  Standard errors in parentheses.
  *** $p<0.01$.
\end{table}

Table \ref{tab:app_balloon_prices_eventstudy} reports the full set of
event-study coefficients for Austrian balloon prices around the first
payment date, with Kennedy corrected percentage effects.  The month-by-month
sequence documents the timing of the price adjustment: a large step at
$t=0$, sustained elevation through $t+5$, and gradual decay through
$t+9$ (December 2024).  These coefficients underpin the discussion in
Section \ref{subsec:balloons} of the main text.

\begin{table}[H]
\caption{Event-study estimates of Austrian balloon prices around the first
  payment date.}
\label{tab:app_balloon_prices_eventstudy}

\setlength{\cellWidtha}{\fulllength/6-2\tabcolsep+0.6in}
\setlength{\cellWidthb}{\fulllength/6-2\tabcolsep-0.3in}
\setlength{\cellWidthc}{\fulllength/6-2\tabcolsep-0.1in}
\setlength{\cellWidthd}{\fulllength/6-2\tabcolsep+0.2in}
\setlength{\cellWidthe}{\fulllength/6-2\tabcolsep-0.3in}
\setlength{\cellWidthf}{\fulllength/6-2\tabcolsep-0.1in}
\scalebox{1}[1]{\begin{tabularx}{\fulllength}{>{\raggedright\arraybackslash}m{\cellWidtha}>{\centering\arraybackslash}m{\cellWidthb}>{\centering\arraybackslash}m{\cellWidthc}>{\raggedright\arraybackslash}m{\cellWidthd}>{\centering\arraybackslash}m{\cellWidthe}>{\centering\arraybackslash}m{\cellWidthf}}
\toprule
 \multirow{2}{*}{\textbf{Event Time}\vspace{-6pt}} & \multicolumn{2}{c}{\textbf{ln(}${\textbf{Average Price}}_{\textbf{\textit{it}}}$\textbf{)}}
 & \multirow{2}{*}{\textbf{Event Time}\vspace{-6pt}} & \multicolumn{2}{c}{\textbf{ln(}$\textbf{Average Price}_{\textbf{\textit{it}}}$\textbf{)}} \\
\cmidrule{2-3}\cmidrule{5-6}
& \textbf{Coefficient} & \textbf{Standard Error}
  & & \textbf{Coefficient} & \textbf{Standard Error} \\
\midrule
Payment month ($t=0$, 2024:03)
  & 0.446 *** & (0.060) & $t+6$ (2024:09) & 0.309 *   & (0.148) \\
  & \multicolumn{2}{c}{\textit{(+55.9\%)}}
  & & \multicolumn{2}{c}{\textit{(+34.7\%)}} \\
$t+1$ (2024:04) & 0.442 *** & (0.108)
  & $t+7$ (2024:10) & 0.207 **  & (0.093) \\
  & \multicolumn{2}{c}{\textit{(+54.8\%)}}
  & & \multicolumn{2}{c}{\textit{(+22.5\%)}} \\
$t+2$ (2024:05) & 0.433 *** & (0.112)
  & $t+8$ (2024:11) & 0.187 **  & (0.067) \\
  & \multicolumn{2}{c}{\textit{(+53.2\%)}}
  & & \multicolumn{2}{c}{\textit{(+20.3\%)}} \\
$t+3$ (2024:06) & 0.516 *** & (0.133)
  & $t+9$ (2024:12) & 0.043    & (0.075) \\
  & \multicolumn{2}{c}{\textit{(+66.2\%)}}
  & & \multicolumn{2}{c}{\textit{(+4.1\%)}} \\
$t+4$ (2024:07) & 0.311 *** & (0.081) & & & \\
  & \multicolumn{2}{c}{\textit{(+36.0\%)}} & & & \\
$t+5$ (2024:08) & 0.248 *** & (0.079) & & & \\
  & \multicolumn{2}{c}{\textit{(+27.8\%)}} & & & \\
\midrule
Observations          & \multicolumn{2}{c}{178}
  & Products included  & \multicolumn{2}{c}{15} \\
Product fixed effects & \multicolumn{2}{c}{Yes}
  & Sample window      & \multicolumn{2}{c}{2020:01--2024:12} \\
Month fixed effects   & \multicolumn{2}{c}{Yes}
  & Treatment month    & \multicolumn{2}{c}{2024:03} \\
Adjusted $R^2$        & \multicolumn{2}{c}{0.964}
  & Final sample month & \multicolumn{2}{c}{2024:12} \\
RMSE                  & \multicolumn{2}{c}{0.118}
  & Standard errors    & \multicolumn{2}{c}{Clustered by product} \\
\bottomrule
\end{tabularx}}

\noindent\footnotesize{Notes:} Event time $t=0$ is March 2024; the omitted category
  is the pre-payment reference period.  The latest observable post-treatment
  horizon is $t+9$ because the sample ends in December 2024.  All
  specifications include product and month fixed effects.  Italicized values
  report Kennedy corrected percentage effects:
  $100\times\!\left[\exp\!\left(\hat{\beta} - \tfrac{1}{2}
  \widehat{\mathrm{Var}}(\hat{\beta})\right)-1\right]$,
  where $\widehat{\mathrm{Var}}(\hat{\beta}) =
  \widehat{\mathrm{SE}}(\hat{\beta})^2$.
  *** $p<0.01$, ** $p<0.05$, * $p<0.10$.
\end{table}

\section[\appendixname \thesection]{Notes on Duration-Weighted Panel Analysis}
\label{app:additional}

Section \ref{sec:additional} of the main text reports four sets of additional analyses using
the aggregated duration-weighted panel.  The corresponding figures
(main text Figures \ref{fig:A1a}--\ref{fig:A5b}) are not reproduced here.  This appendix provides
the panel variable reference used across those analyses and brief notes on
each subsection.

Table \ref{tab:csv_outline} lists the key columns of
\emph{unit\_month\_weighted\_prices.csv} used in Sections \ref{subsec:cat_het}--\ref{subsec:demand}.
\begin{table}[H]
\caption{Key columns of \emph{unit\_month\_weighted\_prices.csv} used in
  the additional analyses (Sections \ref{subsec:cat_het}--\ref{subsec:demand}).}
\label{tab:csv_outline}
\setlength{\cellWidtha}{\textwidth/3-2\tabcolsep+00in}
\setlength{\cellWidthb}{\textwidth/3-2\tabcolsep-1.1in}
\setlength{\cellWidthc}{\textwidth/3-2\tabcolsep+1.1in}
\scalebox{1}[1]{\begin{tabularx}{\textwidth}{>{\raggedright\arraybackslash}m{\cellWidtha}>{\raggedright\arraybackslash}m{\cellWidthb}>{\raggedright\arraybackslash}m{\cellWidthc}}
\toprule
\textbf{Column} & \textbf{Type} & \textbf{Description} \\
\midrule
\multicolumn{3}{l}{\textit{Identifiers}} \\
\textit{sample\_flag}    & str  & Treated/control \\
\textit{unit\_id}        & str  & Retailer--product pair (3219 unique) \\
\textit{product\_id}     & int  & Numeric product id (675 unique) \\
\textit{retailer\_id}    & str  & Retailer slug (314 unique) \\
\textit{month\_date}     & Date & First day of month \\
\midrule
\multicolumn{3}{l}{\textit{Price outcomes}} \\
\textit{price}             & float & Duration-weighted geometric mean price (EUR) \\
\textit{price\_unweighted} & float & Unweighted arithmetic mean price \\
\textit{ln\_price}         & float & $\ln p_{it}$; primary regression outcome \\
\midrule
\multicolumn{3}{l}{\textit{Spell aggregates (demand proxies, Section \ref{subsec:demand})}} \\
\textit{n\_spells\_in\_month}  & int   & Distinct price spells in unit-month \\
\textit{total\_days\_covered}  & int   & $\sum_s d_{s,t}$: active-price days \\
\textit{dur\_days}             & float & Mean raw spell duration (days) \\
\midrule
\multicolumn{3}{l}{\textit{Treatment indicator}} \\
\textit{treated}  & bin. & $D_i=1$ iff SUP product \\
\midrule
\multicolumn{3}{l}{\textit{SUP category flags (9 boolean columns)}} \\
\textit{cat\_luftballons}           & bool & Balloons (\texteuro{}450/t) \\
\textit{cat\_tabak}                 & bool & Tobacco filters (\texteuro{}450/t) \\
\textit{cat\_becher}                & bool & To-go cups (\texteuro{}225/t) \\
\textit{cat\_feuchttuecher}         & bool & Wet wipes (\texteuro{}225/t) \\
\textit{cat\_lebensmittelbehaelter} & bool & Food containers (\texteuro{}225/t) \\
\textit{cat\_tueten\_folien}        & bool & Plastic wrap (\texteuro{}225/t) \\
\textit{cat\_flaschen\_ohne\_pfand} & bool & Non-dep.\ bottles (\texteuro{}225/t) \\
\textit{cat\_flaschen\_mit\_pfand}  & bool & Dep.\ bottles (\texteuro{}225/t) \\
\textit{cat\_plastiktueten}         & bool & Plastic bags (\texteuro{}225/t) \\
\bottomrule
\end{tabularx}}
\noindent\footnotesize{Notes: Estimation window: 43{,}371 rows; 2053 treated
  and 500 control units; 666 products; 231 retailers.
  \textit{ln\_price} is constant within units; all analyses use
  calendar-month fixed effects.  Marketplace classification
  (Section \ref{subsec:seller_type}): \textit{retailer\_id} contains one of am-at, am-de,
  am-uk, eb-uk, eb-de, sh-at, mp-de, rk-de, nk-pl, sz-uk, vk-de,
  gx-de; 203 treated units, 66 retailers.}
\end{table}

\vspace{3pt}\noindent\textit{Section \ref{subsec:cat_het} - Category-level heterogeneity and fee-tier test.} \vspace{3pt}

The joint model (5) is estimated on category-$c$ treated units against the
full control group. Category-specific coefficients $\hat{\beta}_c$ and the
fee-tier comparison (main text Figures \ref{fig:A1a}--\ref{fig:A1c}) are discussed fully in the main
text.  The \texteuro{}450/t tier (balloons, tobacco filters) averages 0.403
log points against $-0.023$ for the \texteuro{}225/t group; the tier
difference is 0.427 ($p=0.035$).

\vspace{3pt}\noindent{Section \ref{subsec:balanced} Balanced-panel robustness.}\vspace{3pt}

The sample is restricted to units observed in at least 25 of the estimation
months.  The payment-due coefficient rises from 0.394 (unbalanced) to 0.541
(balanced), arguing against compositional exit as the primary driver of the
baseline result (main text Figure \ref{fig:A2}).

\vspace{3pt}\noindent{Section \ref{subsec:seller_type} Seller-type heterogeneity.}\vspace{3pt}

The panel is split into marketplace-based sellers (203 treated units) and
standalone e-tailers (1850 treated units).  The payment-due coefficient
is 0.494 for standalone sellers and 0.080 for marketplace sellers, pointing
to differences in compliance cost structure or pricing technology (main text
Figure \ref{fig:A3}).

\vspace{3pt}\noindent{Section \ref{subsec:outcomes} Alternative outcome variables and economic
  magnitude.}\vspace{3pt}
  
Three-period estimates are reported under log price, EUR-level
duration-weighted price, and EUR-level unweighted price.  The qualitative
timing pattern---payment-phase dominant---holds across all three outcomes
(main text Figures \ref{fig:A4a} and \ref{fig:A4b}).

\vspace{3pt}\noindent{Section \ref{subsec:demand} Demand-side proxies.}\vspace{3pt}

Event-study coefficients for $\ln n_{\text{spells},it}$ and
$\ln\mathit{days}_{it}$ both rise post-payment, inconsistent with demand
contraction.  Adding the spell-count control leaves the payment-due price
coefficient unchanged (0.397 vs.\ 0.394), confirming that the estimated
effect reflects within-unit repricing (main text Figures \ref{fig:A5a} and \ref{fig:A5b}).

\end{document}